%% file: axv.tex
\documentclass[12pt,epsf]{article}
\usepackage{amsmath}
\usepackage{amsfonts}
\usepackage{amssymb}
\usepackage{graphicx}

\numberwithin{equation}{subsection}

\newcommand {\dfn} {\stackrel{\Delta} {=}}
\newcommand{\dd}{\mbox{d}}
\newcommand {\exe} {\stackrel{\cdot} {=}}

\newcommand {\reals} {{\rm I\!R}}

\newcommand {\bk} {\mbox{\boldmath $k$}}

\newcommand {\bp} {\mbox{\boldmath $p$}}
\newcommand {\bq} {\mbox{\boldmath $q$}}
\newcommand {\br} {\mbox{\boldmath $r$}}
\newcommand {\bs} {\mbox{\boldmath $s$}}

\newcommand {\bx} {\mbox{\boldmath $x$}}

\newcommand {\bD} {\mbox{\boldmath $D$}}
\newcommand {\bE} {\mbox{\boldmath $E$}}

\newcommand {\bJ} {\mbox{\boldmath $J$}}

\newcommand {\bN} {\mbox{\boldmath $N$}}

\newcommand {\bP} {\mbox{\boldmath $P$}}

\newcommand {\bX} {\mbox{\boldmath $X$}}

\newcommand{\calE}{{\cal E}}
\newcommand{\calF}{{\cal F}}

\newcommand{\calL}{{\cal L}}

\newcommand{\calN}{{\cal N}}

\newcommand{\calX}{{\cal X}}

\allowdisplaybreaks

\begin{document}
\thispagestyle{empty}
\title{Statistical Physics: A Short Course for Electrical Engineering Students}
\author{Neri Merhav}
\date{}
\maketitle

\thispagestyle{empty}

\newpage

\begin{center}
{\bf \Large Statistical Physics: A Short Course for Electrical Engineering Students}

\vspace{2cm}

Neri Merhav\\
Department of Electrical Engineering \\
Technion - Israel Institute of Technology \\
Technion City, Haifa 32000, ISRAEL \\
{\tt merhav@ee.technion.ac.il}
\end{center}
\vspace{1.5\baselineskip}
\setlength{\baselineskip}{1.5\baselineskip}

\begin{abstract}
This is a set of lecture notes of a course on statistical
physics and thermodynamics, which is oriented, to a certain extent, towards
electrical engineering students.
The main body of the lectures is devoted to statistical physics, whereas
much less emphasis is given to the thermodynamics part. In particular, the
idea is to let the most important results of thermodynamics (most notably, the laws of
thermodynamics) to be obtained as conclusions from the derivations in
statistical physics. Beyond the variety of central topics in statistical physics that are
important to the general scientific education of the EE student,
special emphasis is devoted to subjects that are vital to
the engineering education concretely. These include, first of all, quantum statistics,
like the Fermi--Dirac distribution, as well as diffusion processes, 
which are both fundamental for deep
understanding of semiconductor
devices. Another important issue for the EE student is to understand
mechanisms of noise generation and stochastic dynamics 
in physical systems, most notably, in electric
circuitry. Accordingly, the fluctuation--dissipation theorem of statistical
mechanics, which is the theoretical basis for understanding thermal noise
processes in systems,
is presented from a signals--and--systems point of view, in a way that
would hopefully be understandable and useful for an engineering student, and
well connected to other courses in the electrcial engineering curriculum like 
courses on random priocesses.
The quantum regime, in this context, is important too and 
hence provided as well. Finally, we
touch very briefly upon some relationships between statistical mechanics and information
theory, which is the theoretical basis for communications engineering, and
demonstrate how statistical--mechanical approach can be useful in order for
the study of information--theoretic problems. These relationships are further 
explored, and in a much deeper manner, in my previously posted arXiv monograph,
entitled: ``Information Theory and Statistical Physics -- Lecture Notes''
(http://arxiv.org/pdf/1006.1565.pdf).
\end{abstract}

\newpage
\tableofcontents

\newpage
\section{Introduction}

Statistical physics is a branch in physics which deals with systems with a
huge number of particles (or any other elementary units). For example,
{\it Avogadro's number}, which is about $6\times 10^{23}$, is the
number of molecules in 22.4 liters of ideal gas at
standard temperature and pressure.
Evidently, when it comes to systems with such an enormously
large number of particles, there is no hope to keep track of the physical
state (e.g., position and momentum) of each and every individual particle by
means of the classical methods in physics, that is, by solving a gigantic
system of differential equations pertaining to Newton's laws for all particles.
Moreover, even if these differential equations could have been solved somehow
(at least approximately), the information that they would give us would be virtually
useless. What we normally really 
want to know about our physical system boils
down to a fairly short list of {\it macroscopic} parameters, such as energy, heat,
pressure, temperature, volume, magnetization, and the like. In other words, while we
continue to believe in the good old laws of physics that we have known for
some time, even the classical ones,
we no longer use them in the ordinary way that we are familiar with from
elementary physics courses. Instead, we think of the state of the system, at any given
moment, as a
realization of a certain {\it probabilistic ensemble}. This is to say that we
approach the problem from a probabilistic (or a statistical) point of view.
The beauty of statistical physics is that it derives the {\it macroscopic}
theory of thermodynamics
(i.e., the relationships between thermodynamical potentials, temperature,
pressure, etc.) as {\it ensemble averages} that stem from this probabilistic
{\it microscopic} theory, in the limit of
an infinite number of particles, that is, the {\it thermodynamic limit}.

The purpose of this set of lecture notes is to teach statistical
mechanics and thermodynamics, with some degree of orientation towards students in electrical
engineering. The main body of the lectures is devoted to statistical mechanics, whereas
much less emphasis is given to the thermodynamics part. In particular, the
idea is to let the most important results of thermodynamics (most notably, the
laws of thermodynamics) to be obtained as conclusions from the derivations in
statistical mechanics. 

Beyond the variety of central topics in statistical physics that are
important to the general scientific education of the EE student,
special emphasis is devoted to subjects that are vital to
the engineering education concretely.
These include, first of all, quantum
statistics, like the Fermi--Dirac distribution, 
as well as diffusion processes,
which are both fundamental for understanding
semiconductor devices. Another important issue for the EE student is to understand
mechanisms of noise generation and stochastic dynamics 
in physical systems, most notably, in electric
circuitry. Accordingly, the fluctuation--dissipation theorem of statistical
mechanics, which is the theoretical basis for understanding thermal noise
processes and physical systems,
is presented from the standpoint of a system with an input and output, and in a way that
would hopefully be understandable and useful for an engineer, and well related
to other courses in the undergraduate curriculum, like courses in random
processes. This
engineering perspective is typically not available in standard physics textbooks.
The quantum regime, in this context, is important too and hence provided as
well. Finally, we
touch upon some relationships between statistical mechanics and information
theory, which is the theoretical basis for communications engineering, and
demonstrate how statistical--mechanical approach can be useful in order for
the study of information--theoretic problems.
These relationships are further 
explored, and in a much deeper manner, in a previous arXiv paper, entitled:
``Information Theory and Statistical Physics -- Lecture Notes,''
(http://arxiv.org/pdf/1006.1565.pdf).

The reader is assumed to have prior background 
in elementary quantum mechanics and in random processes, both in undergraduate
level. The lecture notes include 
fairly many examples, exercises and figures, which will
hopefully help the student to grasp the material better.
Most of the material in this set of lecture notes is based on well--known,
classical textbooks
(see bibliography), but some of the derivations are original.
Chapters and sections marked by asterisks can be skipped without loss of
continuity.

\newpage
\section{Kinetic Theory and the Maxwell Distribution}

The concept that a gas consists of many small mobile mass particles
is very old -- it dates back to the Greek philosophers. 
It has been periodically rejected and revived throughout many generations of
the history of science. Around the middle of the 19--th century, against the
general trend of rejecting the atomistic approach, Clausius,\footnote{
Rudolf Julius Emanuel Clausius (1822--1888)
was a German physicist and mathematician who is considered
one of the central pioneers of thermodynamics.}
Maxwell\footnote{James Clerk Maxwell (1831--1879)
was a Scottish physicist and mathematician, whose other
prominent achievement was formulating classical electromagnetic theory.}
and Boltzmann\footnote{Ludwig Eduard Boltzmann (1844--1906)
was an Austrian physicist, who has founded
contributions in statistical mechanics and 
thermodynamics. He was one of the advocators of the atomic theory
when it was still very controversial.} succeeded to develop a kinetic theory
for the motion of gas molecules, which was mathematically solid, on the one
hand, and had good agreement with the experimental evidence (at least in
simple cases), on the other hand.

In this part, we will present some elements of Maxwell's formalism
and derivation that builds the kinetic theory of the ideal gas.
It derives some rather useful results from first
principles. While the main results that we shall see in this section can be
viewed as a special case of the more general concepts and principles that will
be provided later on, the purpose here is to give 
a quick snapshot on the taste of this matter and
to demonstrate how the statistical approach to physics, which is based on very
few reasonable assumptions, gives rise to rather far--reaching results and conclusions.

The choice of the ideal gas, as a system of many mobile particles, is a good 
testbed to begin with, as on the one hand, it is simple, and on the
other hand, it is not irrelevant to
electrical engineering and electronics in particular. For example, the free
electrons in a metal can often be considered a ``gas'' (albeit not an ideal
gas), as we shall see later on.

\subsection{The Statistical Nature of the Ideal Gas}

From the statistical--mechanical perspective, and ideal gas is a system of
mobile particles, which interact with one another only via elastic collisions,
whose duration is extremely short compared to the time elapsed between two
consecutive collisions in which a given particle is involved. This basic
assumption is valid as long as the gas is not too dense and the pressure that it
exerts is not too high. As explained in the Introduction, the underlying idea of statistical mechanics in
general, is that instead of hopelessly trying to keep track of the motion of
each individual molecule,
using differential equations that are based on Newton's laws, one treats the
population of molecules as a statistical ensemble using tools from probability
theory, hence the name statistical mechanics (or statistical physics).

What is the probability distribution of the state of the molecules of an
ideal gas in equilibrium? Here, by ``state'' we refer to the positions and the
velocities (or momenta) of all molecules at any given time. As for the
positions, if gravity is neglected, and assuming that the gas is contained in
a given box (container) of volume $V$, there is no apparent reason to believe
that one region is preferable over others, so the distribution of the
locations is assumed uniform across the container, and independently of one
another. Thus, if there are $N$ molecules, the joint probability 
density of their
positions is $1/V^N$ everywhere within the container and zero outside.
It is therefore natural to define the density of particles per unit volume as
$\rho=N/V$.

What about the distribution of velocities? This is slightly more involved,
but as we shall see, still rather simple, and the interesting point is that
once we derive this distribution, we will be able to derive some interesting
relationships between macroscopic quantities pertaining to the equilibrium
state of the system (pressure, density, energy, temperature, etc.).
As for the velocity of each particle, we will make two assumptions:
\begin{enumerate}
\item All possible directions of motion in space are equally likely. In other words,
there are no preferred directions (as gravity is neglected). Thus, the
probability density function (pdf) of the velocity vector
$\vec{v}=v_x\hat{x}+v_y\hat{y}+v_z\hat{z}$ depends on $\vec{v}$ only via its
magnitude, i.e., the speed 
$s=\|\vec{v}\|=\sqrt{v_x^2+v_y^2+v_z^2}$, or in mathematical terms:
\begin{equation}
f(v_x,v_y,v_z)=g(v_x^2+v_y^2+v_z^2)
\end{equation}
for some function $g$.
\item The various components $v_x$, $v_y$ and $v_z$ are identically
distributed and independent, i.e.,
\begin{equation}
f(v_x,v_y,v_z)=f(v_x)f(v_y)f(v_z).
\end{equation}
The rationale behind identical distributions is, like in item 1 above,
namely, the isotropic nature of the pdf.
The rationale behind independence 
is that in each collision between two particles, the total
momentum is conserved and in each component ($x$, $y$, and $z$) separately, 
so there are actually no interactions among the component momenta. Each
three--dimensional particle actually behaves like three independent
one--dimensional particles, as far as the momentum or velocity is concerned.
\end{enumerate}
We now argue that there is only one kind 
of (differentiable) joint pdf
$f(v_x,v_y,v_z)$ that complies with both assumptions at the same time, and
this is the Gaussian density where all three components of $\vec{v}$ are
independent, zero--mean and with the same variance.

To see why this is true, consider the equation
\begin{equation}
f(v_x)f(v_y)f(v_z)=g(v_x^2+v_y^2+v_z^2)
\end{equation}
which combines both requirements.
Let us assume that both $f$ and $g$ are differentiable. 
First, observe that this equality already tells that $f(v_x)=f(-v_x)$,
namely, $f(v_x)$ depends on $v_x$ only via $v_x^2$, and obviously, the same
comment applies to $v_y$ and $v_z$. Let us denote then
$\hat{f}(v_i^2)=f(v_i)$, $i\in\{x,y,z\}$. Further, let us temporarily denote
$u_x=v_x^2$, $u_y=v_y^2$ and $u_z=v_z^2$. The last equation then reads
\begin{equation}
\hat{f}(u_x)\hat{f}(u_y)\hat{f}(u_z)=g(u_x+u_y+u_z).
\end{equation}
Taking now partial derivatives w.r.t.\ $u_x$, $u_y$ and $u_z$, we obtain
\begin{equation}
\hat{f}'(u_x)\hat{f}(u_y)\hat{f}(u_z)=
\hat{f}(u_x)\hat{f}'(u_y)\hat{f}(u_z)=
\hat{f}(u_x)\hat{f}(u_y)\hat{f}'(u_z)=
g'(u_x+u_y+u_z).
\end{equation}
The first two equalities imply that
\begin{equation}
\frac{\hat{f}'(u_x)}{\hat{f}(u_x)}=
\frac{\hat{f}'(u_y)}{\hat{f}(u_y)}=
\frac{\hat{f}'(u_z)}{\hat{f}(u_z)}
\end{equation}
for all $u_x$, $u_y$ and $u_z$,
which means that $\hat{f}'(u)/\hat{f}(u)\equiv
\mbox{d}[\ln\hat{f}(u)]/\mbox{d}u$ must be a constant.
Let us denote this constant by $-\alpha$. Then,
\begin{equation}
\frac{\mbox{d}[\ln\hat{f}(u)]}{\mbox{d}u}=-\alpha
\end{equation}
implies that
\begin{equation}
\ln\hat{f}(u)=\beta-\alpha u
\end{equation}
where $\beta$ is a constant of integration, i.e.,
\begin{equation}
\hat{f}(u)=B\cdot e^{-\alpha u},
\end{equation}
where $B=e^\beta$. Returning to the original variables,
\begin{equation}
f(v_x)=Be^{-\alpha v_x^2},
\end{equation}
and similar relations for $v_y$ and $v_z$.
For $f$ to be a valid pdf, $\alpha$ must be positive and $B$
must be the appropriate constant of normalization, which gives
\begin{equation}
f(v_x)=\sqrt{\frac{\alpha}{\pi}}e^{-\alpha v_x^2}
\end{equation}
and the same applies to $v_y$ and $v_z$. Thus,
we finally obtain
\begin{equation}
f(v_x,v_y,v_z)=\left(\frac{\alpha}{\pi}\right)^{3/2}
e^{-\alpha(v_x^2+v_y^2+v_z^2)}
\end{equation}
and it only remains to determine the constant $\alpha$.

To this end, we adopt the following consideration.
Assume, without essential loss of 
generality, that the container is a box of sizes $L_x\times L_y\times L_z$,
whose walls are parallel to the axes of the coordinate system.
Consider a molecule with velocity $\vec{v}
=v_x\hat{x}+v_y\hat{y}+v_z\hat{z}$ hitting a wall parallel to the
$Y-Z$ plane from its left side. The molecule is elastically reflected
with a new velocity vector $\vec{v'}
=-v_x\hat{x}+v_y\hat{y}+v_z\hat{z}$, and so, the change in momentum,
which is also the impulse $F_x\tau$ 
that the molecule exerts on the wall, is  $\Delta p=2mv_x$, where
$m$ is the mass of the molecule.
For a molecule of velocity $v_x$ in the $x$--direction to hit the wall within
time duration $\tau$, its distance from the wall must 
not exceed $v_x\tau$ in the $x$--direction. Thus, the total average impulse
contributed by all molecules with an $x$--component velocity ranging between
$v_x$ and $v_x+\mbox{d}v_x$, is given by
$$2mv_x\cdot N\cdot\frac{v_x\tau}{L_x}\cdot\sqrt{\frac{\alpha}{\pi}}
e^{-\alpha v_x^2}\mbox{d}v_x.$$
Thus, the total impulse exerted within time $\tau$ is the integral,
given by
$$\frac{2mN\tau}{L_x}\cdot\sqrt{\frac{\alpha}{\pi}}\int_0^\infty
v_x^2 e^{-\alpha v_x^2}\mbox{d}v_x=\frac{mN\tau}{2\alpha L_x}.$$
The total force exerted on the $Y-Z$ wall is then
$mN/(2\alpha L_x)$, and so the pressure is
\begin{equation}
P=\frac{mN}{2\alpha L_xL_yL_z}=\frac{mN}{2\alpha V}=\frac{\rho m}{2\alpha}
\end{equation}
and so, we can determine $\alpha$ in terms of the physical quantities
$P$ and $\rho$ and $m$:
\begin{equation}
\alpha=\frac{\rho m}{2P}.
\end{equation}
From the equation of state of the ideal gas\footnote{Consider 
this as an experimental fact.}
\begin{equation}
P=\rho kT
\end{equation}
where $k$ is Boltzmann's constant ($=1.381\times 10^{-23}$ Joules/degree)
and $T$ is absolute temperature. Thus,
an alternative expression for $\alpha$ is:
\begin{equation}
\alpha=\frac{m}{2kT}.
\end{equation}
On substituting this into the general Gaussian form of the pdf, we finally obtain
\begin{equation}
f(\vec{v})=\left(\frac{m}{2\pi kT}\right)^{3/2}\exp\left[-\frac{m}{2kT}
(v_x^2+v_y^2+v_z^2)\right]=
\left(\frac{m}{2\pi kT}\right)^{3/2}\exp\left[-\frac{\epsilon}{kT}\right],
\end{equation}
where $\epsilon$ is the (kinetic) energy of the molecule.
This form of a pdf, that is proportional to $e^{-\epsilon/(kT)}$, 
where $\epsilon$ is the energy, is
not a coincidence. 
We shall see it again and again later on, and in much greater generality,
as a fact that stems from much deeper and more fundamental principles.
It is called the {\it Boltzmann--Gibbs distribution}.

Having derived the pdf of $\vec{v}$, we can now calculate a few moments. 
Throughout this course, we will denote the expectation operator by
$\left<\cdot\right>$, which is the customary notation used by physicists.
Since
\begin{equation}
\left<v_x^2\right>=
\left<v_y^2\right>=
\left<v_z^2\right>=\frac{kT}{m}
\end{equation}
we readily have
\begin{equation}
\left<\|\vec{v}\|^2\right>=\left<v_x^2+v_y^2+v_z^2\right>=\frac{3kT}{m},
\end{equation}
and so the root mean square (RMS) speed is given by
\begin{equation}
s_{RMS}\dfn\sqrt{\left<\|\vec{v}\|^2\right>}=\sqrt{\frac{3kT}{m}}.
\end{equation}
Other related statistical quantities, 
that can be derived from $f(\vec{v})$, 
are the average speed $\left<s\right>$ and
the most likely speed. Like $s_{RMS}$, they are also proportional to
$\sqrt{kT/m}$ but with different constants of proportionality (see Exercise
below). The
average kinetic energy per molecule is
\begin{equation}
\left<\epsilon\right>=\left<\frac{1}{2}m\|\vec{v}\|^2\right>=\frac{3kT}{2}
\end{equation}
independent of $m$. 
This relation gives to the notion of temperature its basic significance:
at least in the case of the ideal gas, temperature is simply a quantity that
is directly proportional to the average kinetic energy of each particle.
In other words, temperature and kinetic energy are almost synonyms in this
case. In the sequel, we will see a more general definition of temperature.
The factor of 3 at the numerator is due to the fact
that space has three dimensions, and so, each molecule has 3 degrees of
freedom. Every degree of freedom contributes 
an amount of energy given by $kT/2$.
This will turn out later to be a special case of a more general
principle called the {\it equipartition of energy}.

\begin{figure}[h!t!b!]
\centering
\includegraphics[width=8.5cm, height=8.5cm]{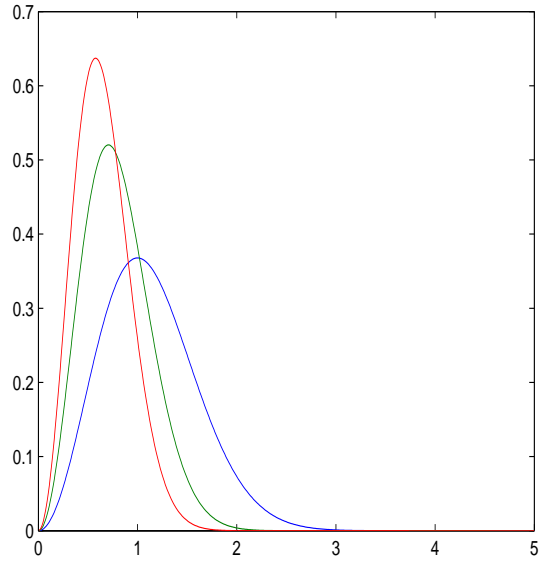}
\caption{\small Demonstration of the Maxwell distribution for various values of
the parameter $kT/m$. The red curve (tall and narrow) 
corresponds to smallest value and the blue
curve (short and wide) -- to the highest value.}
\label{maxwell}
\end{figure}

The pdf of the speed $s=\|\vec{v}\|$ 
can be derived from the pdf of the velocity
$\vec{v}$ using the obvious consideration 
that all vectors $\vec{v}$ of the same
norm correspond to the same speed. Thus, the pdf of $s$ is simply
the pdf of $\vec{v}$ (which depends 
solely on $\|\vec{v}\|=s$) multiplied by the
surface area of a three--dimensional 
sphere of radius $s$, which is $4\pi s^2$, i.e.,
\begin{equation}
f(s)=4\pi s^2 
\left(\frac{m}{2\pi kT}\right)^{3/2}e^{-ms^2/(2kT)}=
\sqrt{\frac{2}{\pi}\left(\frac{m}{kT}\right)^3}\cdot s^2e^{-ms^2/(2kT)}
\end{equation}
This is called the {\it Maxwell distribution} and it is depicted in Fig.\
\ref{maxwell} for various values of the parameter $kT/m$.
To obtain the 
pdf of the energy $\epsilon$, we should change variables according to
$s=\sqrt{2\epsilon/m}$ and $\mbox{d}s=\mbox{d}\epsilon/\sqrt{2m\epsilon}$. 
The result is
\begin{equation}
f(\epsilon)=
\frac{2\sqrt{\epsilon}}{\sqrt{\pi}(kT)^{3/2}}\cdot e^{-\epsilon/(kT)}.
\end{equation}

\vspace{0.2cm}

\noindent
{\small {\it Exercise 2.1:} Use the above to calculate: (i) the average
speed $\left<s\right>$, (ii) the most likely speed $\mbox{argmax}_s f(s)$, and
(iii) the most likely energy $\mbox{argmax}_\epsilon f(\epsilon)$.}

\vspace{0.2cm}

An interesting relation, that will be referred to later on, links between the
average energy per particle $\bar{\epsilon}=\left<\epsilon\right>$, the
density $\rho$,
and the pressure $P$, or equivalently, 
the total energy $E=N\bar{\epsilon}$, the volume $V$
and $P$:
\begin{equation}
P=\rho kT=\frac{2\rho}{3}\cdot\frac{3kT}{2}=
\frac{2\rho}{3}\cdot\bar{\epsilon}.
\end{equation}
which after multiplying by $V$ becomes
\begin{equation}
PV=\frac{2E}{3}.
\end{equation}
It is interesting to note that this relation can be obtained 
directly from the analysis of
the impulse exerted by the particles on the walls, similarly as in the earlier
derivation of the parameter $\alpha$, and without recourse to the equation of
state (see, for example, \cite[Sect.\ 20--4, pp.\ 353--355]{SZY76}). This is
because the parameter $\alpha$ of the Gaussian pdf of each component of
$\vec{v}$ has the obvious meaning of $1/(2\sigma_v^2)$, where
$\sigma_v^2$ is the common variance of each component of $\vec{v}$. 
Thus, $\sigma_v^2=1/(2\alpha)$ and so,
$\left<\|\vec{v}\|^2\right>=3\sigma_v^2=3/(2\alpha)$, which in turn
implies that 
\begin{equation}
\bar{\epsilon}=\left<\frac{m}{2}\|\vec{v}\|^2\right>=\frac{3m}{4\alpha}
=\frac{3m}{4\rho m/(2P)}=\frac{3P}{2\rho},
\end{equation}
which is equivalent to the above.

\subsection{Collisions}

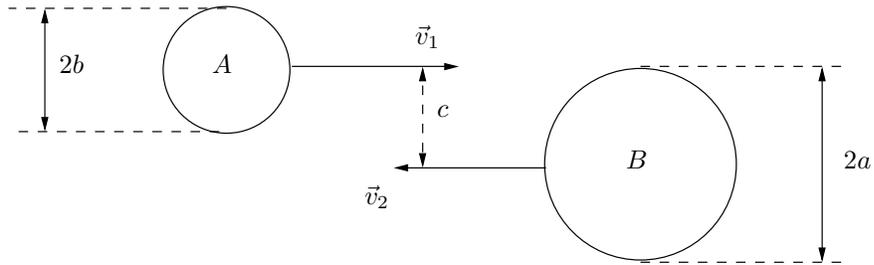
\begin{figure}[ht]
\hspace*{3cm}\input{collision.pstex_t}
\caption{\small Hard sphere collision.}
\label{collision}
\end{figure}

We now take a closer look into the issue of collisions.
We first define the concept of {\it collision cross--section}, which we denote
by $\sigma$. Referring to Fig.\ \ref{collision}, consider a situation, where
two hard spheres, labeled $A$ and $B$, with diameters $2a$ and $2b$,
respectively, are approaching each other, and
let $c$ be the projection of the distance between their centers in the direction
perpendicular to the direction of their relative motion, $\vec{v}_1-\vec{v}_2$.
Clearly, collision will occur if and only if $c < a+b$.
In other words, the two spheres would collide only if the center of $B$
lies in inside a volume 
whose cross sectional area is $\sigma=\pi(a+b)^2$, or for identical spheres,
$\sigma=4\pi a^2$. Let the colliding particles have relative velocity
$\Delta\vec{v}=\vec{v}_1-\vec{v}_2$. Passing to the coordinate system of the
center of mass of the two particles, this is equivalent to the motion of one
particle with the reduced mass $\mu=m_1m_2/(m_1+m_2)$, and so, in the case
of identical particles, $\mu=m/2$. The average relative speed is easily calculated
from the Maxwell distribution, but with $m$ being replaced by $\mu=m/2$, i.e.,
\begin{equation}
\left<\|\Delta\vec{v}\|\right>=
4\pi\left(\frac{m}{4\pi
kT}\right)^{3/2}\int_0^{\infty}(\Delta v)^3e^{-m(\Delta
v)^2/(4kT)}\mbox{d}(\Delta v)=
4\cdot\sqrt{\frac{kT}{\pi m}}=\sqrt{2}\left<s\right>.
\end{equation}
The total number of particles per unit volume that collide with a particular
particle within time $\tau$ is
\begin{equation}
N_{col}(\tau)=\rho\sigma\left<\|\Delta\vec{v}\|\right>\tau=4\rho\sigma\tau\sqrt{\frac{kT}{\pi
m}}
\end{equation}
and so, the collision rate of each particle is
\begin{equation}
\nu=4\rho\sigma\sqrt{\frac{kT}{\pi m}}.
\end{equation}
The mean distance between collisions (a.k.a.\ the mean free path) is therefore
\begin{equation}
\lambda=\frac{\left<\|\vec{v}\|\right>}{\nu}=\frac{1}{\sqrt{2}\rho\sigma}=
\frac{kT}{\sqrt{2}P\sigma}.
\end{equation}

What is the probability distribution of 
the random distance $L$ between two consecutive
collisions of a given particle? In particular, what is $p(l)\dfn\mbox{Pr}\{L
\ge l\}$? Let us assume that the collision process is memoryless in the
sense that the event of not colliding before distance $l_1+l_2$
is the intersection of two {\it independent} events, the first one being
the event of not colliding before distance $l_1$, and the second one being
the event of not colliding before the additional distance $l_2$. That is
\begin{equation}
p(l_1+l_2)=p(l_1)p(l_2). 
\end{equation}
We argue that 
under this assumption, $p(l)$ must be exponential in $l$.
This follows from the following 
consideration.\footnote{Similar idea to the one of the earlier derivation
of the Gaussian pdf of the ideal gas.}
Taking partial derivatives of both sides w.r.t.\ both $l_1$ and $l_2$, we get
\begin{equation}
p'(l_1+l_2)=p'(l_1)p(l_2)=p(l_1)p'(l_2).
\end{equation}
Thus,
\begin{equation}
\frac{p'(l_1)}{p(l_1)}=\frac{p'(l_2)}{p(l_2)}
\end{equation}
for all non-negative $l_1$ and $l_2$. 
Thus, $p'(l)/p(l)$ must be a constant, which
we shall denote by $-a$. This trivial differential equation has only
one solution which obeys with the obvious initial condition 
$p(0)=1$: 
\begin{equation}
p(l)=e^{-al}~~~~l\ge 0
\end{equation}
so it only remains to determine the parameter $a$, which must be positive
since the function $p(l)$ must be monotonically non--increasing by definition.
This can easily found by using the fact that $\left<L\right>=1/a=\lambda$,
and so,
\begin{equation}
p(l)=e^{-l/\lambda}=\exp\left(-\frac{\sqrt{2}P\sigma l}{kT}\right).
\end{equation}

\subsection{Dynamical Aspects}

The discussion thus far focused on the static (equilibrium) behavior of the
ideal gas. In this subsection, we will briefly touch upon dynamical issues
pertaining to non--equilibrium situations. These issues will be further
developed in the second part of the course, and with much greater generality.

\begin{figure}[ht]
\hspace*{4cm}\input{leak.pstex_t}
\caption{\small Gas leakage through a small hole.}
\label{leakage}
\end{figure}
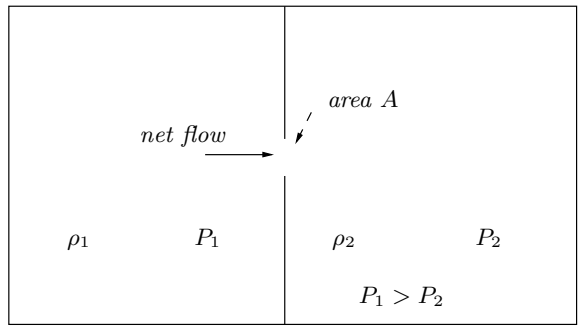

Consider two adjacent containers separated by a wall. Both of them
have the same volume $V$, they both contain the same ideal gas at the
same temperature $T$, but with
different densities $\rho_1$ and $\rho_2$, 
and hence different pressures $P_1$ and
$P_2$. Let us assume that $P_1 > P_2$. At time $t=0$, a small hole is
generated in the separating wall. The area of this hole is $A$ (see Fig.\
\ref{leakage}). 

If the mean free distances $\lambda_1$ and $\lambda_2$ are relatively large
compared to the dimensions of the hole, it is safe to assume that
every molecule that reaches the hole, passes through it.
The mean number of molecules that pass from left to right within time $\tau$ is
given by
\begin{equation}
N_{\rightarrow}
=\rho_1V\cdot\int_0^\infty\mbox{d}v_x\sqrt{\frac{\alpha}{\pi}}e^{-\alpha
v_x^2}\cdot \frac{v_x\tau
A}{V}=\frac{\rho_1\tau A}{2\sqrt{\pi\alpha}}
\end{equation}
and so the number of particles per second, flowing from left to right is
\begin{equation}
\frac{\mbox{d}N_{\rightarrow}}{\mbox{d}t}=\frac{\rho_1 A}{2\sqrt{\pi\alpha}}.
\end{equation}
Similarly, in the opposite direction, we have
\begin{equation}
\frac{\mbox{d}N_{\leftarrow}}{\mbox{d}t}=\frac{\rho_2 A}{2\sqrt{\pi\alpha}},
\end{equation}
and so, the net left--to--right current is
\begin{equation}
I\dfn \frac{\mbox{d}N}{\mbox{d}t}=\frac{(\rho_1-\rho_2)A}
{2\sqrt{\pi\alpha}}=(\rho_1-\rho_2)A\sqrt{\frac{kT}{2\pi m}}.
\end{equation}
An important point here is that the current is proportional to the difference
between densities $(\rho_1-\rho_2)$, and considering the equation of state of the
ideal gas, it is therefore also proportional to the pressure difference, $(P_1-P_2)$.
This rings the bell of the well known analogous 
fact that the electric current is
proportional to the voltage, which in turn is the difference
between the electric potentials at two points.
Considering the fact that $\rho\dfn(\rho_1+\rho_2)/2$ is constant, we obtain a simple
differential equation
\begin{equation}
\frac{\mbox{d}\rho_1}{\mbox{d}t}= (\rho_2-\rho_1)\frac{A}{V}\sqrt{\frac{kT}{2\pi
m}}\dfn C(\rho_2-\rho_1)=2C(\rho-\rho_1)
\end{equation}
whose solution is
\begin{equation}
\rho_1(t)=\rho+[\rho_1(0)-\rho]e^{-2Ct}
\end{equation}
which means that equilibrium is approached exponentially fast with time
constant
\begin{equation}
\tau=\frac{1}{2C}=\frac{V}{2A}\sqrt{\frac{2\pi m}{kT}}.
\end{equation}
Imagine now a situation, where there is a long pipe aligned along the
$x$--direction. The pipe is divided into a chain of cells in a linear fashion.
and in the wall between each two consecutive cells there is a hole
with area $A$. The length of each cell (i.e., the distance between consecutive
walls) is the mean free distance $\lambda$, so that collisions within
each cell can be neglected. Assume further that $\lambda$ is so small that the
density of each cell at time $t$ can be approximated using a continuous
function $\rho(x,t)$.
Let $x_0$ be the location of one of the walls. Then, according to the above
derivation, the current at $x=x_0$ is
\begin{eqnarray}
\label{J1}
I(x_0)&=&\left[\rho\left(x_0-\frac{\lambda}{2},t\right)-
\rho\left(x_0+\frac{\lambda}{2},t\right)\right]A\sqrt{\frac{kT}{2\pi
m}}\nonumber\\
&\approx&-A\lambda\sqrt{\frac{kT}{2\pi
m}}\cdot\frac{\partial \rho(x,t)}{\partial x}\bigg|_{x=x_0}.
\end{eqnarray}
Thus, the current is proportional to the negative gradient of the density.
This is quite a fundamental result which holds with much greater generality. 
In the more general context, it is known as {\it Fick's law}.

Consider next two close points $x_0$ and $x_0+\mbox{d}x$, with possibly
different current densities (i.e., 
currents per unit area) $J(x_0)$ and $J(x_0+\Delta x)$. The difference
$J(x_0)-J(x_0+\Delta x)$ is the rate at which matter accumulates along the
interval $[x_0,x_0+\Delta x]$ per unit area
in the perpendicular plane. Within $\Delta t$ seconds, the number of particles
per unit area
within this interval has grown by $[J(x_0)-J(x_0+\Delta x)]\Delta t$. But
this amount is also $[\rho(x_0,t+\Delta t)-\rho(x_0,t)]\Delta x$, 
Taking the appropriate limits, we get
\begin{equation}
\label{J2}
\frac{\partial J(x)}{\partial x}=-\frac{\partial \rho(x,t)}{\partial t},
\end{equation}
which is a one--dimensional version of the so called {\it equation of continuity}.
Differentiating now eq.\ (\ref{J1}) w.r.t.\ $x$ and comparing with (\ref{J2}),
we obtain the {\it diffusion equation} (in one dimension):
\begin{equation}
\frac{\partial \rho(x,t)}{\partial t}=D\frac{\partial^2 \rho(x,t)}{\partial x^2}
\end{equation}
where the constant $D$, in this case,
\begin{equation}
D=\frac{A\lambda}{S}\cdot \sqrt{\frac{kT}{2\pi m}},
\end{equation}
which is called the {\it diffusion coefficient}.
Here $S$ is the cross--section area.

This is, of course, a toy model -- it is a caricature of a real diffusion
process, but basically, it captures the essence of it.
Diffusion processes are central in irreversible statistical mechanics, since
the solution to the diffusion equation is sensitive to the sign of time. This
is different from the Newtonian equations of frictionless motion, which have
a time reversal symmetry and hence are reversible. We will touch upon these
issues near the end of the course. 

The equation of continuity, Fick's law, the
diffusion equation and its extension,
the Fokker--Planck equation (which will also be discussed), are all very central
in physics in general and in semiconductor physics, in particular, as
they describe processes of propagation of concentrations of electrons and
holes in semiconductor materials. Another branch of physics where these
equations play an important role is fluid mechanics.

\newpage
\section{Elementary Statistical Physics}

In this chapter, we provide 
the formalism and the elementary background in statistical
physics. We first define the basic postulates of statistical mechanics, and
then define various ensembles. Finally, we shall derive some of the
thermodynamic potentials and their 
properties, as well as the relationships among them.
The important laws of thermodynamics will also be pointed out.

\subsection{Basic Postulates}

As explained in the Introduction, statistical physics is about a probabilistic
approach to systems of many particles.
While our discussion here will no longer be specfic to the ideal gas as
before, we will nonetheless
start again with this example in mind, just
for the sake of concreteness, 
Consider then a system with a very large number $N$ of mobile particles, which
are free to move in a given volume. 
The {\it microscopic state} (or {\it microstate}, for short) of the system, at
each time instant $t$,  consists, in this example,
of the position vector $\vec{r}_i(t)$
and the momentum vector $\vec{p}_i(t)$ of each and every particle,
$1\le i\le N$. Since each one of
these is a vector of three components, the microstate is then given by a
$(6N)$--dimensional vector
$\vec{\bx}(t)=\{(\vec{r}_i(t),\vec{p}_i(t)):~i=1,2,\ldots,N\}$, whose trajectory
along the time axis,
in the {\it phase space} $\reals^{6N}$, is called the {\it phase trajectory}.

Let us assume that the system is closed, i.e., {\it isolated} from its environment, in
the sense that no energy flows inside or out.
Imagine that the phase space $\reals^{6N}$ is partitioned into very small hypercubes
(or cells) $\Delta\vec{p}\times\Delta\vec{r}$. One of the basic postulates of statistical
mechanics is the following: In the very long range, the relative
amount of time at which $\vec{\bx}(t)$ spends within each such cell,
converges to a certain number between $0$ and $1$, which can be given the
meaning of the {\it probability} of this cell. Thus, there is an underlying
assumption of equivalence between temporal averages and ensemble averages,
namely, this is the postulate of {\it ergodicity}. Considerable efforts were
dedicated to the proof of the ergodic hypothesis
at least in some cases.
As reasonable and natural as it may seem, the
ergodic hypothesis should not be taken for granted. It does not hold for every system but only if
no other conservation law holds. For example, the ideal gas
in a box is non--ergodic,
as every particle retains its momentum (assuming perfectly elastic collisions
with the walls).

What are then the probabilities of the above--mentioned 
phase--space cells? We would like to derive these
probabilities from first principles, based on as few as possible
basic postulates. Our second postulate is that for an isolated system
(i.e., whose energy is fixed) all microscopic states $\{\vec{\bx}(t)\}$ are
equiprobable. The rationale behind this postulate is twofold:
\begin{itemize}
\item In the absence of additional information, there is no apparent reason
that certain regions in phase space would have preference relative to any
others.
\item This postulate is in harmony with a basic result in kinetic theory of
gases -- {\it the Liouville theorem}, which we will not touch upon in this
course, but in a nutshell, it asserts that the phase trajectories must lie
along hyper-surfaces of constant probability density.\footnote{This is a result of the
energy conservation law along with the fact that probability mass behaves
like an incompressible fluid in the sense that whatever mass that flows into a certain
region from some direction must be equal to the outgoing flow from some other
direction. This is reflected in the equation of continuity, which was
demonstrated earlier.}
\end{itemize}

\subsection{Statistical Ensembles}

\subsubsection{The Microcanonical Ensemble}

Before we proceed, let us slightly broaden the scope of our discussion.
In a more general context, associated with our $N$--particle physical system, 
is a certain instantaneous microstate, generically denoted by
$\bx=(x_1,x_2,\ldots,x_N)$, where each 
$x_i$, $1\le i\le N$, may itself be a vector of several
physical quantities associated particle number 
$i$, e.g., its position, momentum,
angular momentum, magnetic moment, spin, and so on,
depending on the type and the nature 
of the physical system. For each possible value of
$\bx$, there is a certain
{\it Hamiltonian} (i.e., energy function) that assigns to
$\bx$ a certain energy $\calE(\bx)$.\footnote{For example, in the case of an {\it ideal gas},
$\calE(\bx)=\sum_{i=1}^N\|\vec{p}_i\|^2/(2m)$, where $m$ is the mass of
each molecule,
namely, it accounts for the contribution of the kinetic energies only. 
In more complicated situations, there might be additional contributions of potential
energy, which depend on the positions.} Now, let us denote by $A(E)$ 
the volume of the shell of energy about $E$. This means
\begin{equation}
A(E)=\mbox{Vol}\{\bx:~E\le\calE(\bx)\le
E+\mbox{d}E\}=\int_{\{\bx:~E\le\calE(\bx)\le E+\mbox{d}E\}}\mbox{d}\bx,
\end{equation}
where $\mbox{d}E$ is a very small (but fixed) energy increment, which is
immaterial when $N$ is large.
Then, our above postulate concerning the ensemble of an isolated system, which
is called the {\it microcanonincal ensemble}, is that the probability density
$P(\bx)$ is given by
\begin{equation}
P(\bx)=\left\{\begin{array}{ll}
\frac{1}{A(E)} & E\le \calE(\bx)\le E+
\mbox{d}E\\
0 & \mbox{elsewhere}\end{array}\right.
\end{equation}
In the discrete case, things are simpler, of course: Here, $A(E)$ is
the number of microstates with $\calE(\bx)=E$ (exactly) and $P(\bx)$ is
the uniform probability mass function over this set of states. 

Back to the general case,
we next define the notion of the {\it density of states} $\Omega(E)$, which is
intimately related to $A(E)$, but with a few minor corrections. The first
correction has to do with the fact that
$A(E)$ is, in general, not dimensionless: 
In the above example of a gas, it has the
physical units
of $[\mbox{length}\times\mbox{momentum}]^{3N}=[\mbox{Joule}\cdot\mbox{sec}]^{3N}$, but we must 
eliminate these physical units because very soon we are going to apply non--linear
functions like the logarithmic function. To this end, we 
normalize the volume $A(E)$ by the volume of an elementary 
reference volume. In the gas example, this reference volume
is taken to be $h^{3N}$, where $h$ is {\it Planck's constant} ($h\approx 6.62\times 10^{-34}$ 
Joules$\cdot$sec). Informally, the
intuition comes from the fact that $h$ is our best available ``resolution'' in the plane
spanned by each component of $\vec{r}_i$ and the corresponding component of
$\vec{p}_i$, owing to the {\it uncertainty principle} in quantum mechanics,
which tells that the product of the standard deviations $\Delta p_a\cdot\Delta
r_a$ of each component $a$ ($a=x,y,z$) is lower bounded by $\hbar/2$, where
$\hbar=h/(2\pi)$. More
formally, this reference volume is obtained in a natural manner 
from quantum statistical mechanics: by changing
the integration variable $\vec{p}$ to $\vec{k}$ using the relation
$\vec{p}=\hbar\vec{k}$, where $\vec{k}$ is the wave vector.
This is a well--known relationship 
(one of the de Broglie relationships)
pertaining to particle--wave duality. The second correction that is needed to
pass from $A(E)$ to $\Omega(E)$ is applicable only when the particles are
indistinguishable:\footnote{In the example of 
the ideal gas, since the particles are mobile and since
they have no colors and no
identity certificates, there is no distinction between a state where particle
no.\ 15 has position $\vec{r}$ and momentum $\vec{p}$ while particle no.\ 437
has position $\vec{r}'$ and momentum $\vec{p}'$ and a state where these two
particles are swapped.}
in these cases, we don't consider permutations between
particles in a given configuration as distinct microstates. Thus, we have to
divide also by
$N!$. Thus, taking into account both corrections, we find that in the example of the ideal gas, 
\begin{equation}
\Omega(E)=\frac{A(E)}{N!h^{3N}}.
\end{equation}
Once again, it should be understood that both of these corrections are optional and their
appicability depends on the system in question: 
The first correction is applicable only if $A(E)$ has physical units and the
second corection is applicable only if the particles are indistinguishable. For example, if $\bx$ is
discrete, in which case the integral defining $A(E)$ is replaced by a sum (that
counts $\bx$'s with $\calE(\bx)=E$), and the particles are distinguishable,
then no corrections are needed at all, i.e.,
\begin{equation}
\Omega(E)=A(E).
\end{equation}
Now, the {\it entropy} is
defined as
\begin{equation}
S(E)=k\ln\Omega(E),
\end{equation}
where $k$ is {\it Boltzmann's constant}.
We will see later what is the relationship between $S(E)$ and the
classical thermodynamical entropy, due to Clausius (1850), as well as the
information--theoretic entropy, due to Shannon (1948).
As it will turn out,
all three are equivalent to one another.
Here, a comment on the notation is in order: The entropy $S$ may depend on
additional quantities, other than the energy $E$, like the volume $V$ and the
number of particles $N$. When this dependence will be relevant and important,
we will use the more complete form of notation $S(E,V,N)$. If only the
dependence on $E$ is relevant in a certain context, we use the simpler
notation $S(E)$.

To get some feeling of this, it should be noted that normally,
$\Omega(E)$ behaves as an exponential function of $N$ (at least
asymptotically), and so, $S(E)$ is roughly linear in $N$. For example,
if $\calE(\bx)=\sum_{i=1}^N\frac{\|\vec{p}_i\|^2}{2m}$, then $\Omega(E)$
is the volume of a shell or surface of a $(3N)$--dimensional sphere with
radius $\sqrt{2mE}$, divided by $N!h^{3N}$, which is proportional 
to $(2mE)^{3N/2}V^N/N!h^{3N}$, where $V$ is the volume,
More precisely, we get
\begin{align}
\label{entropyidealgas}
S(E,V,N)&=k\ln\left[\left(\frac{4\pi
mE}{3N}\right)^{3N/2}\cdot\frac{V^N}{N!h^{3N}}\right]+\frac{3}{2}Nk\nonumber\\
&\approx
Nk\ln\left[\left(\frac{4\pi
mE}{3N}\right)^{3/2}\cdot\frac{V}{Nh^{3}}\right]+\frac{5}{2}Nk.
\end{align}
Assuming that $E$ and $V$ are both proportional to $N$ ($E=N\epsilon$ and
$V=N/\rho$), it is readily seen
that $S(E,V,N)$ is also proportional to $N$.
A physical quantity that has a linear dependence on the size of the
system $N$, is called an {\it extensive quantity}. Energy, volume and
entropy are then
extensive quantities. 
Other quantities, which are not extensive, i.e., independent of the
system size, like temperature and pressure, are called {\it intensive}.

It is interesting to point out that from the function 
$S(E,V,N)$, one can obtain the entire information about the relevant
macroscopic physical quantities of the system, e.g., temperature,
pressure, and so on. Specifically,
the {\it temperature} $T$ of the system is defined according to:
\begin{equation}
\frac{1}{T}=\left[\frac{\partial S(E,V,N)}{\partial E}\right]_{V,N}
\end{equation}
where $[\cdot]_{V,N}$ emphasizes that the derivative is taken while keeping
$V$ and $N$ constant.
One may wonder, at this point, what is the justification
for {\it defining} temperature this way. We will get back to this point a
bit later, but for now, we can easily see that this is indeed true
at least for the ideal gas, as by taking the derivative of
(\ref{entropyidealgas})
w.r.t.\ $E$, we get
\begin{equation}
\label{idealgastemp}
\frac{\partial S(E,V,N)}{\partial E}=\frac{3Nk}{2E}=\frac{1}{T},
\end{equation}
where the second equality has been shown already in Chapter 2.

Intuitively, in most situations,
we expect that $S(E)$ would be an increasing
function of $E$ for fixed $V$ and $N$ (although this is not strictly always the case), which means
$T \ge 0$. But $T$ is also expected to be increasing with $E$ (or
equivalently, $E$ is increasing with $T$, as otherwise, the heat capacity
$\dd E/\dd T < 0$). Thus, $1/T$ should decrease with $E$, which means that the
increase of $S$ in $E$ slows down as $E$ grows. In other words, we expect
$S(E)$ to be a concave function of $E$. In the above example, indeed, $S(E)$
is logarithmic in $E$ and
$E=3NkT/2$, as we have seen.

How can we convince ourselves, in mathematical terms,
that under ``conceivable conditions'', $S(E)$ is
concave function in $E$? 
The answer may be given by a simple superadditivity argument:
As both $E$ and $S$ are extensive quantities, let
us define $E=N\epsilon$ and for a given density $\rho$,
\begin{equation}
s(\epsilon)=\lim_{N\to\infty}\frac{S(N\epsilon)}{N},
\end{equation}
i.e., the per--particle entropy as a function of the per--particle energy,
where we assume that the limit exists. 
Consider the case where the Hamiltonian is additive, i.e.,
\begin{equation}
\calE(\bx)=\sum_{i=1}^N\calE(x_i)
\end{equation}
just like in the above example 
where $\calE(\bx)=\sum_{i=1}^N\frac{\|\vec{p}_i\|^2}{2m}$. Then,
the inequality
\begin{equation}
\Omega(N_1\epsilon_1+N_2\epsilon_2)\ge
\Omega(N_1\epsilon_1)\cdot\Omega(N_2\epsilon_2),
\end{equation}
expresses the simple fact that 
if our system is partitioned into two parts,\footnote{This argument works
for distinguishable particles. We will see later on a more general argument
that holds for indistingusihable particles too.} one with $N_1$ particles,
and the other with $N_2=N-N_1$ particles, then
every combination of individual microstates with energies
$N_1\epsilon_1$ and $N_2\epsilon_2$ corresponds to a combined microstate with
a total energy of $N_1\epsilon_1+N_2\epsilon_2$ (but there are more ways to
split this total energy between the two parts). 
Thus,
\begin{align}
\frac{k\ln \Omega(N_1\epsilon_1+N_2\epsilon_2)}{N_1+N_2}&\ge
\frac{k\ln \Omega(N_1\epsilon_1)}{N_1+N_2}+\frac{k\ln
\Omega(N_2\epsilon_2)}{N_1+N_2}\nonumber\\
&=\frac{N_1}{N_1+N_2}\cdot\frac{k\ln \Omega(N_1\epsilon_1)}{N_1}+\nonumber\\
&  \frac{N_2}{N_1+N_2}\cdot\frac{k\ln
\Omega(N_2\epsilon_2)}{N_2}.
\end{align}
and so, by taking $N_1$ and $N_2$ to $\infty$, with
$N_1/(N_1+N_2)\to\lambda\in(0,1)$, we get:
\begin{equation}
s(\lambda\epsilon_1+(1-\lambda)\epsilon_2)\ge
\lambda s(\epsilon_1)+(1-\lambda)s(\epsilon_2),
\end{equation}
which establishes the concavity of $s(\cdot)$ at least in the case of an
additive Hamiltonian, which means that the entropy of mixing two systems of
particles is greater than the total entropy before they are mixed.
A similar proof
can be generalized to the
case where $\calE(\bx)$ includes also a limited degree of 
interactions (short range interactions), e.g.,
$\calE(\bx)=\sum_{i=1}^N\calE(x_i,x_{i+1})$, but this requires somewhat more
caution. In general, however, concavity may no longer hold when there are long range
interactions, e.g., where some terms of $\calE(\bx)$ depend on a linear
subset of particles. 

\vspace{0.5cm}

\noindent
{\small {\it Example 3.1 -- Schottky defects.}
In a certain crystal, the atoms are located in a lattice, and at any positive
temperature there may be defects, where some of the atoms are dislocated (see
Fig.\ \ref{schottky}).
Assuming that defects are sparse enough, such that around each dislocated atom
all neighbors are in place, the activation energy, $\epsilon_0$, required for dislocation
is fixed. Denoting the total number of atoms by $N$ and the
number of defected ones by $n$, the total energy is then $E=n\epsilon_0$, and
so,
\begin{equation}
\Omega(E)=\left(\begin{array}{cc} N \\ n
\end{array}\right)=\frac{N!}{n!(N-n)!},
\end{equation}
or, equivalently,
\begin{align}
S(E)&=k\ln\Omega(E)=k\ln\left[\frac{N!}{n!(N-n)!}\right]\nonumber\\
&\approx k[N\ln N-n\ln n -(N-n)\ln(N-n)]
\end{align}
where in the last passage we have used the Stirling approximation.
An important comment to point out is that here, unlike in the example
of the ideal gas we have not divided $\Omega(E)$ by $N!$. The reason is
that we do distinguish between two different configurations where the same number of 
particles were dislocated but the sites of dislocation are different. Yet we
do not distinguish between two microstates whose only difference is that two
(identical) particles that were not dislocated are swapped. This is the reason
for the denominator $n!(N-n)!$ in the expression of $\Omega(E)$. 
Now,\footnote{Here and in the sequel, the reader might wonder about the
meaning of taking derivatives of, and with respect to, integer valued
variables, like the number of dislocated particles, $n$. To this end, imagine
an approximation where
$n$ is interpolated to be a continuous valued variable.}
\begin{equation}
\frac{1}{T}=\frac{\partial S}{\partial E}= \frac{\dd S}{\dd n}\cdot\frac{\dd
n}{\dd E}=
\frac{1}{\epsilon_0}\cdot k\ln\frac{N-n}{n},
\end{equation}
which gives the number of defects as
\begin{equation}
\label{numofdefects}
n=\frac{N}{\exp(\epsilon_0/kT)+1}.
\end{equation}
\begin{figure}[ht]
\hspace*{5cm}\input{schottky.pstex_t}
\caption{\small Schottky defects in a crystal lattice.}
\label{schottky}
\end{figure}
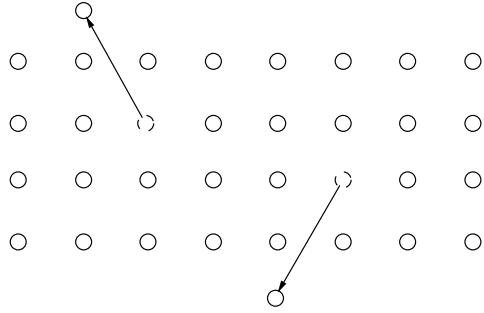
At $T=0$, there are no defects, but their number increases gradually with $T$,
approximately according to $\exp(-\epsilon_0/kT)$. Note also that
\begin{align}
S(E)&=k\ln \left(\begin{array}{cc} N \\ n
\end{array}\right)
\approx kN h_2\left(\frac{n}{N}\right)\nonumber\\
&=kN
h_2\left(\frac{E}{N\epsilon_0}\right)
=kN h_2\left(\frac{\epsilon}{\epsilon_0}\right),
\end{align}
where 
$$h_2(x)\dfn -x\ln x-(1-x)\ln(1-x),~~~~~0\le x\le 1$$
is the called the binary entropy function. Note also that
$s(\epsilon)=kh_2(\epsilon/\epsilon_0)$ is indeed concave in this example.}
$\Box$

What happens if we have two independent subsystems with total energy $E$, which
are both isolated from the environment and they reside in equilibrium with each other?
What is the temperature $T$ and how does the energy split between them?
The number of combined microstates where subsystem no.\ 1 has energy $E_1$ and
subsystem no.\ 2 has energy $E_2=E-E_1$ is $\Omega_1(E_1)\cdot\Omega_2(E-E_1)$.
As the combined system is isolated, the probability of such a combined
macrostate is proportional to $\Omega_1(E_1)\cdot\Omega_2(E-E_1)$.
Keeping in mind that normally, $\Omega_1$ and $\Omega_2$ are exponential in
$N$, then for large $N$, this product is dominated by the value of $E_1$
for which it is maximum, or equivalently, the sum of logarithms,
$S_1(E_1)+S_2(E-E_1)$, is maximum, i.e., it is a {\bf maximum entropy} 
situation, which is {\bf the second law of thermodynamics}, asserting that
an isolated system (in this case, combined of two subsystems) 
achieves its maximum possible entropy in equilibrium.
This maximum is normally achieved at the
value of $E_1$ for which the derivative vanishes, i.e.,
\begin{equation}
S_1'(E_1)-S_2'(E-E_1)=0
\end{equation}
or
\begin{equation}
S_1'(E_1)-S_2'(E_2)=0
\end{equation}
which means
\begin{equation}
\frac{1}{T_1}\equiv S_1'(E_1)=S_2'(E_2)\equiv\frac{1}{T_2}.
\end{equation}
Thus, in equilibrium, which is the maximum entropy situation, 
the energy splits in a way that temperatures are the same.
Now, we can understand the concavity of entropy more generally: $\lambda
s(\epsilon_1)+(1-\lambda)s(\epsilon_2)$ was the total entropy per particle when two
subsystems (with the same entropy function) were isolated from one another, whereas
$s(\lambda\epsilon_1+(1-\lambda)\epsilon_2)$ is the equilibrium entropy per
particle after we let them interact thermally.

At this point, we are ready to justify why $S'(E)$
is equal to $1/T$ in general, as was promised earlier.
Although it is natural to expect
that equality between
$S_1'(E_1)$ and $S_2'(E_2)$, in thermal equilibrium, is related
equality between $T_1$ and $T_2$, this does not automatically
mean that the derivative of each entropy is given by one over
its temperature. On the face of it,
for the purpose of this implication,
this derivative could have
been equal any one--to--one function
of temperature $f(T)$.
To see why $f(T)=1/T$ indeed,
imagine that we have a system with an entropy function
$S_0(E)$ and that we let it interact thermally with an ideal gas whose entropy
function, which we shall denote now by $S_g(E)$, is given as in
eq.\ (\ref{entropyidealgas}).
Now, at equilibrium $S_0'(E_0)=S_g'(E_g)$, but as we have seen already,
$S_g'(E_g)=1/T_g$, where $T_g$ is the temperature of the ideal gas. But in
thermal equilibrium the
temperatures equalize, i.e., $T_g=T_0$, where $T_0$ is the
temperature of the system of interest. It then follows eventually that
$S_0'(E_0)=1/T_0$,
which now means that
in equilibrium,
the derivative of entropy of the system of interest is equal to the reciprocal
of its temperature {\it in general}, and not only for the ideal gas!
At this point,
the fact that our system has interacted and equilibrated with an ideal gas
is not important anymore and
it does not limit the generality this statement. In simple words,
our system does not `care' what kind system it has
interacted with, whether ideal gas or any other.
This follows from a fundamental principle in thermodynamics,
called {\bf the zero--th law of thermodynamics}, which states that thermal equilibrium has a
transitive property: If system $A$ is in equilibrium with system $B$ and system $B$ is in
equilibrium with
system $C$, then $A$ is in equilibrium with $C$.

So we have seen that $\partial S/\partial E=1/T$, or equivalently,
$\delta S=\delta E/T$. But in
the absence of any mechanical work ($V$ is fixed) applied to the system
and any chemical energy injected into the system ($N$ is fixed), any change in
energy must be in the form of {\it heat}, thus we denote
$\delta E=\delta Q$, where $Q$ is the heat intake. Consequently,
\begin{equation}
\delta S=\frac{\delta Q}{T},
\end{equation}
This is exactly the definition of the classical
thermodynamical entropy due to Clausius. Thus,
at least for the case where no mechanical work is involved,
we have demonstrated the equivalence of
the two notions of entropy, the statistical notion due to Boltzmann
$S=k\ln\Omega$, and the thermodynamical entropy due to Clausius,
$S=\int\mbox{d}Q/T$, where the integration should be understood to be taken
along a slow (quasi--static) process, where after each small increase in the
heat intake, the system is allowed to equilibrate, which means that $T$ is
given enough time to adjust before more heat is further added. For a given $V$
and $N$, the difference $\Delta S$ between the entropies $S_A$ and $S_B$
associated with two temperatures
$T_A$ and $T_B$ (pertaining to internal energies $E_A$ and $E_B$, respectively)
is given by $\Delta S=\int_{A}^{B}\mbox{d}Q/T$ along such a quasi--static
process. This is a rule that defines entropy differences, but not absolute
levels. A reference value is determined by
the {\bf third law of thermodynamics}, 
which asserts that as $T$ tends to zero, the
entropy tends to zero as well.\footnote{In this context, it should be
understood that the results we derived for the ideal gas hold only for high
enough temperatures: Since $S$ was found proportional to $\ln E$ and $E$ is
proportional to $T$, then $S$ is proportional to $\ln T$, but this cannot be
true for small $T$ as it contradicts (among other things) the third law.} 

We have seen what is the meaning of the partial derivative of $S(E,V,N)$
w.r.t.\ $E$. Is there also a simple meaning to the partial derivative w.r.t.\ $V$?
Again, let us begin by examining the ideal gas. Differentiating the expression
of $S(E,V,N)$ of the ideal gas w.r.t.\ $V$, we obtain
\begin{equation}
\frac{\partial S(E,V,N)}{\partial V}=\frac{Nk}{V}=\frac{P}{T},
\end{equation}
where the second equality follows again from the equation of state.
So at least for the ideal gas, this partial derivative is related to the
pressure $P$. For similar considerations as before, the relation
\begin{equation}
\frac{\partial S(E,V,N)}{\partial V}=\frac{P}{T}
\end{equation}
is true not only for the ideal gas, but in general.
Consider again an isolated system that consists of two
subsystems, with a wall (or a piston) separating between them. Initially, this
wall is fixed such that the volumes are $V_1$ and $V_2$. At a certain moment,
this wall is released and allowed to be pushed in either direction. How would
the total volume $V=V_1+V_2$ divide between the two subsystems in equilibrium?
Again, the total entropy $S_1(E_1,V_1)+S_2(E-E_1,V-V_1)$ would tend to its
maximum for the same reasoning as before. The maximum will be reached when the
partial derivatives of this sum w.r.t.\ both $E_1$ and $V_1$ would vanish.
The partial derivative w.r.t.\ $E_1$ has already been addressed. The partial
derivative w.r.t.\ $V_1$ gives
\begin{equation}
\frac{P_1}{T_1}=\frac{\partial S_1(E_1,V_1)}{\partial V_1}
=\frac{\partial S_2(E_2,V_2)}{\partial V_2}=\frac{P_2}{T_2}
\end{equation}
Since $T_1=T_2$ by the thermal equilibrium pertaining to derivatives w.r,t.\
energies, it follows that $P_1=P_2$, which means mechanical equilibrium: the
wall will be pushed to the point where the pressures from both sides are
equal. We now have the following differential relationship:
\begin{eqnarray}
\delta S &=&\frac{\partial S}{\partial E}\delta E+
\frac{\partial S}{\partial V}\delta V\nonumber\\
&=&\frac{\delta E}{T}+
\frac{P\delta V}{T}
\end{eqnarray}
or
\begin{equation}
\label{1stlaw}
\delta E = T\delta S-P\delta V=\delta Q-\delta W,
\end{equation}
which is the {\bf the first law of thermodynamics}, asserting that the change
in the energy $\delta E$ of a system with a fixed number of particles is equal
to the difference between the incremental heat intake $\delta Q$ and the
incremental mechanical work
$\delta W$ carried out by the system. This is nothing but a restatement of the
law of energy conservation.  

Finally, we should consider the partial derivative of $S$ w.r.t.\ $N$.
This is given by
\begin{equation}
\frac{\partial S(E,V,N)}{\partial N}=-\frac{\mu}{T},
\end{equation}
where $\mu$ is called the {\it chemical potential}.
If we now consider again the isolated system, which consists of two subsystems
that are allowed to exchange, not only heat and volume, but also 
particles (of the same kind), whose total number
is $N=N_1+N_2$, then again, maximum entropy considerations would yield 
an additional equality between the chemical potentials, $\mu_1=\mu_2$ (chemical
equilibrium).\footnote{Equity of chemical potentials is, in fact, the general
principle of chemical equilibrium, and not equity of concentrations or
densities. In
Section 2.3, we saw equity of densities, because in the case of the ideal
gas, the chemical potential is a function of the density, so equity of
chemical potentials happens to be equivalent to equity of densities in this
case.} The chemical
potential should be understood as a kind of a force
that controls the ability to inject particles into the system. For example, if the particles
are electrically charged, then the chemical potential has a simple analogy to the
electrical potential. The first law is now extended to have an additional
term, pertaining to an increment of chemical energy, and it now reads:
\begin{equation}
\label{g1stlaw}
\delta E = T\delta S-P\delta V+\mu\delta N.
\end{equation}

\noindent
{\small {\it Example 3.2 -- compression of ideal gas.} 
Consider again an ideal gas of $N$ particles
at constant temperature $T$. The energy is $E=3NkT/2$ regardless of the
volume. This means that if we (slowly) compress the gas from volume $V_1$ to volume
$V_2$ ($V_2 < V_1$), the energy remains the same, in spite of the fact that
we injected energy by applying mechanical work}
\begin{equation}
W=-\int_{V_1}^{V_2}P\mbox{d}V
=-NkT\int_{V_1}^{V_2}\frac{\mbox{d}V}{V}=NkT\ln\frac{V_1}{V_2}.
\end{equation}
{\small What happened to that energy? The answer is that it was transformed into heat
as the entropy of the system (which is proportional to $\ln V$) has changed by 
the amount $\Delta S =-Nk\ln (V_1/V_2)$, and so, the heat intake $\Delta Q=T\Delta
S=-NkT\ln(V_1/V_2)$ exactly balances the work.} $\Box$

\subsubsection{The Canonical Ensemble}

So far we have assumed that our system is isolated, and therefore has a strictly
fixed energy $E$. Let us now relax this assumption and assume instead that our system
is free to exchange energy 
with its very large environment (heat bath) and that the total energy of
the heat bath $E_0$ is by far larger than the typical energy of the system. The
combined system, composed of our original system plus the heat bath, is now an
isolated system at temperature $T$.

Similarly as before, since the combined 
system is isolated, it is governed by the microcanonical
ensemble. The only difference is that now we assume that one of the systems
(the heat bath) is very large compared to the other (our test system).
This means that if our small system is in microstate $\bx$
(for whatever definition of the microstate vector) 
with energy $\calE(\bx)$, then the heat bath must have energy $E_0-\calE(\bx)$
to complement the total energy to $E_0$. The number of ways that the heat bath
may have energy $E_0-\calE(\bx)$ is $\Omega_{B}(E_0-\calE(\bx))$,
where $\Omega_{B}(\cdot)$ is the density--of--states function pertaining to the
heat bath. In other words, the number of microstates of the {\it combined}
system for which the small subsystem is in microstate $\bx$ 
is $\Omega_{B}(E_0-\calE(\bx))$. Since the combined system is governed by the
microcanonical ensemble, the probability of this is proportional to 
$\Omega_{B}(E_0-\calE(\bx))$. More precisely:
\begin{equation}
P(\bx)=\frac{\Omega_{B}(E_0-\calE(\bx))}
{\sum_{\bx'}\Omega_{B}(E_0-\calE(\bx'))}.
\end{equation}
Let us focus on the numerator for now, and  normalize the result at the end.
Then,
\begin{align}
P(\bx) &\propto \Omega_{B}(E_0-\calE(\bx))\nonumber\\
&= \exp\{S_{B}(E_0-\calE(\bx))/k\}\nonumber\\
&\approx\exp\left\{\frac{S_{B}(E_0)}{k}-\frac{1}{k}\frac{\partial
S_{B}(E)}{\partial E}\bigg|_{E=E_0}\cdot\calE(\bx)\right\}\nonumber\\
&=\exp\left\{\frac{S_{B}(E_0)}{k}-\frac{1}{kT}
\cdot\calE(\bx)\right\}\nonumber\\
&\propto \exp\{-\calE(\bx)/(kT)\}.
\end{align}
It is customary to work with the so called {\it inverse temperature}:
\begin{equation}
\beta=\frac{1}{kT}
\end{equation}
and so,
\begin{equation}
P(\bx)\propto e^{-\beta\calE(\bx)},
\end{equation}
as we have already seen in the example of the ideal gas (where $\calE(\bx)$
was the kinetic energy), but now it is much more general.
Thus, all that remains to do is to normalize, and 
we then obtain the {\it Boltzmann--Gibbs} (B--G) distribution, 
or the {\it canonical ensemble}, which
describes the underlying probability law in equilibrium:
\begin{equation}
P(\bx)=\frac{\exp\{-\beta\calE(\bx)\}}{Z(\beta)}
\end{equation}
where $Z(\beta)$ is the normalization factor:
\begin{equation}
Z(\beta)=\sum_{\bx}\exp\{-\beta \calE(\bx)\}
\end{equation}
in the discrete case, or
\begin{equation}
Z(\beta)=\int \dd \bx\exp\{-\beta \calE(\bx)\}
\end{equation}
in the continuous case. This is called the canonical ensemble. While the
microcanonical ensemble was defined in terms of the extensive variables $E$, $V$ and
$N$, in the canonical ensemble, we replaced the variable $E$ by the intensive
variable that controls it, namely, $\beta$ (or $T$). Thus, the full notation
of the partition function should be $Z_N(\beta,V)$ or $Z_N(T,V)$.

\vspace{0.2cm}

\noindent
{\small {\it Exercise 3.1:} Show that for the ideal gas
\begin{equation}
Z_N(T,V)=\frac{1}{N!}\left(\frac{V}{\lambda^3}\right)^N
\end{equation}
where
\begin{equation}
\lambda\dfn\frac{h}{\sqrt{2\pi mkT}}.
\end{equation}
$\lambda$ is called the {\it thermal de Broglie wavelength}.}

The formula of the B--G distribution 
is one of the most fundamental results in statistical mechanics, which
was obtained solely from the energy conservation law and the postulate that
in an isolated system the distribution is uniform. The function $Z(\beta)$
is called the {\it partition function}, and as we shall see, its meaning is by
far deeper than just being a normalization constant. Interestingly, a great
deal of the macroscopic physical quantities, like the internal energy, the
free energy, the entropy, the heat capacity, the pressure, etc., can be
obtained from the partition function. This is in analogy to the fact that in
the microcanonical ensemble, $S(E)$ (or, more generally, $S(E,V,N)$) was
pivotal to the derivation of all macroscopic physical quantities of interest.

The B--G distribution tells us then that the system ``prefers'' to visit its
low energy states more than the high energy states, and what counts
is only energy differences, not absolute energies: If we add to all states
a fixed amount of energy $E_0$, this will result in an extra factor of
$e^{-\beta E_0}$ both in the numerator and in the denominator of the B--G
distribution, which will, of course, cancel out. Another obvious observation
is that when the Hamiltonian is additive, that is,
$\calE(\bx)=\sum_{i=1}^N\calE(x_i)$, the various particles are statistically
independent: Additive Hamiltonians correspond to
non--interacting particles. In other words, the $\{x_i\}$'s
behave as if they were drawn from a i.i.d.\ probability distribution. By the law of
large numbers $\frac{1}{N}\sum_{i=1}^N\calE(x_i)$ will tend (almost surely)
to $\epsilon=\left<\calE(X_i)\right>$. Thus, the average energy of the system
is about $N\cdot\epsilon$, not only on the average, but moreover, with an
overwhelmingly {\it high probability} for large $N$.
Nonetheless, this is different from the
microcanonical ensemble where $\frac{1}{N}\sum_{i=1}^N\calE(x_i)$ was held
{\it strictly} at the value of $\epsilon$. 

One of the important principles of statistical mechanics is that
the microcanonical ensemble and the canonical ensemble 
(with the corresponding temperature)
are asymptotically equivalent (in the thermodynamic limit) as far as
macroscopic quantities go. 
They continue to be such even in cases of interactions,
as long as these are short range\footnote{This is 
related to the concavity of $s(\epsilon)$ \cite{BMR01}, \cite{Mukamel08}.} 
and the same is true with the other
ensembles that we will encounter later in this chapter. This is an important
and useful fact, because more often than not, it is more convenient to analyze things in
one ensemble rather than in others, so it is OK to pass to another ensemble for the
purpose of the analysis, even though the ``real system'' is in the other ensemble.
We will use this {\it ensemble equivalence principle} many times later on. The important thing, however, is to
be consistent and not to mix up two ensembles or more. Once you moved to
the other ensemble, stay there.

\vspace{0.2cm}

\noindent
{\small {\it Exercise 3.2:} Consider the ideal gas with gravitation, where the
Hamiltonian includes, in addition to the kinetic energy term for each
molecule, also an
additive term of
potential energy $mgz_i$ for the $i$--th molecule ($z_i$ being its height).
Suppose that an ideal
gas of $N$ molecules of mass
$m$ is confined to a room
whose floor and ceiling
areas are both $A$ and whose height is $h$: (i) Write an expression for the
joint pdf of the location $\vec{\br}$ and the momentum $\vec{\bp}$ of each
molecule. (ii) Use this expression to show that the gas pressures at the floor and the
ceiling are given by
\begin{equation}
P_{floor}=\frac{mgN}{A(1-e^{-mgh/kT})};~~~~
P_{ceiling}=\frac{mgN}{A(e^{mgh/kT}-1)}.
\end{equation}}

\vspace{0.2cm}

It is instructive to point out that the B--G distribution could
have been obtained also in a different manner, owing to the maximum--entropy
principle that stems from the second law. Specifically, 
define the {\it Gibbs entropy} (which is also the {\it Shannon entropy} of
information theory -- see Chapter 9 later on) of a given distribution $P$ as
\begin{equation}
H(P) =-\sum_{\bx}P(\bx)\ln P(\bx)=-\left<\ln P(\bX)\right>
\end{equation}
and consider the
following optimization problem:
\begin{align}
&\max~H(P) \nonumber\\
&\mbox{s.t.}~\langle \calE(\bX)\rangle =E
\end{align}
By formalizing the equivalent Lagrange problem, where $\beta$ now plays the
role of a Lagrange multiplier:
\begin{equation}
\max~\left\{H(P)+\beta\left[E-\sum_{\bx}P(\bx)\calE(\bx)\right]\right\},
\end{equation}
or equivalently,
\begin{equation}
\min~\left\{\sum_{\bx}P(\bx)\calE(\bx)-\frac{H(P)}{\beta}\right\}
\end{equation}
one readily verifies that the solution to this problem is the B-G distribution
where the choice of $\beta$ {\bf controls} the average energy $E$.\footnote{
At this point, one may ask what is the relationship 
between the Boltzmann entropy
as we defined it, $S=k\ln\Omega$, and the Shannon entropy,
$H=-\left<\ln P(\bX)\right>$, where $P$ is the B--G distribution.
It turns out that at least asymptotically,
$S=kH$, as we shall see shortly. For now, let us continue under the assumption
that this is true.} In many
physical systems, the Hamiltonian is a quadratic (or ``harmonic'') function,
e.g., $\frac{1}{2}mv^2$, $\frac{1}{2}kx^2$,
$\frac{1}{2}CV^2$, $\frac{1}{2}LI^2$, $\frac{1}{2}I\omega^2$, etc., in which
case the resulting B--G distribution turns out to be Gaussian. This is at
least part of the explanation why the Gaussian distribution is so frequently
encountered in Nature. 

\subsubsection*{\bf Properties of the Partition Function and the Free Energy}

Let us now examine more closely the partition function and make a few
observations about its basic properties. For simplicity, we shall assume
that $\bx$ is discrete. First, let's look at the limits:
Obviously, $Z(0)$ is equal to the size of the entire set of microstates, which is also
$\sum_E\Omega(E)$, This is the high temperature limit, where
all microstates are equiprobable. At the other extreme, we have:
\begin{equation}
\lim_{\beta\to\infty} \frac{\ln Z(\beta)}{\beta}=-\min_{\bx} \calE(\bx)\dfn
-E_{GS}
\end{equation}
which describes the situation where the system is frozen to the absolute zero.
Only states with minimum energy -- the {\it ground--state energy}, prevail.

Another important property of $Z(\beta)$, or more precisely, of $\ln
Z(\beta)$,
is that it is a cumulant generating function: By taking derivatives of $\ln
Z(\beta)$, we can obtain cumulants of $\calE(\bX)$. 
For the first cumulant, we have
\begin{equation}
\left<\calE(\bX)\right>
=\frac{\sum_{\bx}\calE(\bx)e^{-\beta\calE(\bx)}}{\sum_{\bx}e^{-\beta\calE(\bx)}}=
-\frac{\dd \ln Z(\beta)}{\dd \beta}.
\end{equation}
For example, referring to Exercise 3.1, for the ideal gas, 
\begin{equation}
Z_N(\beta,V)=\frac{1}{N!}\left(\frac{V}{\lambda^3}\right)^N=
\frac{1}{N!}\frac{V^N}{h^{3N}}\cdot\left(\frac{2\pi m}{\beta}\right)^{3N/2},
\end{equation}
thus, $\left<\calE(\bX)\right>=-\mbox{d}\ln
Z_N(\beta,V)/\mbox{d}\beta=3N/(2\beta)=3NkT/2$ in agreement with the result we
already got both in Chapter 2 and in the microcanonical ensemble, thus
demonstrating the ensemble equivalence principle.
Similarly, it is easy to show that
\begin{equation}
\mbox{Var}\{\calE(\bX)\}=\langle\calE^2(\bX)\rangle -
\langle\calE(\bX)\rangle^2=\frac{\dd^2\ln Z(\beta)}{\dd\beta^2}.
\end{equation}
This in turn implies that 
\begin{equation}
\frac{\dd^2\ln Z(\beta)}{\dd\beta^2}\ge 0,
\end{equation}
which means that $\ln Z(\beta)$ must always be a convex function.
Note also that
\begin{align}
\label{heatcapacity}
\frac{\dd^2\ln Z(\beta)}{\dd\beta^2}&=-\frac{\dd \langle
\calE(\bX)\rangle}{\dd\beta}\nonumber\\
&=-\frac{\dd \langle \calE(\bX)\rangle}{\dd T}\cdot\frac{\dd T}{\dd \beta}\nonumber\\
&=kT^2C(T)
\end{align}
where $C(T)=\dd\langle \calE(\bX)\rangle/\dd T$ is the heat capacity (at constant volume).
Thus, the convexity of $\ln Z(\beta)$ is intimately related to the physical
fact that the heat capacity of the system is positive.

Next, we look at the function $Z(\beta)$ slightly differently. Instead of summing
the terms $\{e^{-\beta \calE(\bx)}\}$ over all states individually, 
we sum them by energy levels, in a collective manner.
This amounts to:
\begin{align}
\label{freeenergy}
Z(\beta)&=\sum_{\bx}e^{-\beta\calE(\bx)}\nonumber\\
&=\sum_{E}\Omega(E)e^{-\beta E}
\nonumber\\
&\approx\sum_\epsilon e^{Ns(\epsilon)/k}\cdot e^{-\beta
N\epsilon}\nonumber\\
&=\sum_\epsilon \exp\{-N\beta[\epsilon-Ts(\epsilon)]\}\nonumber\\
&\exe\max_\epsilon \exp\{-N\beta[\epsilon-Ts(\epsilon)]\}\nonumber\\
&=\exp\{-N\beta\min_\epsilon[\epsilon-Ts(\epsilon)]\}\nonumber\\
&\dfn\exp\{-N\beta[\epsilon^*-Ts(\epsilon^*)]\}\nonumber\\
&\dfn e^{-\beta F},
\end{align}
where here and throughout the sequel, the notation $\exe$ means asymptotic
equivalence in the exponential scale. More precisely, $a_N\exe b_N$ for
two positive sequences $\{a_N\}$ and $\{b_N\}$, means that
$\lim_{N\to\infty}\frac{1}{N}\ln\frac{a_N}{b_N}=0$.

The quantity $f\dfn \epsilon^*-Ts(\epsilon^*)$ is the (per--particle) {\it free
energy}. Similarly, the entire free energy, $F$, is defined as
\begin{equation}
\label{FreeEnergy}
F\dfn E-TS=-\frac{\ln Z(\beta)}{\beta}= -kT\ln Z(\beta).
\end{equation}
Once again, due to the exponentiality of (\ref{freeenergy}) in $N$, with {\it very high probability}
the system would be found in a microstate $\bx$ whose normalized energy
$\epsilon(\bx)=\calE(\bx)/N$ is very close to $\epsilon^*$, the normalized
energy that minimizes $\epsilon-Ts(\epsilon)$ and hence achieves $f$.
We see then that
equilibrium in the canonical ensemble amounts to {\bf minimum free energy.}
This extends the second law of thermodynamics from
isolated systems to non--isolated ones.
While in an isolated system, the second law asserts the principle of maximum
entropy, when it comes to a non--isolated system, this rule is 
replaced by the
principle of minimum free energy.

\vspace{0.2cm}

\noindent
{\small {\it Exercise 3.3:} 
Show that the canonical average pressure is given by
$$P=-\frac{\partial F}{\partial V}=kT\cdot\frac{\partial\ln Z_N(\beta,V)}{\partial V}.$$
Examine this formula for the canonical ensemble of the ideal gas. Compare to
the equation of state.}\\

The physical meaning of the free energy, or more precisely,
the difference between two free energies $F_1$ and $F_2$,
is the minimum amount of work that
it takes to transfer the system from equilibrium state 1 to another
equilibrium state 2 in an isothermal (fixed temperature) process. This
minimum is achieved when the process is {\it quasi--static}, i.e., so slow that
the system is always almost in equilibrium. Equivalently, $-\Delta F$ is the
maximum amount energy in the system, that is {\it free} and useful
for performing work
(i.e., not dissipated as heat) in fixed temperature.

To demonstrate
this point, let us consider the case where $\calE(\bx)$ includes a term of a
potential energy that is given by the (scalar) product of a certain external force and
the conjugate physical variable at which this force is exerted (e.g., pressure
times volume, gravitational force times height, moment times angle, magnetic field
times magnetic moment, voltage times electric charge, etc.), i.e.,
\begin{equation}
\calE(\bx)=\calE_0(\bx)-\lambda\cdot L(\bx)
\end{equation}
where $\lambda$ is the force and $L(\bx)$ is the conjugate physical variable, which
depends on (some coordinates of) the microstate. The partition function then
depends on both $\beta$ and $\lambda$ and hence will be denoted
$Z(\beta,\lambda)$. It
is easy to see (similarly as before) that $\ln Z(\beta,\lambda)$ is convex in
$\lambda$
for fixed $\beta$. Also,
\begin{equation}
\label{gibbs1}
\left<L(\bX)\right>=kT\cdot\frac{\partial\ln Z(\beta,\lambda)}{\partial
\lambda}.
\end{equation}
The free energy is given by\footnote{At this point, there is a distinction
between the {\it Helmholtz free energy} and the {\it Gibbs free energy}. The
former is defined as $F=E-TS$ in general, as mentioned earlier. The latter is defined as
$G=E-TS-\lambda L=-kT\ln Z$, where $L$ is shorthand notation for $\left<L(\bX)\right>$
(the quantity $H=E-\lambda L$ is called the enthalpy).
The physical significance of the Gibbs free energy is similar to that of the
Helmholtz free energy, except that it refers to the total work of all other external forces
in the system (if there are any), except the work contributed by the force
$\lambda$ (Exercise 3.4: show this!). The passage to the Gibbs ensemble, which replaces a fixed value of
$L(\bx)$ (say, constant volume of a gas) by the control of the conjugate
external force $\lambda$,
(say, pressure in the example of a gas)
can be carried out by another Legendre transform (see, e.g., \cite[Sect.\
1.14]{Kubo61}) as well as Subsection 3.2.3 in the sequel.}
\begin{align}
\label{gibbs2}
F&=E-TS\nonumber\\
&=-kT\ln Z+\lambda\left<L(\bX)\right>\nonumber\\
&=
kT\left(\lambda\cdot\frac{\partial\ln Z}{\partial\lambda}-\ln Z\right).
\end{align}
Now, let $F_1$ and $F_2$ be the equilibrium free energies pertaining to two
values of $\lambda$, denoted $\lambda_1$ and $\lambda_2$. Then,
\begin{align}
\label{gibbs3}
F_2-F_1&=\int_{\lambda_1}^{\lambda_2}\dd \lambda\cdot\frac{\partial
F}{\partial \lambda}\nonumber\\
&= kT\cdot\int_{\lambda_1}^{\lambda_2}\dd\lambda\cdot\lambda\cdot\frac{\partial^2\ln Z}{\partial
\lambda^2}\nonumber\\
&=\int_{\lambda_1}^{\lambda_2}\dd \lambda\cdot \lambda
\cdot\frac{\partial\left<L(\bX)\right>}{\partial \lambda}\nonumber\\
&=
\int_{\langle L(X)\rangle_{\lambda_1}}^{\langle L(X)\rangle_{\lambda_2}}
\lambda \cdot\dd\left<L(\bX)\right>
\end{align}
The product $\lambda\cdot\dd\left<L(\bX)\right>$ designates an infinitesimal amount of
(average) work performed by the force $\lambda$ on a small change in
the average of the conjugate variable $\left<L(\bX)\right>$, where the
expectation is taken w.r.t.\ the actual value of $\lambda$. Thus, the last integral
expresses the total work along a slow process of changing the force $\lambda$ in
small steps and letting
the system adapt and equilibrate after this small change every time. 
On the other hand, it is easy to show (using the convexity of $\ln Z$ in
$\lambda$),
that if $\lambda$ varies in large steps, the resulting amount of work will always be
larger.

Returning to the definition of $f$,
as we have said, the value $\epsilon^*$ of $\epsilon$ that minimizes $f$, dominates
the partition function and hence captures most of the probability for large
$N$. Note that the Lagrange minimization problem that we
formalized before, i.e.,
\begin{equation}
\min~\left\{\sum_{\bx}P(\bx)\calE(\bx)-\frac{H(P)}{\beta}\right\},
\end{equation}
is nothing but minimization of the free energy, provided that we identify
$H$ with the physical entropy $S$ (to be done soon) and the Lagrange
multiplier $1/\beta$ with $kT$. Thus, the B--G distribution minimizes the
free energy for a given temperature.

Let us define
\begin{equation}
\phi(\beta)=\lim_{N\to\infty}\frac{\ln Z(\beta)}{N}
\end{equation}
and, in order to avoid dragging the constant $k$, let us define
\begin{equation}
\Sigma(\epsilon)=\lim_{N\to\infty}\frac{\ln\Omega(N\epsilon)}{N}=\frac{s(\epsilon)}{k}.
\end{equation}
Then, 
the chain of equalities (\ref{freeenergy}), written slightly differently, gives
\begin{align}
\phi(\beta)&=\lim_{N\to\infty}\frac{\ln Z(\beta)}{N}\nonumber\\
&=\lim_{N\to\infty}\frac{1}{N}\ln\left\{\sum_\epsilon 
e^{N[\Sigma(\epsilon)-\beta \epsilon]}\right\}\nonumber\\
&=\max_\epsilon[\Sigma(\epsilon)-\beta\epsilon].\nonumber
\end{align}
Thus, $\phi(\beta)$ is (a certain variant of) the {\it Legendre
transform}\footnote{More precisely, the 1D Legendre transform of a real function
$f(x)$ is defined as $g(y)=\sup_x[xy-f(x)]$. If $f$ is convex, it can readily
be shown that: (i) The inverse transform has the very same form, i.e.,
$f(x)=\sup_y[xy-g(y)]$, and (ii) The derivatives $f'(x)$ and $g'(y)$ are inverses of each other.}
of $\Sigma(\epsilon)$. As $\Sigma(\epsilon)$ is (normally) a concave,
monotonically increasing function,
then it can readily be shown\footnote{Should be done in a recitation.} that the inverse transform is:
\begin{equation}
\label{entropy1}
\Sigma(\epsilon)=\min_\beta[\beta\epsilon+\phi(\beta)].
\end{equation}
The achiever, $\epsilon^*(\beta)$, of $\phi(\beta)$ in the forward transform is obtained by
equating the derivative to zero, i.e., it is the solution to the equation
\begin{equation}
\beta=\Sigma'(\epsilon),
\end{equation}
where $\Sigma'(\epsilon)$ is the derivative of $\Sigma(\epsilon)$.
In other words, $\epsilon^*(\beta)$ the inverse function of $\Sigma'(\cdot)$. By the same
token, the achiever, $\beta^*(\epsilon)$, of $\Sigma(\epsilon)$ in the backward transform
is obtained by
equating the other derivative to zero, i.e., it is the solution to the equation
\begin{equation}
\epsilon=-\phi'(\beta)
\end{equation}
or in other words, the inverse function of $-\phi'(\cdot)$.\\
This establishes a relationship between the
typical per--particle energy $\epsilon$ and the inverse temperature $\beta$
that gives rise to $\epsilon$ (cf.\ the Lagrange interpretation above, where we said
that $\beta$ controls the average energy).
Now, observe that whenever $\beta$ and $\epsilon$ are related as explained
above, we have:
\begin{equation}
\label{phitosigma}
\Sigma(\epsilon)=\beta\epsilon+\phi(\beta)=\phi(\beta)-\beta\cdot\phi'(\beta).
\end{equation}
The Gibbs entropy per particle
is defined in its normalized for as
\begin{equation}
\bar{H}=-\lim_{N\to\infty}\frac{1}{N}\sum_{\bx}P(\bx)\ln P(\bx)=
-\lim_{N\to\infty}\frac{1}{N}\left<\ln P(\bx)\right>,
\end{equation}
which in the case of the B--G distribution amounts to
\begin{align}
\bar{H}&=\lim_{N\to\infty}\frac{1}{N}\left<\ln\frac{Z(\beta)}{e^{-\beta\calE(\bX)}}
\right>\nonumber\\
&=\lim_{N\to\infty}\left[\frac{\ln
Z(\beta)}{N}+\frac{\beta\left<\calE(\bX)\right>}{N}\right]\nonumber\\
&=\phi(\beta)-\beta\cdot\phi'(\beta).\nonumber
\end{align}
but this is exactly the same expression as in (\ref{phitosigma}),
and so, $\Sigma(\epsilon)$ and $\bar{H}$ are identical whenever $\beta$ and
$\epsilon$ are related accordingly. The former, as we recall, we defined as
the normalized logarithm of the number of microstates with per--particle 
energy $\epsilon$. Thus, we have learned that the number of such microstates
is of the exponential order of $e^{N\bar{H}}$.
Another look at this relation is
the following:
\begin{align}
1&\geq\sum_{\bx:~\calE(\bx)\approx N\epsilon} P(\bx)=
\sum_{\bx:~\calE(\bx)\approx N\epsilon} \frac{\exp\{-\beta\sum_i\calE(x_i)\}}
{Z^N(\beta)}\nonumber\\
&\exe\sum_{\bx:~\calE(\bx)\approx N\epsilon} 
\exp\{-\beta N\epsilon-N\phi(\beta)\}\nonumber\\
&=\Omega(N\epsilon)\cdot\exp\{-N[\beta\epsilon+\phi(\beta)]\}
\end{align}
which means that
\begin{equation}
\Omega(N\epsilon)\le 
\exp\{N[\beta\epsilon+\phi(\beta)]\} 
\end{equation}
for all $\beta$, and so,
\begin{equation}
\Omega(N\epsilon)\le 
\exp\{N\min_\beta[\beta\epsilon+\phi(\beta)]\}=
e^{N\Sigma(\epsilon)}=e^{N\bar{H}}.
\end{equation}
A compatible lower bound is obtained by observing that the minimizing $\beta$
gives rise to $\left<\calE(X_1)\right>=\epsilon$, which makes the event
$\{\bx:~\calE(\bx)\approx N\epsilon\}$ a high--probability event, by the
weak law of large numbers.
A good reference for further study, and
from a more general perspective, is the article by Hall \cite{Hall99}.
See also \cite{Ellis85}.

Note also that eq.\ (\ref{phitosigma}), 
which we will rewrite, with a slight abuse
of notation as
\begin{equation}
\label{diffeq}
\phi(\beta)-\beta\phi'(\beta)=\Sigma(\beta)
\end{equation}
can be viewed in two ways. The first suggests to take derivatives of both
sides w.r.t.\ $\beta$ and then obtain
$\Sigma'(\beta)=-\beta\phi''(\beta)$ and so,
\begin{eqnarray}
s(\beta)&=&k\Sigma(\beta)\nonumber\\
&=&k\int_\beta^\infty\tilde{\beta}\phi''(\tilde{\beta})\mbox{d}\tilde{\beta}~~~~~~~\mbox{3rd
law}\nonumber\\
&=&k\int_0^T\frac{1}{k\tilde{T}}\cdot k\tilde{T}^2c(\tilde{T})
\cdot\frac{\mbox{d}\tilde{T}}{k\tilde{T}^2}~~~~~\mbox{$c(\tilde{T})\dfn$ 
heat capacity per
particle}\nonumber\\
&=&\int_0^T\frac{c(\tilde{T})\mbox{d}\tilde{T}}{\tilde{T}}
\end{eqnarray}
recovering the Clausius entropy as $c(\tilde{T})\mbox{d}\tilde{T}$ is the
increment of heat intake per particle $\mbox{d}q$.
The second way to look at eq.\ (\ref{diffeq}) is as a first order 
differential equation in $\phi(\beta)$, whose solution is
easily found to be
\begin{equation}
\phi(\beta)=-\beta \epsilon_{GS}+\beta\cdot\int_{\beta}^\infty\frac{\dd
\hat{\beta}\Sigma(\hat{\beta})}{\hat{\beta}^2},
\end{equation}
where $\epsilon_{GS}=\lim_{N\to\infty}E_{GS}/N$.
Equivalently,
\begin{equation}
Z(\beta)\exe\exp\left\{-\beta E_{GS}+N\beta\cdot\int_{\beta}^\infty\frac{\dd
\hat{\beta}\Sigma(\hat{\beta})}{\hat{\beta}^2}\right\},
\end{equation}
namely, the partition function at a certain 
temperature can be expressed as a functional of the entropy
pertaining to all temperatures lower 
than that temperature. Changing the integration variable from $\beta$ to
$T$, this readily gives the relation 
\begin{equation}
F=E_{GS}-\int_0^T S(T')\dd T'. 
\end{equation}
Since
$F=E-ST$, we have 
\begin{equation}
E=E_{GS}+ST-\int_0^T S(T')\dd T'=E_{GS}+\int_0^S T(S')\dd S', 
\end{equation}
where the second term amounts to the
heat $Q$ that accumulates in the system, as the temperature is raised from $0$ to $T$. 
This is a special case of the first law of thermodynamics. The more
general form, as said, takes into account also possible work performed on (or by) the system.

Let us now summarize the main properties of the partition function that we
have seen thus far:
\begin{enumerate}
\item $Z(\beta)$ is a continuous function. $Z(0)=|\calX^n|$ and
$\lim_{\beta\to\infty}\frac{\ln Z(\beta)}{\beta}=-E_{GS}$.
\item Generating cumulants: $\langle\calE(\bX)\rangle =-\dd\ln Z/\dd\beta$,
$\mbox{Var}\{\calE(\bX)\}=\dd^2\ln Z/\dd\beta^2$, which implies convexity of
$\ln Z$, and hence also of $\phi(\beta)$.
\item $\phi$ and $\Sigma$ are a Legendre--transform pair. $\Sigma$ is
concave.
\end{enumerate}
We have also seen that Boltzmann's entropy is not only equivalent to the
Clausius entropy, but also to the Gibbs/Shannon entropy. Thus, there are
actually three different forms of the expression of entropy.

\noindent
{\bf Comment:} Consider $Z(\beta)$ for an {\it imaginary temperature}
$\beta=i\omega$, where $i=\sqrt{-1}$, and define $z(E)$ as the inverse
Fourier transform of $Z(j\omega)$. It can readily be seen that $z(E)=\Omega(E)$ is the density of states,
i.e., for $E_1 < E_2$, the number of states with energy between $E_1$ and
$E_2$ is given by $\int_{E_1}^{E_2}z(E)\dd E$.
Thus, $Z(\cdot)$ can be related to energy enumeration in two different ways:
one is by the Legendre 
transform of $\ln Z(\beta)$ for real $\beta$, and the other
is by the inverse Fourier transform of $Z(\beta)$ for imaginary $\beta$.
It should be kept in mind, however, that
while the latter relation holds for every system size $N$, the former is true
only in the thermodynamic limit, as mentioned.

\vspace{0.25cm}

\noindent
{\small {\it Example 3.3 -- two level system.} Similarly to the
earlier example of Schottky defects, which was previously given in the context
of the microcanonical ensemble,
consider now a system of $N$ independent
particles, each having two possible states: state $0$ of zero energy and
state $1$, whose energy is $\epsilon_0$, i.e., $\calE(x)=\epsilon_0x$,
$x\in\{0,1\}$. The $x_i$'s are independent, each having a
marginal:\footnote{Note that the expected number of `activated' particles
$\left<n\right>=NP(1)=Ne^{-\beta\epsilon_0}/(1+e^{-\beta\epsilon_0})=N/(e^{\beta\epsilon_0}+1)$,
in agreement with the result of Example 3.1 (eq.\ (\ref{numofdefects})). This demonstrates the ensemble
equivalence principle.}
\begin{equation}
P(x)=\frac{e^{-\beta\epsilon_0x}}{1+e^{-\beta\epsilon_0}}~~~x\in\{0,1\}.
\end{equation}
In this case,
\begin{equation}
\phi(\beta )=\ln(1+e^{-\beta\epsilon_0})
\end{equation}
and 
\begin{equation}
\Sigma(\epsilon)=\min_{\beta\ge
0}[\beta\epsilon+\ln(1+e^{-\beta\epsilon_0})].
\end{equation}
To find $\beta^*(\epsilon)$, we take the derivative and equate to zero:
\begin{equation}
\epsilon-\frac{\epsilon_0e^{-\beta\epsilon_0}}{1+e^{-\beta\epsilon_0}}=0
\end{equation}
which gives
\begin{equation}
\beta^*(\epsilon)=\frac{\ln(\epsilon_0/\epsilon-1)}{\epsilon_0}.
\end{equation}
On substituting this back into the above expression of $\Sigma(\epsilon)$, we
get:
\begin{equation}
\Sigma(\epsilon)=\frac{\epsilon}{\epsilon_0}\ln\left(\frac{\epsilon}{\epsilon_0}-1\right)
+\ln\left[1+\exp\left\{-\ln\left(\frac{\epsilon}{\epsilon_0}-1\right)\right\}\right],
\end{equation}
which after a short algebraic manipulation, becomes
\begin{equation}
\Sigma(\epsilon)=h_2\left(\frac{\epsilon}{\epsilon_0}\right),
\end{equation}
just like in the Schottky example. In the other direction:
\begin{equation}
\phi(\beta)=\max_\epsilon\left[h_2\left(\frac{\epsilon}{\epsilon_0}\right)-
\beta\epsilon\right],
\end{equation}
whose achiever $\epsilon^*(\beta)$ solves the zero--derivative equation:
\begin{equation}
\frac{1}{\epsilon_0}\ln\left[\frac{1-\epsilon/\epsilon_0}{\epsilon/\epsilon_0}\right]=\beta
\end{equation}
or equivalently,
\begin{equation}
\epsilon^*(\beta)=\frac{\epsilon_0}{1+e^{-\beta\epsilon_0}},
\end{equation}
which is exactly the inverse function of $\beta^*(\epsilon)$ above, and
which when plugged back into the expression of $\phi(\beta)$, indeed gives
\begin{equation}
\phi(\beta)=\ln(1+e^{-\beta\epsilon_0}).~~~~\Box
\end{equation}
}

\vspace{0.25cm}

\noindent
{\it Comment:} A very similar model 
(and hence with similar results)
pertains to non--interacting spins
(magnetic moments), where the only difference is that $x\in\{-1,+1\}$ rather
than $x\in\{0,1\}$. Here, the meaning of the parameter $\epsilon_0$
becomes that of a magnetic field, which is more customarily denoted by 
$B$ (or $H$), and which is either parallel or anti-parallel to
that of the spin, and so the potential energy 
(in the appropriate physical units), $\vec{B}\cdot\vec{x}$, is either
$Bx$ or $-Bx$. Thus,
\begin{equation}
P(x)=\frac{e^{\beta Bx}}{2\cosh(\beta B)};~~~Z(\beta)=2\cosh(\beta B).
\end{equation}
The net {\it magnetization} per--spin is defined as
\begin{equation}
m\dfn \left< \frac{1}{N}\sum_{i=1}^NX_i\right> = \langle X_1\rangle=
\frac{\partial \phi}{\partial (\beta
B)}=\tanh(\beta B).
\end{equation}
This is the paramagnetic characteristic of the magnetization as a function of
the magnetic field: As $B\to\pm\infty$, the magnetization $m\to\pm 1$
accordingly. When the magnetic field is removed ($B=0$), the magnetization
vanishes too.
We will get back to this model and its extensions in Chapter 6. $\Box$

\subsubsection*{The Energy Equipartition Theorem}

It turns out that in the case of
a quadratic Hamiltonian, $\calE(x)=\frac{1}{2}\alpha x^2$, which means
that $x$ is
Gaussian, the average per--particle energy, 
is always given by $1/(2\beta)=kT/2$,
independently 
of $\alpha$. If we have $N$ such quadratic terms, then of course, we
end up with $NkT/2$. In the case of the ideal gas, we have three such terms (one
for each dimension) per particle, thus a total of $3N$ terms, and so,
$E=3NkT/2$, which is exactly the expression we obtained also from the microcanonical
ensemble as well as in the previous chapter. 
In fact, we observe that in the canonical ensemble,
whenever we have an Hamiltonian of the form $\frac{\alpha}{2}x_i^2$ plus
some arbitrary terms that do not depend on $x_i$, 
then $x_i$ is Gaussian
(with variance $kT/\alpha$) and independent
of the other variables, i.e.,
$p(x_i)\propto
e^{-\alpha x_i^2/(2kT)}$.
Hence it contributes an amount of 
\begin{equation}
\left<\frac{1}{2}\alpha X_i^2\right> =\frac{1}{2}\alpha\cdot
\frac{kT}{\alpha}=
\frac{kT}{2}
\end{equation}
to the total
average energy, independently of $\alpha$. It is more precise to refer
to this $x_i$ as a
{\it degree of freedom} rather than a particle. This is because in
the three--dimensional world, the kinetic energy, for example, is given by
$p_x^2/(2m)+p_y^2/(2m)+p_z^2/(2m)$, 
that is, each particle contributes {\it three}
additive quadratic terms rather than one 
(just like three independent one--dimensional
particles) and so, it contributes $3kT/2$. This principle is called the
{\it the energy equipartition theorem}. 

Below is a direct derivation of the equipartition theorem:
\begin{align}
\left<\frac{1}{2}aX^2\right> &= \frac{\int_{-\infty}^{\infty}\dd
x(\alpha x^2/2) e^{-\beta\alpha x^2/2}}{\int_{-\infty}^{\infty}\dd
xe^{-\beta\alpha x^2/2)}}\nonumber\\
&=-\frac{\partial}{\partial \beta}\ln\left[\int_{-\infty}^{\infty}\dd xe^{-\beta\alpha x^2/2}
\right]\nonumber\\
&=-\frac{\partial}{\partial \beta}\ln\left[\frac{1}{\sqrt{\beta}}
\int_{-\infty}^{\infty}\dd(\sqrt{\beta}x)e^{-\alpha(\sqrt{\beta}x)^2/2}\right]\nonumber\\
&=-\frac{\partial}{\partial \beta}\ln\left[\frac{1}{\sqrt{\beta}}
\int_{-\infty}^{\infty}\dd ue^{-\alpha u^2/2}\right]\nonumber\\
&=\frac{1}{2}\frac{\dd
\ln\beta}{\dd\beta}=\frac{1}{2\beta}=\frac{kT}{2}.\nonumber
\end{align}
Note that although we could have used closed--form expressions for both the
numerator and the denominator of the first line, we have deliberately taken a somewhat different
route in the second line, where we have presented it as the derivative of the
denominator of the first line. Also, rather than calculating the Gaussian
integral explicitly, we only figured out how it scales with $\beta$, because
this is the only thing that matters after taking the derivative relative to
$\beta$. The reason for using this trick of
bypassing the need to calculate integrals, is that it
can easily be extended in two directions at least:

\vspace{0.2cm}

\noindent
1. Let $\bx\in\reals^N$ and let $\calE(\bx)=\frac{1}{2}\bx^TA\bx$, where $A$ is
a $N\times N$ positive definite matrix. This corresponds to a physical system
with a quadratic Hamiltonian, which includes also interactions between pairs (e.g.,
harmonic oscillators or springs, which are coupled because they are tied
to one another). It turns out that here, regardless of $A$, we get:
\begin{equation}
\langle \calE(\bX)\rangle = \left<\frac{1}{2}\bX^TA\bX\right> = N\cdot\frac{kT}{2}.
\end{equation}

\vspace{0.2cm}

\noindent
2. Back to the case of a scalar $x$, but suppose now a more general power--law
Hamiltonian, $\calE(x)=\alpha|x|^\theta$. In this case, we get
\begin{equation}
\langle \calE(X)\rangle = \left<\alpha|X|^\theta\right> = \frac{kT}{\theta}.
\end{equation}
Moreover, if $\lim_{x\to\pm\infty} xe^{-\beta\calE(x)}=0$ for all $\beta > 0$,
and we denote $\calE'(x)\dfn \dd\calE(x)/\dd x$, then
\begin{equation}
\langle X\cdot\calE'(X)\rangle =kT.
\end{equation}
It is easy to see that the earlier power--law result is obtained as a special case of
this, as $\calE'(x)=\alpha\theta|x|^{\theta-1}\mbox{sgn}(x)$ in this case.

\vspace{0.25cm}

\noindent
{\small {\it Example 3.4 -- ideal gas with gravitation}: Let}
\begin{equation}
\calE(x)= \frac{p_x^2+p_y^2+p_z^2}{2m}+mgz.
\end{equation}
{\small The average kinetic energy of each particle is $3kT/2$, as said before.
The contribution of the average potential energy is $kT$ (one degree of
freedom with $\theta=1$). Thus, the total is $5kT/2$, where $60\%$ come from
kinetic energy and $40\%$ come from potential energy, universally,
that is, independent of $T$, $m$, and $g$.} $\Box$

\subsubsection{The Grand--Canonical Ensemble and the Gibbs Ensemble}
\label{grandcanonical}

A brief summary of what we have
done thus far, is the following: we started with the microcanonical
ensemble, which was very restrictive in the sense that the energy was held
strictly fixed to the value of $E$, the number of particles was held strictly
fixed to the value of $N$, and at least in the example of a gas, the volume
was also held strictly fixed to a certain value $V$. In the passage from the
microcanonical ensemble to the canonical one, we slightly relaxed the first of
these parameters, $E$: Rather than insisting on a fixed value of $E$, we
allowed energy to be exchanged back and forth with the environment, and thereby to
slightly fluctuate (for large $N$)
around a certain average value, which was controlled by temperature,
or equivalently, by the choice of $\beta$. This was done while keeping in mind
that the total energy of both system and heat bath must be kept fixed, by the
law of energy conservation, which allowed us to look at the combined system as
an isolated one, thus obeying the microcanonical ensemble.
We then had a one--to--one
correspondence between the extensive quantity $E$ and the intensive variable
$\beta$, that adjusted its average value. But the other extensive
variables, like $N$ and $V$ were still kept strictly fixed.

It turns out, that we can continue in this spirit, and `relax' also either one
of the other variables $N$ or $V$ (but not both at the same time),
allowing it to fluctuate around a typical average
value, and controlling it by a corresponding intensive variable.
Like $E$, both $N$ and $V$ are also subjected to conservation laws
when the combined system is considered.
Each one of these relaxations, leads to a new ensemble in addition to the
microcanonical
and the canonical ensembles that we have already seen.
In the case where it is the variable $V$ that is allowed to be flexible, this
ensemble is called the {\it Gibbs ensemble}.
In the case where it is the variable $N$, this
ensemble is called the {\it grand--canonical ensemble}. 
There are, of course, additional ensembles based on this principle,
depending on the kind of the physical system. 

\subsubsection*{The Grand--Canonical Ensemble}

The fundamental idea is essentially the very same as the one we used to derive
the canonical ensemble:
Let us get back to our (relatively small) subsystem, which is in contact
with a heat bath, and this time, let us allow this subsystem to exchange
with the heat bath,
not only energy, but also matter, i.e., particles. The heat bath consists of
a huge reservoir of energy and particles. The total energy is $E_0$ and the
total number of particles is $N_0$. Suppose that we can calculate the density
of states of the heat bath as function of both its energy $E'$ and amount of
particles $N'$, call it $\Omega_{B}(E',N')$. A microstate is now a
combination $(\bx,N)$, where $N$ is the (variable) number of particles
in our subsystem and $\bx$ is as before for a given $N$.
From the same considerations as before, whenever our subsystem
is in state $(\bx,N)$, the heat bath can be in any one of
$\Omega_{B}(E_0-\calE(\bx),N_0-N)$ microstates of its own. Thus,
owing to the microcanonical ensemble,
\begin{align}
P(\bx,N)&\propto \Omega_{B}(E_0-\calE(\bx),N_0-N)\nonumber\\
&= \exp\{S_{B}(E_0-\calE(\bx),N_0-N)/k\}\nonumber\\
&\approx \exp\left\{\frac{S_{B}(E_0,N_0)}{k}-\frac{1}{k}
\frac{\partial S_{B}}{\partial E}\cdot\calE(\bx)-
\frac{1}{k}\frac{\partial S_{B}}{\partial N}\cdot N
\right\}\nonumber\\
&\propto \exp\left\{-\frac{\calE(\bx)}{kT}+\frac{\mu N}{kT}\right\}
\end{align}
where $\mu$ is the chemical potential of the heat
bath.
Thus, we now have the grand--canonical distribution:
\begin{equation}
P(\bx,N)=\frac{e^{\beta[\mu N-\calE(\bx)]}}{\Xi(\beta,\mu)},
\end{equation}
where the denominator is called the {\it grand partition function}:
\begin{equation}
\Xi(\beta,\mu)\dfn \sum_{N=0}^\infty e^{\beta\mu
N}\sum_{\bx}e^{-\beta\calE(\bx)}\dfn \sum_{N=0}^\infty e^{\beta\mu
N}Z_N(\beta).
\end{equation}

\vspace{0.2cm}

\noindent
{\small {\it Example 3.5 -- grand partition function of the ideal gas.}
Using the result of Exercise 3.1, we have for the ideal gas:
\begin{eqnarray}
\label{gcidealgas}
\Xi(\beta,\mu)&=&\sum_{N=0}^\infty e^{\beta\mu
N}\cdot\frac{1}{N!}\left(\frac{V}{\lambda^3}\right)^N\nonumber\\
&=&\sum_{N=0}^\infty
\frac{1}{N!}\left(e^{\beta\mu}\cdot\frac{V}{\lambda^3}\right)^N\nonumber\\
&=&\exp\left(e^{\beta\mu}\cdot\frac{V}{\lambda^3}\right).
\end{eqnarray}
}
It is sometimes convenient to change variables and to define $z=e^{\beta\mu}$
(which is called the {\it fugacity}) and then, define
\begin{equation}
\tilde{\Xi}(\beta,z)=\sum_{N=0}^\infty z^N Z_N(\beta).
\end{equation}
This notation emphasizes the fact that for a given $\beta$, $\tilde{\Xi}(z)$
is actually the $z$--transform of the sequence $Z_N$.
A natural way to think about $P(\bx,N)$ is as $P(N)\cdot P(\bx|N)$, where
$P(N)$ is proportional to $z^N Z_N(\beta)$ and $P(\bx|N)$ corresponds to the
canonical ensemble as before.

Using the grand partition function, it is now easy to obtain moments of the
random variable
$N$. For
example, the first moment is:
\begin{equation}
\label{avgN}
\langle N \rangle =\frac{\sum_N Nz^N Z_N(\beta)}{\sum_N z^N Z_N(\beta)}=
z\cdot \frac{\partial \ln \tilde{\Xi}(\beta,z)}{\partial z}.
\end{equation}
Thus, we have replaced the fixed number of particles $N$ by a random number of
particles, which concentrates around an average controlled by the parameter
$\mu$, or
equivalently, $z$. The dominant value of $N$ is the one that maximizes the
product $z^NZ_N(\beta)$, or equivalently, $\beta\mu N+\ln Z_N(\beta)
=\beta(\mu N-F)$. Thus,
$\ln\tilde{\Xi}$ is related to $\ln Z_N$ by another kind of a Legendre
transform. 

Note that by passing to the grand--canonical ensemble, we have replaced two
extensive quantities, $E$ and $N$, be their respective conjugate intensive
variables, $T$ and $\mu$. This means that the grand partition function depends
only on one remaining extensive variable, which is $V$, and so, under ordinary
conditions, $\ln\Xi(\beta,z)$, or in its more complete notation, 
$\ln\Xi(\beta,z,V)$, depends linearly on $V$ at least in the 
thermodynamic limit, namely, $\lim_{V\to\infty}[\ln\Xi(\beta,z,V)]/V$ tends
to a constant that depends only on $\beta$ and $z$. What is this constant?
Let us examine again the first law in its more general form, as it appears in
eq.\ (\ref{g1stlaw}). For fixed $T$ and $\mu$, we have the following:
\begin{eqnarray}
P\delta V&=&\mu\delta N+T\delta S-\delta E\nonumber\\
&=&\delta(\mu N+TS-E)\nonumber\\
&=&\delta(\mu N-F)\nonumber\\
&\approx&kT\cdot\delta[\ln\Xi(\beta,z,V)]~~~~\mbox{$V$ large}
\end{eqnarray}
Thus, the constant of proportionality must be $P$. In other words,
the grand--canonical formula of the pressure is:
\begin{equation}
\label{gcpressure}
P=kT\cdot\lim_{V\to\infty}\frac{\ln\Xi(\beta,z,V)}{V}.
\end{equation}

\vspace{0.2cm}

\noindent
{\small {\it Example 3.6 -- more on the ideal gas.}
Applying formula (\ref{avgN}) on eq.\ (\ref{gcidealgas}), we readily obtain
\begin{equation}
\left<N\right>= \frac{zV}{\lambda^3}=\frac{e^{\mu/kT}V}{\lambda^3}.
\end{equation}
We see then that the grand--canonical factor $e^{\mu/kT}$ has the physical
meaning of the average number of ideal gas atoms in a cube of size
$\lambda\times\lambda\times\lambda$, where $\lambda$ is the thermal de Broglie
wavelength. Now,
applying eq.\ (\ref{gcpressure}) on (\ref{gcidealgas}), we get
\begin{equation}
P=\frac{kT\cdot e^{\mu/kT}}{\lambda^3}=\frac{\left<N\right>\cdot kT}{V},
\end{equation}
recovering again the equation of state of the ideal gas.
This is also demonstrates the principle of ensemble equivalence.}

\vspace{0.2cm}

Once again, it should be pointed out that beyond the obvious physical
significance of the grand--canonical ensemble, sometimes it proves
useful to work with it from the reason of pure mathematical convenience,
using the principle of enemble equivalence.
We will see this very clearly in the next chapters
on quantum statistics.

\subsubsection*{The Gibbs Ensemble}

Consider next the case where $T$ and $N$ are fixed, but $V$ is allowed to
fluctuate around an average volume controlled by the pressure $P$.
Again, we can analyze our relatively small test system surrounded by a heat
bath. The total energy is $E_0$ and the
total volume of the system and the heat bath is $V_0$. Suppose that we can calculate the density
of states of the heat bath as function of both its energy $E'$ and the volume
$V'$, call it $\Omega_{B}(E',V')$. A microstate is now a
combination $(\bx,V)$, where $V$ is the (variable) volume of
our subsystem. Once again, the same line of thought is used:
whenever our subsystem
is at state $(\bx,V)$, the heat bath can be in any one of
$\Omega_{B}(E_0-\calE(\bx),V_0-V)$ microstates of its own. Thus,
\begin{align}
P(\bx,V)&\propto \Omega_{B}(E_0-\calE(\bx),V_0-V)\nonumber\\
&= \exp\{S_{B}(E_0-\calE(\bx),V_0-V)/k\}\nonumber\\
&\approx \exp\left\{\frac{S_{B}(E_0,V_0)}{k}-\frac{1}{k}
\frac{\partial S_{B}}{\partial E}\cdot\calE(\bx)-
\frac{1}{k}\frac{\partial S_{B}}{\partial V}\cdot V
\right\}\nonumber\\
&\propto \exp\left\{-\frac{\calE(\bx)}{kT}-\frac{PV}{kT}\right\}\nonumber\\
&= \exp\{-\beta[\calE(\bx)+PV]\}.
\end{align}
The corresponding partition function that normalizes this probability function
is given by
\begin{equation}
Y_N(\beta,P)=\int_0^\infty e^{-\beta PV}Z_N(\beta,V)\mbox{d}V=
\int_0^\infty e^{-\beta PV}\mbox{d}V\sum_{\bx} e^{-\beta\calE(\bx)}.
\end{equation}
For a given $N$ and $\beta$, the function $Y_N(\beta,P)$ can be thought of as
the Laplace transform of $Z_N(\beta,V)$ as a function of $V$.
In the asymptotic regime (the thermodynamic limit),
$\lim_{N\to\infty}\frac{1}{N}\ln Y_N(\beta,P)$ is the Legendre transform of
$\lim_{N\to\infty}\frac{1}{N}\ln Z_N(\beta,V)$ for fixed $\beta$,
similarly to the Legendre relationship between the entropy and the canonical
log--partition function.
Note that analogously to eq.\ (\ref{avgN}), here the Gibbs partition function
serves as a cumulant generating function for the random variable $V$, thus,
for example,
\begin{equation}
\left<V\right>=-kT\cdot\frac{\partial \ln Y_N(\beta,P)}{\partial P}.
\end{equation}

As mentioned in an earlier footnote,
\begin{equation}
G=-kT\ln Y_N(\beta,P)=E-TS+PV=F+PV
\end{equation}
is the Gibbs free energy of the system, and for the case considered here, the force is
pressure and the conjugate variable it controls is the volume.
In analogy to the grand--canonical ensemble, here too, there is only one
extensive variable, this time, the variable $N$. 
Thus, $G$ should be (at least asymptotically) proportional to
$N$ with a constant of proportionality
that depends on the fixed values of $T$ and $P$. 

\vspace{0.2cm}

\noindent
{\small {\it Exercise 3.5:} Show that this constant is the chemical potential
$\mu$.}

All this is, of course, relevant when the physical system is a gas in a container. In
general, the Gibbs ensemble is obtained by a similar Legendre transform
replacing an extensive physical quantity of the canonical ensemble by its conjugate force.
For example, magnetic field is conjugate to magnetization, electric field is
conjugate to electric charge, mechanical force is conjugate to displacement, moment is
conjugate to angular shift, and so on. By the same token, the chemical
potential is a `force' that is conjugate to the number of particles in grand--canonical
ensemble, and (inverse) temperature is a `force' that controls the heat energy.

\begin{figure}[ht]
\hspace*{2cm}\input{legendre.pstex_t}
\caption{\small Diagram of Legendre relations between the various ensembles.}
\label{legendre}
\end{figure}
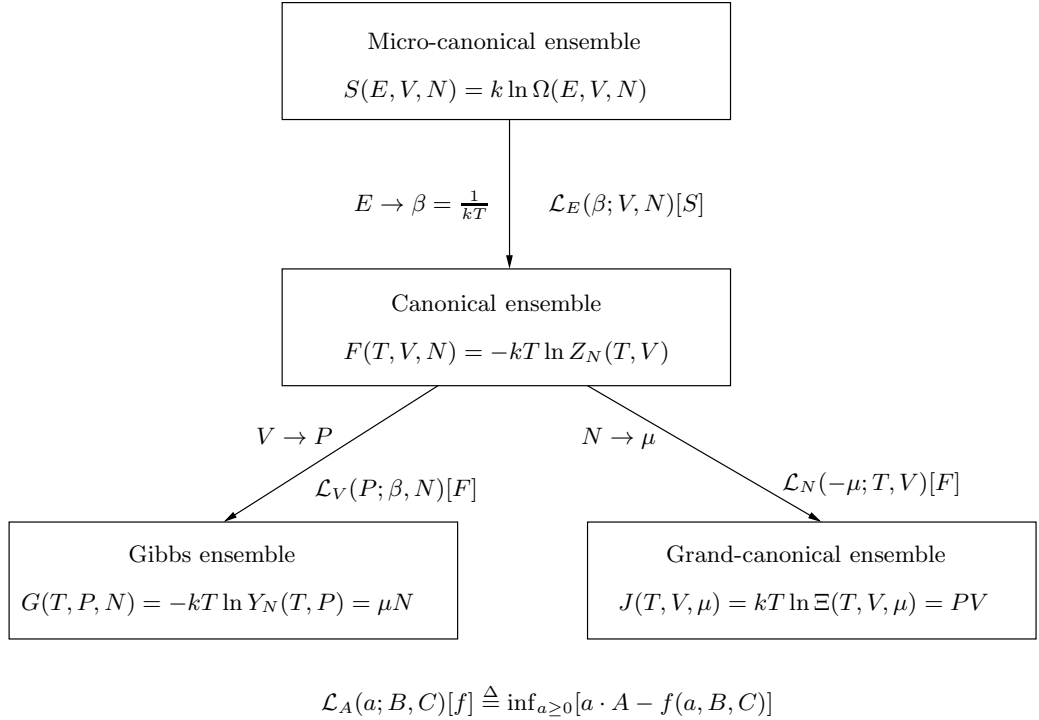

Fig.\ \ref{legendre} summarizes the thermodynamic potentials associated
with the various statistical ensembles.
The arrow between each two connected blocks in the diagram designates a passage from one
ensemble to another by a Legendre transform operator $\calL$ that is defined
generically at the bottom of the figure. In each passage, it is also indicated
which extensive variable is replaced by its conjugate intensive variable.

It should be noted, that at least mathematically, one could have defined
three more ensembles that would complete the picture of Fig.\ \ref{legendre} in a
symmetric manner. Two of the additional ensembles can be obtained by applying
Legendre transforms on $S(E,V,N)$, other than the transform that takes us
to the canonical ensemble. The first Legendre transform is w.r.t.\ the
variable $V$, replacing it by $P$, and the second additional ensemble is
w.r.t.\ the variable $N$, replacing it by $\mu$. Let us denote the new
resulting `potentials' (minus $kT$ times log--partition functions) by 
$A(E,P,N)$ and $B(E,V,\mu)$, respectively.
The third ensemble, with potential $C(E,P,\mu)$, whose
only extensive variable is $E$, could obtained by yet another Legendre
transform, either on $A(E,P,N)$ or $B(E,V,\mu)$ w.r.t.\ the appropriate
extensive variable. Of course, $A(E,P,N)$ and $B(E,V,\mu)$ are also connected directly 
to the Gibbs ensemble and to the
grand--canonical ensemble, respectively, both by Legendre--transforming w.r.t.\
$E$. While these three ensembles are not really used in physics, they might
prove useful to work with them for the purpose of calculating certain physical
quantities, by taking advantage of the principle of ensemble equivalence.

\vspace{0.2cm}

\noindent
{\small {\it Exercise 3.6:} Complete the diagram of Fig.\ \ref{legendre} by
the three additional ensembles just defined. 
Can you give physical meanings to $A$, $B$ and $C$?
Also, as said, $C(E,P,\mu)$ has only
$E$ as an extensive variable. 
Thus, $\lim_{E\to\infty}C(E,P,\mu)/E$ should be
a constant. What this constant is?}

Even more generally, we could start from a system model, whose micro--canonical
ensemble consists of many extensive variables $L_1,\ldots, L_n$, in addition
to the internal energy $E$ (not just $V$ and $N$). The entropy function is
then $S(E,L_1,\ldots,L_n,N)$. Here, $L_i$ can be, for example, volume, mass, electric
charge, electric polarization in each one of the three axes, 
magnetization in each one of three axes, and so on. The first Legendre
transform takes us from the micro--canonical ensemble to the canonical one upon
replacing $E$ by $\beta$. Then we can think of various Gibbs ensembles
obtained by replacing any subset of extensive variables $L_i$ by their
respective conjugate
forces $\lambda_i=T\partial S/\partial L_i$, $i=1,...,n$ (in the above
examples, pressure, gravitational force (weight), voltage (or electric
potential), electric fields,
and magnetic fields in the corresponding axes, respectively). In the extreme case, all
$L_i$ are replaced by $\lambda_i$ upon applying successive Legendre transforms, or
equivalently, a multi--dimensional Legendre transform:
\begin{equation}
G(T,\lambda_1,\ldots,\lambda_n,N)=-kT\sup_{L_1,\ldots,L_n}\left[kT\ln
Z_N(\beta,L_1,\ldots,L_n)-\lambda_1L_1-\ldots-\lambda_nL_n\right].
\end{equation}
Once again, there must be at least one extensive variable.

\newpage
\section{Quantum Statistics -- the Fermi--Dirac Distribution}

In our discussion thus far, we have largely taken for granted the assumption
that our system can be analyzed in the classical regime, where quantum
effects are negligible. This is, of course, not always the case, especially at
very low temperatures. Also, if radiation plays a role in the physical
system, then at very high frequency 
$\nu$, the classical approximation also breaks
down. Roughly speaking, $kT$ should be much larger than $h\nu$ for the
classical regime to be well 
justified.\footnote{One well--known example is black--body
radiation. According to the classical theory, the radiation density per unit
frequency grows proportionally to $kT\nu^2$, a function whose integral over
$\nu$, from zero to infinity, diverges (``the ultraviolet catastrophe''). This absurd is resolved
by quantum mechanical considerations, according to which the factor $kT$
should be replaced by $h\nu/[e^{h\nu/(kT)}-1]$, which 
is close to $kT$ at low frequencies, but decays
exponentially for $\nu > kT/h$.}
It is therefore necessary to address quantum effects in
statistical physics issues, most notably,
the fact that certain quantities, like energy and angular momentum (or spin), 
no longer take on values in the
continuum, but only in a discrete set, which depends on the system in
question.

Consider a gas of identical particles with discrete single--particle quantum
states, $1,2,\ldots,r,\ldots$, corresponding to energies
$$\epsilon_1\le\epsilon_2\le\ldots\le\epsilon_r\le\ldots.$$ 
Since the particles are assumed indistinguishable, then for a gas of $N$
particles, a micro--state is defined by the combination of occupation numbers
$$N_1,N_2,\ldots,N_r,\ldots,$$\
where $N_r$ is the number of particles at single
state $r$.

The first fundamental question is the following: What values can the occupation
numbers $N_1,N_2,\ldots$ assume? According to quantum mechanics, there
might be certain restrictions on these numbers. In particular, there are two
kinds of situations that may arise, which divide the various particles in the
world into two mutually exclusive classes. 

For the first class of particles, there are no
restrictions at all. The occupation numbers can assume any non--negative
integer value ($N_r=0,1,2,\ldots$), Particles of this class are called 
{\it Bose--Einstein (BE) particles}\footnote{Bosons were first introduced by
Bose (1924) in order to derive Planck's radiation law, and Einstein applied
this finding
in the same year to a perfect gas of particles.}
or {\it bosons} for short.
Another feature of bosons is that their spins are always integral multiples
of $\hbar$, namely, $0$, $\hbar$, $2\hbar$, etc.
Examples of bosons are
photons, $\pi$ mesons and $K$ mesons. 
We will focus on them in the next
chapter.

In the second class of particles, the occupation numbers are restricted by the
{\it Pauli exclusion principle} (discovered in 1925),
according to which no more than one particle
can occupy a given quantum state $r$ (thus $N_r$ is either $0$ or $1$
for all $r$), since the wave function of two such
particles is anti--symmetric and thus vanishes if they assume the same quantum
state (unlike bosons for which the wave function is symmetric).
Particles of this kind are called {\it Fermi--Dirac (FD)
particles}\footnote{Introduced independently by Fermi and Dirac in 1926.}
or {\it fermions} for short. Another characteristic of fermions is that
their spins are always odd multiples 
of $\hbar/2$, namely, $\hbar/2$, $3\hbar/2$, $5\hbar/2$, etc.
Examples of fermions are electrons, positrons, protons,
and neutrons. The statistical mechanics of fermions will be discussed in this chapter.

\subsection{Combinatorial Derivation of the FD Statistics}

Consider a gas of $N$ fermions in volume $V$ and temperature $T$. In the
thermodynamic limit, where the dimensions of the system are large, 
the discrete single--particle energy levels $\{\epsilon_r\}$ are very close to one another.
Therefore, instead of considering each one of them 
individually, we shall consider groups of neighboring states.
Since the energy levels in each group are very close, we will approximate all
of them by a single energy value. Let us label these groups by $s=1,2,\ldots$.
Let group no.\ $s$ contain $G_s$ single--particle states and let the
representative energy level be $\hat{\epsilon}_s$. Let us assume that $G_s \gg
1$. A microstate of the gas is now defined by the occupation numbers
$$\hat{N}_1,\hat{N}_2,\ldots,\hat{N}_s,\ldots,$$
$\hat{N}_s$ being the total number of particles in group no.\ $s$, where, of
course $\sum_s\hat{N}_s=N$.

To derive the equilibrium behavior of this system, we analyze the Helmholtz
free energy $F$ as a function of the occupation numbers, and use the fact
that in equilibrium, it should be minimum. Since
$E=\sum_s\hat{N}_s\hat{\epsilon}_s$ and $F=E-TS$, this boils down to the
evaluation of the entropy $S=k\ln\Omega(\hat{N}_1,\hat{N}_2,\ldots)$.
Let $\Omega_s(\hat{N}_s)$ be the number of ways of putting $\hat{N}_s$ particles
into $G_s$ states of group no.\ $s$. Now,
for fermions each one of the $G_s$ states is either empty or occupied by one
particle. Thus,
\begin{equation}
\Omega_s(\hat{N}_s)=\frac{G_s!}{\hat{N}_s!(G_s-\hat{N}_s)!}
\end{equation}
and
\begin{equation}
\Omega(\hat{N}_1,\hat{N}_2,\ldots)=\prod_s\Omega_s(\hat{N}_s).
\end{equation}
Therefore,
\begin{eqnarray}
F(\hat{N}_1,\hat{N}_2,\ldots)&=&\sum_s[\hat{N}_s\hat{\epsilon}_s-kT\ln\Omega_s(\hat{N}_s)]
\nonumber\\
&\approx&\sum_s\left[\hat{N}_s\hat{\epsilon}_s-kTG_sh_2\left(\frac{\hat{N}_s}{G_s}\right)\right].
\end{eqnarray}
As said, we wish to minimize $F(\hat{N}_1,\hat{N}_2,\ldots)$ s.t.\ the
constraint $\sum_s\hat{N}_s=N$. Consider then the minimization of the Lagrangian
\begin{equation}
L=\sum_s\left[\hat{N}_s\hat{\epsilon}_s-kTG_s
h_2\left(\frac{\hat{N}_s}{G_s}\right)\right]-
\lambda\left(\sum_s\hat{N}_s-N\right).
\end{equation}
The solution is readily obtained to read
\begin{equation}
\hat{N}_s=\frac{G_s}{e^{(\hat{\epsilon}_s-\lambda)/kT}+1}
\end{equation}
where the Lagrange multiplier $\lambda$ is determined to satisfy
the constraint
\begin{equation}
\sum_s\frac{G_s}{e^{(\hat{\epsilon}_s-\lambda)/kT}+1}=N.
\end{equation}

\vspace{0.2cm}

\noindent
{\small {\it Exercise 4.1:} After showing the general relation
$\mu=(\partial F/\partial N)_{T,V}$, convince yourself that
$\lambda=\mu$, namely, the Lagrange multiplier $\lambda$ has
the physical meaning of the chemical potential. From now on, then
we replace the notation $\lambda$ by $\mu$.}

\vspace{0.2cm}

Note that $\hat{N}_s/G_s$ is the mean
occupation number $\bar{N}_r$ of a single state $r$ within group no.\ $s$.
I.e.,
\begin{equation}
\bar{N}_r=\frac{1}{e^{(\epsilon_r-\mu)/kT}+1}
\end{equation}
with the constraint
\begin{equation}
\sum_r\frac{1}{e^{(\epsilon_r-\mu)/kT}+1}=N.
\end{equation}
The nice thing is that this result no longer depends on the partition into
groups. This is the FD distribution.

\begin{figure}[h!t!b!]
\centering
\includegraphics[width=8.5cm, height=8.5cm]{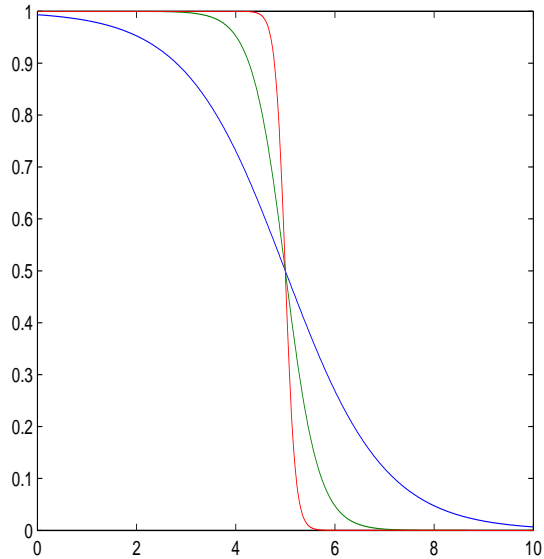}
\caption{\small Illustration of the FD distribution. As $T$ decreases,
the curve becomes closer to $\bar{n}_r=u(\mu-\epsilon_r)$,
where $u(\cdot)$ is the unit step function.}
\label{fermidirac}
\end{figure}

\newpage
\subsection{FD Statistics From the Grand--Canonical
Ensemble}

Thanks to the principle of ensemble equivalence,
an alternative, simpler derivation
of the FD distribution results from the use of the
grand--canonical ensemble. Beginning from the canonical partition function
\begin{equation}
\label{qg}
Z_N(\beta)=\sum_{N_1=0}^1\sum_{N_2=0}^1\ldots
\delta\left(\sum_rN_r=N\right)e^{-\beta\sum_rN_r\epsilon_r},
\end{equation}
we pass to the grand--canonical ensemble in the following manner:
\begin{eqnarray}
\Xi(\beta,\mu)&=&\sum_{N=0}^\infty e^{\beta\mu N}
\sum_{N_1=0}^1\sum_{N_2=0}^1\ldots
\delta\left(\sum_rN_r=N\right)e^{-\beta\sum_rN_r\epsilon_r}\nonumber\\
&=&\sum_{N=0}^\infty
\sum_{N_1=0}^1\sum_{N_2=0}^1\ldots e^{\beta\mu \sum_rN_r}
\delta\left(\sum_rN_r=N\right)e^{-\beta\sum_rN_r\epsilon_r}\nonumber\\
&=&\sum_{N=0}^\infty
\sum_{N_1=0}^1\sum_{N_2=0}^1\ldots
\delta\left(\sum_rN_r=N\right)e^{\beta\sum_rN_r(\mu-\epsilon_r)}\nonumber\\
&=&\sum_{N_1=0}^1\sum_{N_2=0}^1\ldots
\left[\sum_{N=0}^\infty\delta\left(
\sum_rN_r=N\right)\right]e^{\beta\sum_rN_r(\mu-\epsilon_r)}\nonumber\\
&=&\sum_{N_1=0}^1\sum_{N_2=0}^1\ldots
e^{\beta\sum_rN_r(\mu-\epsilon_r)}\nonumber\\
&=&\prod_r\left[\sum_{N_r=0}^1e^{\beta N_r(\mu-\epsilon_r)}\right]\nonumber\\
&=&\prod_r\left[1+e^{\beta(\mu-\epsilon_r)}\right].
\end{eqnarray}
Note that this product form of the grand partition function means that under
the grand--canonical ensemble the binary random variables $\{N_r\}$ are
statistically independent,
i.e.,
\begin{equation}
P(N_1,N_2,\ldots)=\prod_rP_r(N_r)
\end{equation}
where
\begin{equation}
P_r(N_r)=\frac{e^{\beta N_r(\mu-\epsilon_r)}}{1+e^{\beta(\mu-\epsilon_r)}},
~~~~~N_r=0,1,~~~~~r=1,2,\ldots.
\end{equation}
Thus,
\begin{equation}
\bar{N}_r=\mbox{Pr}\{N_r=1\}=\frac{e^{
(\mu-\epsilon_r)/kT}}{1+e^{(\mu-\epsilon_r)/kT}}=
\frac{1}{e^{(\epsilon_r-\mu)/kT}+1}.
\end{equation}

\subsection{The Fermi Energy}

Let us now examine what happens if the system is cooled
to the absolute zero ($T\to 0$). It should be kept in mind that the chemical
potential $\mu$ depends on $T$, so let $\mu_0$ be the chemical
potential at $T=0$. It is readily seen that $\bar{N}_r$ approaches as
step function (see Fig.\ \ref{fermidirac}), namely, all energy levels $\{\epsilon_r\}$ below $\mu_0$
are occupied ($\bar{N}_r\approx 1$) 
by a fermion, whereas all those that are above $\mu_0$
are empty ($\bar{N}_r\approx 0$). The explanation is simple:
Pauli's exclusion principle does not allow all particles to
reside at the ground state at $T=0$ since then many of them
would have occupied the same quantum state. The minimum energy 
of the system that can possibly be achieved
is when all energy levels are filled up, one by one, starting 
from the ground state up to some maximum level, 
which is exactly $\mu_0$. This explains why even at the absolute zero, fermions
have energy.\footnote{Indeed, free
electrons in a metal continue
to be mobile and free even at $T=0$.} The maximum occupied energy level 
in a gas of non--interacting fermions at the absolute zero is called the
{\it Fermi energy}, which we shall denote by $\epsilon_F$.
Thus, $\mu_0=\epsilon_F$, and then the 
FD distribution at very low temperatures is
approximately
\begin{equation}
\bar{N}_r=\frac{1}{e^{(\epsilon_r-\epsilon_F)/kT}+1}.
\end{equation}
We next take a closer look on the FD distribution, taking into account
the density of states.
Consider a metal box of dimensions $L_x\times L_y\times L_z$ and
hence volume $V=L_xL_yL_z$. The
energy level associated with quantum number $(l_x,l_y,l_z)$ is given by
\begin{equation}
\label{energies}
\epsilon_{l_x,l_y,l_z}=\frac{\pi^2\hbar^2}{2m}\left(\frac{l_x^2}{L_x^2}+
\frac{l_y^2}{L_y^2}+\frac{l_z^2}{L_z^2}\right)
=\frac{\hbar^2}{2m}(k_x^2+k_y^2+k_z^2),
\end{equation}
where the $k$'s are the wave numbers pertaining to the various solutions of the
Schr\"odinger equation. First, we would like to count how many quantum states
$\{(l_x,l_y,l_z)\}$
give rise to energy between $\epsilon$ and $\epsilon+\mbox{d}\epsilon$.
We denote this number by $g(\epsilon)\mbox{d}\epsilon$, where $g(\epsilon)$ is
the density of states. 
\begin{eqnarray}
g(\epsilon)\mbox{d}\epsilon&=&\sum_{l_x,l_y,l_z}
1\left\{\frac{2m\epsilon}{\hbar^2}\le\frac{\pi^2l_x^2}{L_x^2}+
\frac{\pi^2l_y^2}{L_y^2}+\frac{\pi^2l_z^2}{L_z^2}\le
\frac{2m(\epsilon+\mbox{d}\epsilon)}{\hbar^2}\right\}\nonumber\\
&\approx&\frac{L_xL_yL_z}{\pi^3}\cdot\mbox{Vol}\left\{\vec{\bk}:~
\frac{2m\epsilon}{\hbar^2}\le\|\vec{\bk}\|^2\le
\frac{2m(\epsilon+\mbox{d}\epsilon)}{\hbar^2}\right\}\nonumber\\
&=&\frac{V}{\pi^3}\cdot
\mbox{Vol}\left\{\vec{\bk}:~\frac{2m\epsilon}{\hbar^2}\le\|\vec{\bk}\|^2\le
\frac{2m(\epsilon+\mbox{d}\epsilon)}{\hbar^2}\right\}.
\end{eqnarray}
A volume element pertaining to a given value of $K=\|\vec{\bk}\|$ 
($\vec{\bk}$ being $k_x\hat{x}+k_y\hat{y}+k_z\hat{z}$) is 
given by $\mbox{d}K$ times 
the surface
area of sphere of radius $K$, namely, 
$4\pi K^2\mbox{d}K$, but it has to be divided by $8$, to
account for the fact that the
components of $\vec{\bk}$ are positive. I.e., it is
$\frac{\pi}{2}K^2\mbox{d}K$. From eq.\ (\ref{energies}), we have
$\epsilon=\hbar^2K^2/2m$, and so
\begin{equation}
K^2\mbox{d}K=
\frac{2m\epsilon}{\hbar^2}\cdot\frac{1}{2\hbar}\sqrt{\frac{2m}{\epsilon}}\mbox{d}\epsilon
=\frac{\sqrt{2\epsilon m^3}\mbox{d}\epsilon}{\hbar^3}
\end{equation}
Therefore, combining the above, we get
\begin{equation}
g(\epsilon)= 
\frac{\sqrt{2\epsilon m^3}V}{2\pi^2\hbar^3}.
\end{equation}
For electrons, spin values of $\pm 1/2$ are allowed, so this density 
should be doubled, and so
\begin{equation}
g_e(\epsilon)= 
\frac{\sqrt{2\epsilon m^3}V}{\pi^2\hbar^3}.
\end{equation}
Approximating the equation of the constraint on the total number of electrons, we get
\begin{eqnarray}
\label{ef}
N_e&=&\sum_r\frac{1}{e^{(\epsilon_r-\epsilon_F)/kT}+1}\nonumber\\
&\approx&\frac{\sqrt{2m^3}V}{\pi^2\hbar^3}\cdot
\int_0^\infty\frac{\sqrt{\epsilon}\mbox{d}\epsilon}{e^{(\epsilon-\epsilon_F)/kT}+1}\nonumber\\
&\approx&\frac{\sqrt{2m^3}V}{\pi^2\hbar^3}\cdot
\int_0^{\epsilon_F}\sqrt{\epsilon}\mbox{d}\epsilon\nonumber~~~~~T\approx 0\\
&=&\frac{\sqrt{2m^3}V}{\pi^2\hbar^3}\cdot\frac{2\epsilon_F^{3/2}}{3}
\end{eqnarray}
which easily leads to the following simple formula for the Fermi energy:
\begin{equation}
\epsilon_F=\frac{\hbar^2}{2m}\left(\frac{3\pi^2N_e}{V}\right)^{2/3}
=\frac{\hbar^2}{2m}(3\pi^2\rho_e)^{2/3},
\end{equation}
where $\rho_e$ is the electron density.
In most metals $\epsilon_F$ is about the order of 5--10 electron--volts (eV's), whose
equivalent temperature $T_F=\epsilon_F/k$ (the Fermi temperature) is of the
order of magnitude of $100,000^o$K. 
So the Fermi energy is much larger than $kT$ in laboratory conditions. In
other words,
electrons in a metal behave like a gas at an extremely high temperature.
This means that the internal pressure in metals (the Fermi pressure) is huge
and that's a reason why metals are almost incompressible.
This kind of pressure also stabilizes a neutron star (a Fermi gas of
neutrons) or a white dwarf star (a Fermi gas of electrons) against the inward
pull of gravity, which would ostensibly collapse the star into a Black Hole.
Only when a star is sufficiently massive to overcome the degeneracy pressure
can it collapse into a singularity.

\vspace{0.2cm}

\noindent
{\small {\it Exercise 4.2:} Derive an expression for $\left<\epsilon^n\right>$ 
of an electron near
$T=0$, in terms of $\epsilon_F$.}

\subsection{Useful Approximations of Fermi Integrals}
\label{approx}

Before considering applications, it will be instructive to develop some 
useful approximations for integrals associated with the Fermi function
\begin{equation}
f(\epsilon)\dfn\frac{1}{e^{(\epsilon-\mu)/kT}+1}.
\end{equation}
For example, if we wish to calculate the average energy, we have to
deal with an integral like
$$\int_0^\infty\epsilon^{3/2}f(\epsilon)\mbox{d}\epsilon.$$
Consider then, more generally, an integral of the form
$$I_n=\int_0^\infty\epsilon^{n}f(\epsilon)\mbox{d}\epsilon.$$
Upon integrating by parts, we readily have
\begin{eqnarray}
I_n&=&f(\epsilon)\cdot\frac{\epsilon^{n+1}}{n+1}\bigg|_0^\infty-\frac{1}{n+1}\int_0^\infty\epsilon^{n+1}
f'(\epsilon)\mbox{d}\epsilon\nonumber\\
&=&-\frac{1}{n+1}\int_0^\infty\epsilon^{n+1}
f'(\epsilon)\mbox{d}\epsilon
\end{eqnarray}
Changing variables to $x=(\epsilon-\mu)/kT$,
\begin{eqnarray}
I_n&=&-\frac{1}{n+1}\int_{-\mu/kT}^\infty(\mu+kTx)^{n+1}
\phi'(x)\mbox{d}x\nonumber\\
&\approx&-\frac{\mu^{n+1}}{n+1}\int_{-\infty}^\infty\left(1+\frac{kTx}{\mu}\right)^{n+1}
\phi'(x)\mbox{d}x,
\end{eqnarray}
where we have introduced the scaled version of $f$, which is $\phi(x)=f(\mu+kTx)=1/(e^x+1)$
and $\phi'$ is its derivative, and where in the second line we are assuming
$\mu \gg kT$.
Applying the Taylor series expansion to the binomial term (recall that $n$ is
not necessarily integer), and using the symmetry of $\phi'$ around the origin, we have
\begin{eqnarray}
I_n&=&-\frac{\mu^{n+1}}{n+1}\int_{-\infty}^\infty\left[1+(n+1)\frac{kTx}{\mu}+\frac{n(n+1)}{2}
\left(\frac{kTx}{\mu}\right)^2+\ldots\right]\phi'(x)\mbox{d}x\nonumber\\
&=&-\frac{\mu^{n+1}}{n+1}\left[\int_{-\infty}^\infty\phi'(x)\mbox{d}x+
\frac{n(n+1)}{2}\left(\frac{kT}{\mu}\right)^2
\int_{-\infty}^{\infty}x^2\phi'(x)\mbox{d}x+\cdot\cdot\cdot\right]\nonumber\\
&\approx&
\frac{\mu^{n+1}}{n+1}\left[1+
\frac{n(n+1)\pi^2}{6}\left(\frac{kT}{\mu}\right)^2\right]
\end{eqnarray}
where the last line was obtained by calculating the integral of $x^2\phi'(x)$
using a power series expansion. Note that this series contains only even
powers of $kT/\mu$, thus the convergence is rather fast. 
Let us now repeat the calculation of eq.\ (\ref{ef}), this time at $T>0$.
\begin{eqnarray}
\label{ef1}
\rho_e&\approx&\frac{\sqrt{2m^3}}{\pi^2\hbar^3}\cdot
\int_0^\infty\frac{\sqrt{\epsilon}\mbox{d}\epsilon}{e^{(\epsilon-\mu)/kT}+1}\nonumber\\
&=&\frac{\sqrt{2m^3}}{\pi^2\hbar^3}\cdot I_{1/2}\nonumber\\
&=&\frac{\sqrt{2m^3}}{\pi^2\hbar^3}\cdot
\frac{2}{3}\mu^{3/2}\left[1+\frac{\pi^2}{8}\left(\frac{kT}{\mu}\right)^2\right]
\end{eqnarray}
which gives
\begin{eqnarray}
\epsilon_F&=&\mu\left[1+\frac{\pi^2}{8}\left(\frac{kT}{\mu}\right)^2\right]^{2/3}\nonumber\\
&\approx&\mu\left[1+\frac{\pi^2}{12}\left(\frac{kT}{\mu}\right)^2\right]\nonumber\\
&=&\mu+\frac{(\pi kT)^2}{12\mu}.
\end{eqnarray}
This relation between $\mu$ and $\epsilon_F$ can be easily inverted by
solving a simple quadratic equation, which yields
\begin{eqnarray}
\mu&\approx&\frac{\epsilon_F+\epsilon_F\sqrt{1-(\pi
kT/\epsilon_F)^2/3}}{2}\nonumber\\
&\approx&\epsilon_F\left[1-\frac{\pi^2}{12}\cdot\left(\frac{kT}{\epsilon_F}\right)^2\right]\nonumber\\
&=&\epsilon_F\left[1-\frac{\pi^2}{12}\cdot\left(\frac{T}{T_F}\right)^2\right]
\end{eqnarray}
Since $T/T_F \ll 1$ for all $T$ in the interesting range, we observe
that the chemical potential depends extremely weakly on $T$. In other words,
we can safely approximate $\mu\approx\epsilon_F$ for all relevant temperatures
of interest. The assumption that $kT \ll \mu$ was found self--consistent with
the result $\mu\approx\epsilon_F$.

Having established the approximation $\mu\approx\epsilon_F$,
we can now calculate the average energy of the electron at
an arbitrary temperature $T$:
\begin{eqnarray}
\left<\epsilon\right>&=&\frac{\sqrt{2m^3}}{\pi^2\hbar^3\rho_e}
\int_0^\infty\frac{\epsilon^{3/2}\mbox{d}\epsilon}{e^{(\epsilon-\epsilon_F)/kT}
+1}\nonumber\\
&=&\frac{\sqrt{2m^3}}{\pi^2\hbar^3\rho_e}\cdot I_{3/2}\nonumber\\
&\approx&\frac{\sqrt{2m^3}}{\pi^2\hbar^3\rho_e}\cdot
\frac{2\epsilon_F^{5/2}}{5}
\left[1+\frac{5\pi^2}{8}\left(\frac{T}{T_F}\right)^2
\right]\nonumber\\
&=&\frac{3\hbar^2}{10m}\cdot(3\pi^2\rho_e)^{2/3}\cdot
\left[1+\frac{5\pi^2}{8}\left(\frac{T}{T_F}\right)^2
\right]\nonumber\\
&=&\frac{3\epsilon_F}{5}\cdot
\left[1+\frac{5\pi^2}{8}\left(\frac{T}{T_F}\right)^2
\right]
\end{eqnarray}
Note that the dependence of the average per--particle 
energy on the temperature is
drastically different from that of the ideal gas. While in the idea
gas it was linear ($\left<\epsilon\right>=3kT/2$), 
here it is actually almost a constant, independent of the temperature
(just like the chemical potential).

The same technique can be used, of course, to calculate any moment of
the electron energy.

\subsection{Applications of the FD Distribution}

The FD distribution is at the heart of modern solid--state physics and
semiconductor physics 
(see also, for example, 
\cite[Section 4.5]{Beck76}) 
and indeed frequently encountered in related courses, like
courses on semiconductor devices.
It is also useful in understanding the physics of white
dwarfs. We next briefly touch upon 
the very basics of conductance in solids
(as a much more comprehensive
treatment is given, of course, in other courses),
as well as on two other applications: thermionic emission and photoelectric
emission.

\subsubsection{Electrons in a Solid}

The structure of the electron energy levels in a solid
are basically obtained using quantum--mechanical considerations. In the case
of a crystal,
this amounts to solving the Schr\"odinger equation in a periodic potential,
stemming from the corresponding periodic lattice structure. Its
idealized form, which ignores the size of each atom, is given by a 
train of equispaced Dirac 
delta functions. This is an extreme case of the so called
{\it Kronig--Penney model}, where the potential function is a periodic
rectangular on--off function (square wave function),
and it leads to a certain band structure. In particular,
bands of allowed energy levels
are alternately interlaced with bands of forbidden energy levels. The Fermi energy
level $\epsilon_F$, which depends on the overall concentration of electrons, may either
fall in an allowed band or in a forbidden band. The former case is the case of
a metal, whereas the latter case is the case of an insulator or a
semiconductor (the difference being only how wide is the forbidden band in
which $\epsilon_F$ lies). While in metals it is impossible to change
$\epsilon_F$, in semiconductors, it is possible by doping.

A semiconductor can then be thought of as a system with electron orbitals grouped into
two\footnote{We treat both bands as single bands for our purposes. It does not
matter that both may be themselves groups of (sub)bands with additional gaps
within each group.} energy
bands separated by an energy gap. The lower band is the {\it valence band}
(where electrons are tied to their individual atoms) and
the upper band is the {\it conduction band}, where they are free. In a pure semiconductor at $T=0$,
all valence orbitals are occupied with electrons and all conduction orbitals
are empty. A full band cannot carry any current so a pure semiconductor at
$T=0$ is an insulator. In a pure semiconductor the Fermi energy is exactly in
the middle of the gap between the valence band (where $f(\epsilon)$ is very close 1) 
and the conduction band (where $f(\epsilon)$ is very close to 0).
Finite conductivity in a semiconductor follows either
from the presence of electrons in the conduction band (conduction electrons)
or from unoccupied orbitals in the valence band (holes).

Two different mechanisms give rise to conduction electrons and holes: The
first is thermal excitation of electrons from the valence band to the
conduction band, and the second is the presence of impurities that change the
balance between the number of orbitals in the valence band and the number of
electrons available to fill them.

We will not delve into this too much beyond this point, since this material is
well--covered in other 
courses solid state physics. Here we just demonstrate the use of the FD
distribution in order to calculate the density of charge carriers.
The density of charge carriers $n$ of the conduction band is found by
integrating up, from the conduction band edge $\epsilon_C$, the product of the
density of states $g_e(\epsilon)$ and the FD distribution $f(\epsilon)$,
i.e.,
\begin{equation}
n=\int_{\epsilon_C}^\infty\mbox{d}\epsilon\cdot g_e(\epsilon)f(\epsilon)=
\frac{\sqrt{2m^3}}{\pi^2\hbar^3}\int_{\epsilon_C}^\infty\frac{\sqrt{\epsilon-\epsilon_C}\mbox{d}\epsilon}{
e^{(\epsilon-\epsilon_F)/kT}+1},
\end{equation}
where here $m$ designates the {\it effective mass} of the electron\footnote{The
effective mass is
obtained by a second order Taylor series expansion of the energy as a function
of the wavenumber (used to obtain the density of states), 
and thinking of the coefficient of the quadratic term as
$\hbar^2/2m$.} and where
we have taken the density of states to be proportional to
$\sqrt{\epsilon-\epsilon_C}$ since $\epsilon_C$ is now the reference energy
and only the difference $\epsilon-\epsilon_C$
goes for kinetic energy.\footnote{Recall that earlier we calculated the
density of states for a simple potential well, not for a periodic potential
function. Thus, the earlier expression of $g_e(\epsilon)$ is not correct here.}
For a semiconductor at room temperature, $kT$ is much smaller than the gap, and so
\begin{equation}
f(\epsilon)\approx e^{-(\epsilon-\epsilon_F)/kT}
\end{equation}
which yields the approximation
\begin{eqnarray}
n&\approx&\frac{\sqrt{2m^3}}{\pi^2\hbar^3}\cdot
e^{\epsilon_F/kT}\int_{\epsilon_C}^\infty\mbox{d}\epsilon\cdot\sqrt{\epsilon-\epsilon_C}\cdot
e^{-\epsilon/kT}\nonumber\\
&=& \frac{\sqrt{2m^3}}{\pi^2\hbar^3}\cdot
e^{-(\epsilon_C-\epsilon_F)/kT}\int_0^\infty
\mbox{d}\epsilon\cdot\sqrt{\epsilon}e^{-\epsilon/kT}\nonumber\\
&=&\frac{\sqrt{2(mkT)^3}}{\pi^2\hbar^3}\cdot
e^{-(\epsilon_C-\epsilon_F)/kT}\int_0^\infty
\mbox{d}x\cdot\sqrt{x}e^{-x}\nonumber\\
&=&\frac{\sqrt{\pi}}{2}
\cdot\frac{\sqrt{2(mkT)^3}}{\pi^2\hbar^3}\cdot
e^{-(\epsilon_C-\epsilon_F)/kT}\nonumber\\
&=&\frac{1}{4}\cdot\left(\frac{2mkT}{\pi\hbar^2}\right)^{3/2}\cdot
e^{-(\epsilon_C-\epsilon_F)/kT}.
\end{eqnarray}
We see then that the density of conduction electrons, and hence also the conduction
properties, depend critically on the
gap between $\epsilon_C$ and $\epsilon_F$. A similar calculation holds for the
holes, of course.

\subsubsection{Thermionic Emission$^*$}

If a refractory metal, like tungsten, is heated to a temperature 
of a few hundred degrees below the melting point, an electron emission can be
drawn from it to a positive anode. From a quantum--mechanical 
viewpoint, the heated metal can be regarded as a potential well with finitely
high walls determined by the surface potential barrier. Thus, some of the
incident particles will have sufficient energy to surmount the surface barrier
(a.k.a.\ the surface work function) and hence will be emitted. The work
function $\phi$ varies between 2eV and 6eV for pure metals. The electron will
not be emitted unless the energy component normal to the surface would
exceed $\epsilon_F+\phi$. The excess energy beyond this threshold is in the
form of translational kinetic energy which dictates the velocity away from the
surface.

The analysis of this effect is made by transforming the energy distribution
into a distribution in terms of the three component velocities $v_x$, $v_y$,
$v_z$. We begin with the expression of the energy of a single
electron\footnote{We are assuming that the potential barrier $\phi$ is
fairly large (relative to $kT$), such that the relationship between energy
and quantum numbers is reasonably well approximated by that of a particle in
a box.}
\begin{equation}
\epsilon= \frac{1}{2}m(v_x^2+v_y^2+v_z^2)=\frac{\pi^2\hbar^2}{2m}\left(
\frac{l_x^2}{L_x^2}+\frac{l_y^2}{L_y^2}+\frac{l_z^2}{L_z^2}\right).
\end{equation}
Thus, $\mbox{d}v_x=h\mbox{d}l_x/(2mL_x)$ and similar relations hold for
two the other components, which together yield
\begin{equation}
\mbox{d}l_x\mbox{d}l_y\mbox{d}l_z=
\left(\frac{m}{h}\right)^3V\mbox{d}v_x\mbox{d}v_y\mbox{d}v_z,
\end{equation}
where we have divided by 8 since every quantum state can be occupied by only
one out of 8 combinations of the signs of the three component veocities.
Thus, we can write the distribution function of the number of electrons in
a cube $\mbox{d}v_x\times\mbox{d}v_y\times\mbox{d}v_z$ as
\begin{equation}
\mbox{d}N=2V\left(\frac{m}{h}\right)^3\frac{\mbox{d}v_x\mbox{d}v_y\mbox{d}v_z}
{1+\exp\left\{\frac{1}{kT}\left[\frac{1}{2}m(v_x^2+v_y^2+v_z^2)-\epsilon_F\right]\right\}},
\end{equation}
where we have taken the chemical potential of the electron gas to be $\epsilon_F$, independently
of temperature,  as was justified in the previous subsection.
Assuming that the surface is parallel to the YZ plane, the minimum escape
velocity in the $x$--direction is $v_0=\sqrt{\frac{2}{m}(\epsilon_F+\phi)}$
and there are no restrictions on $v_y$ and $v_z$. 
The current along the $x$--direction is 
\begin{eqnarray}
I&=&\frac{\mbox{d}q}{\mbox{d}t} 
=\frac{q_e\mbox{d}N[\mbox{leaving the surface}]}{\mbox{d}t}\nonumber\\
&=&\frac{q_e}{\mbox{d}t}\int_{v_0}^\infty\int_{-\infty}^{+\infty}
\int_{-\infty}^{+\infty}\frac{v_x\mbox{d}t}{L_x}\cdot
2V\left(\frac{m}{h}\right)^3\frac{\mbox{d}v_x\mbox{d}v_y\mbox{d}v_z}
{1+\exp\left\{\frac{1}{kT}\left[\frac{1}{2}m(v_x^2+v_y^2+v_z^2)-\epsilon_F\right]\right\}}\nonumber\\
&=&2L_yL_zq_e\left(\frac{m}{h}\right)^3
\int_{v_0}^\infty v_x\mbox{d}v_x\int_{-\infty}^{+\infty}
\frac{\mbox{d}v_y\mbox{d}v_z}
{1+\exp\left\{\frac{1}{kT}\left[\frac{1}{2}m(v_x^2+v_y^2+v_z^2)-\epsilon_F\right]\right\}},
\end{eqnarray}
where the factor $v_x\mbox{d}t/L_x$ in the second line is the fraction of electrons close enough
to the sufrace so as to be emitted within time $\mbox{d}t$.
Thus, the current density (current per unity area) is
\begin{equation}
J=2q_e\left(\frac{m}{h}\right)^3\int_{v_0}^\infty\mbox{d}v_x\cdot v_x
\int_{-\infty}^{+\infty}\int_{-\infty}^{+\infty}\frac{\mbox{d}v_y\mbox{d}v_z}
{1+\exp\left\{\frac{1}{kT}\left[\frac{1}{2}m(v_x^2+v_y^2+v_z^2)-\epsilon_F\right]\right\}}.
\end{equation}
As for the inner double integral, transform to polar coordinates to obtain
\begin{eqnarray}
&&\int_{-\infty}^{+\infty}\int_{-\infty}^{+\infty}\frac{\mbox{d}v_y\mbox{d}v_z}
{1+\exp\left\{\frac{1}{kT}\left[\frac{1}{2}m(v_x^2+v_y^2+v_z^2)-\epsilon_F\right]\right\}}\nonumber\\
&=&2\pi\int_0^\infty\frac{v_{yz}\mbox{d}v_{yz}}
{1+e^{mv_{yz}^2/2kT}\cdot
\exp\left[\frac{1}{kT}\left(\frac{1}{2}mv_x^2-\epsilon_F\right)\right]}\nonumber\\
&=&\frac{2\pi
kT}{m}\int_0^\infty\frac{\mbox{d}u}{1+
\exp\left[\frac{1}{kT}\left(\frac{1}{2}mv_x^2-\epsilon_F\right)\right]\cdot
e^u}~~~~~~~~~~~~~u=mv_{yz}^2/2kT\nonumber\\
&=&\frac{2\pi
kT}{m}\ln\left\{1+\exp\left[\frac{1}{kT}\left(\epsilon_F-\frac{1}{2}mv_x^2\right)\right]\right\}
\end{eqnarray}
which yields
\begin{equation}
J=\frac{4\pi m^2q_ekT}{h^3}\int_{v_0}^\infty\mbox{d}v_x\cdot v_x\ln
\left\{1+\exp\left[\frac{1}{kT}\left(\epsilon_F-\frac{1}{2}mv_x^2\right)\right]\right\}.
\end{equation}
Now, since normally $\phi \gg kT$, the exponent in the integrand is very small
throughout the entire range of integration
and so, it is safe to approximate it by $\ln(1+x)\approx x$, i.e.,
\begin{eqnarray}
J&\approx&\frac{4\pi m^2q_ekT}{h^3}e^{\epsilon_F/kT}\int_{v_0}^\infty\mbox{d}v_x
\cdot v_xe^{-mv_x^2/2kT}\nonumber\\
&=&\frac{4\pi
mq_e(kT)^2}{h^3}\exp\left\{\frac{1}{kT}\left(\epsilon_F-\frac{1}{2}mv_0^2\right)\right\}\nonumber\\
&=&\frac{4\pi
mq_e(kT)^2}{h^3}e^{-\phi/kT},
\end{eqnarray}
and thus we have obtained a simple expression
for the density current as function of temperature.
This result, which is known as the {\it Richardson--Dushman equation},
is in very good agreement with experimental evidence.
Further discussion on this result can be found in \cite{Beck76} and
\cite{Pathria96}. 

\subsubsection{Photoelectric Emission$^*$}

An analysis based on a similar line of thought
applies also to the photoelectric emission, an effect
where electrons are emitted from a metal as a result of radiation
at frequency beyond a certain critical frequency $\nu_0$ (the Einstein
threshold frequency), whose
corresponding photon energy $h\nu_0$ is equal to the work function
$\phi$. 
Here, the electron
gains an energy amount of $h\nu$ from a photon, which
helps to pass the energy barrier, and so the minimum velocity of emission,
after excitation by a photon of energy $h\nu$ is given by
\begin{equation}
h\nu+\frac{1}{2}mv_0^2=\epsilon_F+\phi=\epsilon_F+h\nu_0.
\end{equation}
Let $\alpha$ denote the probability that a photon actually excites an
electron. Then, similarly as in the previous subsection,
\begin{equation}
J=\alpha\cdot\frac{4\pi m^2q_ekT}{h^3}\int_{v_0}^\infty\mbox{d}v_x\cdot v_x
\ln\left\{1+\exp\left[\frac{1}{kT}\left(\epsilon_F-\frac{1}{2}mv_x^2\right)\right]\right\}.
\end{equation}
where this time
\begin{equation}
v_0=\sqrt{\frac{2}{m}[\epsilon_F+h(\nu_0-\nu)]}.
\end{equation}
Changing the integration variable to
$$x=\frac{1}{kT}\left[\frac{1}{2}mv_x^2+h(\nu-\nu_0)-\epsilon_F\right],$$
we can write the last integral as
\begin{equation}
J=\alpha\cdot\frac{4\pi mq_e(kT)^2}{h^3}\int_0^\infty\mbox{d}x
\ln\left\{1+\exp\left[\frac{h(\nu-\nu_0)}{kT}-x\right]\right\}\mbox{d}x.
\end{equation}
Now, let us denote
\begin{equation}
\Delta=\frac{h(\nu-\nu_0)}{kT}.
\end{equation}
Integrating by parts (twice), we have
\begin{eqnarray}
\int_0^\infty\mbox{d}x\ln(1+
e^{\Delta-x})&=&\int_0^\infty\frac{x\mbox{d}x}{e^{x-\Delta}+1}\nonumber\\
&=&\frac{1}{2}\int_0^\infty\frac{x^2e^{x-\Delta}\mbox{d}x}
{(e^{x-\Delta}+1)^2}
\dfn f(e^{\Delta}).
\end{eqnarray}
For $h(\nu-\nu_0)\gg kT$, we have $e^{\Delta}\gg 1$, and then
it can be shown (using the same technique as in Subsection \ref{approx})
that $f(e^{\Delta})\approx \Delta^2/2$, which gives
\begin{equation}
J=\alpha\cdot\frac{2\pi mq_e}{h}(\nu-\nu_0)^2
\end{equation}
independently of $T$. In other words, when the energy
of light quantum is much larger than the thermal energy $kT$, temperature
becomes irrelevant. At the other extreme of very low frequency, where
$h(\nu_0-\nu)\gg kT$, and then $e^{\Delta}\ll 1$, we have
$f(e^{\Delta})\approx e^{\Delta}$, and then
\begin{equation}
J=\alpha\cdot\frac{4\pi mq_e(kT)^2}{h^3}e^{(h\nu-\phi)/kT}
\end{equation}
which is like the thermionic current density, enhanced by a photon factor
$e^{h\nu/kT}$. 

\newpage
\section{Quantum Statistics -- the Bose--Einstein Distribution}

The general description of bosons was provided in the introductory
paragraphs of Chapter 4. As said, the crucial difference between bosons and
fermions is that in the case of bosons, Pauli's exclusion principle does not
apply. In this chapter, we study the statistical mechanics 
of bosons.

\subsection{Combinatorial Derivation of the BE Statistics}

We use the same notation as in Chapter 4. Again, we are partitioning the
energy levels $\epsilon_1,\epsilon_2,\ldots$ into groups, labeled by
$s=1,2,\ldots$, where in group no.\ $s$,
where the representative energy is $\hat{\epsilon}_s$, 
there are $G_s$ quantum states. As before, a microstate is defined in terms of
the combination of occupation numbers $\{\hat{N}_s\}$ and
$\Omega(\hat{N}_1,\hat{N}_2,\ldots)=\prod_s
\Omega_s(\hat{N}_s)$, but now we need a different estimate of each factor
$\Omega_s(\hat{N}_s)$,
since now there are no restrictions on 
the occupation numbers of the quantum states.

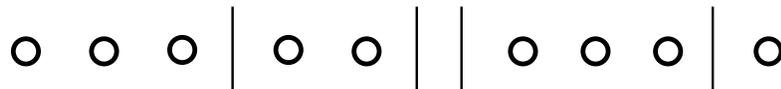
\begin{figure}[ht]
\hspace*{3cm}\input{combinatorics.pstex_t}
\caption{\small $\hat{N}_s$ particles and $G_s-1$ partitions.}
\label{combinatorics}
\end{figure}

In how many ways can one partition $\hat{N}_s$ particles among $G_s$
quantum states? Imagine that the $\hat{N}_s$ particles of group no.\ $s$
are arranged along a straight line. By means of $G_s-1$ partitions
we can divide these particles into $G_s$ different subsets 
corresponding to the various states in that group. In fact, we have
a total of $(\hat{N}_s+G_s-1)$ elements, $\hat{N}_s$ of them are particles
and the remaining $(G_s-1)$ are partitions (see Fig.\ \ref{combinatorics}). 
In how many distinct ways can
we configure them? The answer is simple:
\begin{equation}
\Omega_s(\hat{N}_s)=\frac{(\hat{N}_s+G_s-1)!}{\hat{N}_s!(G_s-1)!}.
\end{equation}
On the account that $G_s \gg 1$, the $-1$ can be safely 
neglected, and we might as
well write instead
\begin{equation}
\Omega_s(\hat{N}_s)=\frac{(\hat{N}_s+G_s)!}{\hat{N}_s!G_s!}.
\end{equation}
Repeating the same derivation as in Subsection 4.1, but with the
above assignment for $\Omega_s(\hat{N}_s)$,
we get by Stirling's formula:
\begin{equation}
\ln\Omega_s(\hat{N}_s)\approx (\hat{N}_s+G_s)h_2\left(\frac{G_s}{
\hat{N}_s+G_s}\right),
\end{equation}
and so the free energy is now
\begin{equation}
F\approx \sum_s\left[\hat{N}_s\hat{\epsilon}_s-kT
(\hat{N}_s+G_s)h_2\left(\frac{\hat{N}_s}{
\hat{N}_s+G_s}\right)\right],
\end{equation}
which again, should be minimized s.t.\ the constraint
$\sum_s\hat{N}_s=N$. Upon carrying out the minimization of the
corresponding Lagrangian (see Subsection 4.1), we arrive\footnote{Exercise
5.1: fill in the detailed derivation.} at the
following result for the most probable occupation numbers:
\begin{equation}
\hat{N}_s=\frac{G_s}{e^{\beta(\hat{\epsilon}_s-\mu)}-1}
\end{equation}
or, moving back to the original occupation numbers,
\begin{equation}
\label{be}
\bar{N}_r=\frac{1}{e^{\beta(\epsilon_r-\mu)}-1},
\end{equation}
where $\mu$ is again the Lagrange multiplier, which has the
meaning of the chemical potential. This is Bose--Einstein
(BE) distribution. As we see, the formula is very similar to that of the FD
distribution, the only difference is that in the denominator, $+1$ is
replaced by $-1$. Surprisingly enough, 
this is a crucial difference that makes the
behavior of bosons drastically different from that of fermions.
Note that for this expression to make sense, $\mu$ must be smaller than
the ground energy $\epsilon_1$, 
otherwise the denominator either vanishes or becomes
negative. If the ground--state energy is zero, this means $\mu < 0$.

\subsection{Derivation Using the Grand--Canonical Ensemble}

As in Subsection 4.2, an alternative derivation can be carried out using the
grand--canonical ensemble. The only difference is that now, the summations 
over $\{N_r\}$, are not only over $\{0,1\}$ but over all non--negative
integers. In particular,
\begin{equation}
\Xi(\beta,\mu)=\prod_r\left[\sum_{N_r=0}^\infty e^{\beta
N_r(\mu-\epsilon_r)}\right],
\end{equation}
Of course, here too, for convergence of each geometric series, we must assume
$\mu < \epsilon_1$, and then the result is
\begin{equation}
\Xi(\beta,\mu)=\prod_r\frac{1}{1-e^{\beta(\mu-\epsilon_r)}}.
\end{equation}
Here, under the grand--canonical ensemble, $N_1,N_2,\ldots$ are independent
geometric random variables with distributions
\begin{equation}
P_r(N_r)=[1-e^{\beta(\mu-\epsilon)}]e^{\beta N_r(\mu-\epsilon_r)}
~~~~~N_r=0,1,2,\ldots,~~~r=1,2,\ldots
\end{equation}
Thus, $\bar{N}_r$ is just the expectation of this geometric random
variable, which is readily found\footnote{Exercise 5.2: show this.} to be
as in eq.\ (\ref{be}).

\subsection{Bose--Einstein Condensation}

In analogy to the FD case, here too, the chemical potential $\mu$
is determined from the constraint on the total number of particles.
In this case, it reads
\begin{equation}
\sum_r\frac{1}{e^{\beta(\epsilon_r-\mu)}-1}=N.
\end{equation}
Taking into account the density of states in a potential well
of sizes $L_x\times L_y\times L_z$ (as was done in Chapter 4), 
in the continuous limit, this yields
\begin{equation}
\label{bee}
\rho= \frac{\sqrt{2m^3}}{2\pi^2\hbar^3}\cdot
\int_0^\infty\frac{\sqrt{\epsilon}\mbox{d}\epsilon}{e^{(\epsilon-\mu)/kT}-1}.
\end{equation}
At this point, an important peculiarity should be discussed. 
Consider eq.\ (\ref{bee}) and
suppose that
we are cooling the system. As $T$ decreases, $\mu$ must adjust in order
to keep eq.\ (\ref{bee}) holding since the number of particles must be
preserved. In particular, as $T$ decreases, $\mu$ must increase, yet it must be
negative. The point is that even for $\mu=0$, which is the maximum allowed
value of $\mu$, the integral at the r.h.s.\ of (\ref{bee}) is
finite\footnote{Exercise 5.3: show this.}
as the density of states is proportional to $\sqrt{\epsilon}$ and hence balances
the `explosion' of the BE integral near $\epsilon=0$. Let us define then
\begin{equation}
\varrho(T)\dfn
\frac{\sqrt{2m^3}}{2\pi^2\hbar^3}\cdot
\int_0^\infty\frac{\sqrt{\epsilon}\mbox{d}\epsilon}{e^{\epsilon/kT}-1}
\end{equation}
and let $T_c$ be the solution to the equation $\varrho(T)=\rho$, which can be
found as follows. By changing the integration variable to $z=\epsilon/kT$,
we can rewrite the r.h.s.\ as
\begin{equation}
\varrho(T)=\left(\frac{mkT}{2\pi\hbar^2}\right)^{3/2}\left\{
\frac{2}{\sqrt{\pi}}\int_0^\infty\frac{\sqrt{z}\mbox{d}z}{e^z-1}\right\}
\approx 2.612\cdot\left(\frac{mkT}{2\pi\hbar^2}\right)^{3/2},
\end{equation}
where the constant $2.612$ is the numerical value of the expression in the
curly brackets. Thus,
\begin{equation}
T_c\approx 0.5274\cdot\frac{2\pi\hbar^2}{mk}\cdot\rho^{2/3}=
3.313\cdot\frac{\hbar^2\rho^{2/3}}{mk}
\end{equation}
The problem is that for $T < T_c$, eq.\ (\ref{bee}) can no longer be solved
by any non--positive value of $\mu$. So what happens below $T_c$?

The root of the problem is in the passage from the discrete sum over
$r$ to the integral over $\epsilon$. The paradox is resolved when it is
understood that below $T_c$, the contribution of $\epsilon=0$ should be
separated from the integral. That is, the
correct form is
\begin{equation}
N=\frac{1}{e^{-\mu/kT}-1}+
\frac{\sqrt{2m^3}V}{2\pi^2\hbar^3}\cdot
\int_0^\infty\frac{\sqrt{\epsilon}\mbox{d}\epsilon}{e^{(\epsilon-\mu)/kT}-1}.
\end{equation}
or, after dividing by $V$,
\begin{equation}
\rho=\rho_0+
\frac{\sqrt{2m^3}}{2\pi^2\hbar^3}\cdot
\int_0^\infty\frac{\sqrt{\epsilon}\mbox{d}\epsilon}{e^{(\epsilon-\mu)/kT}-1},
\end{equation}
where $\rho_0$ is the density of ground--state particles, and now
the integral accommodates the contribution of all particles with
strictly positive energy. Now, for $T<T_c$, we simply have
$\rho_0=\rho-\varrho(T)$, which means that a {\it macroscopic} fraction of the
particles condensate at the ground state. This phenomenon is called
{\it Bose--Einstein condensation}. Note that for $T< T_c$,
\begin{eqnarray}
\rho_0&=&\rho-\varrho(T)\nonumber\\
&=&\varrho(T_c)-\varrho(T)\nonumber\\
&=&\varrho(T_c)\left[1-\frac{\varrho(T)}{\varrho(T_c)}\right]\nonumber\\
&=&\varrho(T_c)\left[1-\left(\frac{T}{T_c}\right)^{3/2}\right]\nonumber\\
&=&\rho\left[1-\left(\frac{T}{T_c}\right)^{3/2}\right]
\end{eqnarray}
which gives a precise characterization of the condensation as a function
of temperature. It should be pointed out that $T_c$ is normally extremely
low.\footnote{
In 1995 the first gaseous condensate was produced by Eric Cornell and Carl
Wieman at the University of Colorado, using a gas
of rubidium atoms cooled to 170 nanokelvin. For their
achievements Cornell, Wieman, and Wolfgang Ketterle at MIT received the 2001
Nobel Prize in Physics. In November 2010 the first photon BEC was
observed.}

One might ask why does the point $\epsilon=0$ require special caution
when $T<T_c$, but doesn't require such caution for $T> T_c$? The
answer is that for $T> T_c$, $\rho_0=1/V[e^{-\mu/kT}-1]$ tends to 
zero in the thermodynamic limit ($V\to\infty$) 
since $\mu< 0$. However, as $T\to T_c$,
$\mu\to 0$, and $\rho_0$ becomes singular.

It is instructive to derive the pressure exerted by the ideal Boson gas for $T
< T_c$. This can be obtained from the grand partition function
\begin{eqnarray}
\ln\Xi&=&-\sum_r\ln(1-e^{-\epsilon_r/kT})~~~~~~~~~~~~~~~~~(\mu=0)\nonumber\\
&\sim&-\frac{\sqrt{2m^3}V}{2\pi^2\hbar^3}\int_0^\infty\mbox{d}\epsilon\cdot
\sqrt{\epsilon}\ln(1-e^{-\epsilon/kT})\nonumber\\
&=&-\frac{\sqrt{2m^3}(kT)^{3/2}V}{2\pi^2\hbar^3}\int_0^\infty\mbox{d}x\cdot
\sqrt{x}\ln(1-e^{-x}),
\end{eqnarray}
where integral over $x$ (including the minus sign) is just a positive 
constant $C$ 
that we won't bother to calculate here. Now,
\begin{equation}
P=\lim_{V\to\infty}\frac{kT\ln\Xi}{V}=
\frac{C\sqrt{2m^3}(kT)^{5/2}}{2\pi^2\hbar^3}.
\end{equation}
We see that the pressure is independent of the density $\rho$ (compare with
the ideal gas where $P=\rho kT$). This is because the condensed particles
do not contribute to the pressure. What matters 
is only the density of those 
with positive energy, and this density in turn depends only on $T$.

\vspace{0,2cm}

\noindent
{\small {\it Exercise 5.4:} Why don't fermions condensate?
What changes in the last derivation?}

\vspace{0,2cm}

\noindent
{\small {\it Exercise 5.5:} The last derivation was in three dimensions
($d=3$). Modify the derivation of the BE 
statistics to apply to a general dimension
$d$, taking into account the dependence of the density of states upon $d$.
For which values of $d$ bosons condensate?}

\subsection{Black--Body Radiation}

One of the important applications of the BE statistics is to investigate the
equilibrium properties of black--body radiation. All bodies emit
electromagnetic radiation whenever at positive temperature, but normally, this
radiation is not in thermal equilibrium. If we consider the radiation within
an opaque enclosure whose walls are maintained at temperature $T$ then
radiation and walls together arrive at thermal equilibrium and in this state, the radiation
has certain important properties. In order to study this equilibrium radiation, one
creates a small hole in the walls of the enclosure, so that it will not
disturb the equilibrium of the cavity and then the emitted radiation
will have the same properties as the cavity radiation, which in turn are the
same as the radiation properties of a {\it black body} -- a body that perfectly
absorbs all the radiation falling on it. The temperature of the black body is
$T$ as well, of course. In this section, we study these radiation properties
using BE statistics.

We consider then a radiation
cavity of volume $V$ and temperature $T$. Historically, Planck (1900) viewed
this system as an assembly of harmonic oscillators with quantized energies
$(n+1/2)\hbar\omega$, $n=0,1,2,\ldots$, where $\omega$ is the angular
frequency of the oscillator. An alternative point of view is as an ideal gas of
identical and indistinguishable photons, each one with energy $\hbar\omega$.
Photons have integral spin and hence are bosons, but they have
zero mass and zero chemical potential when
they interact with a black--body. The reason is that there is no constraint
that their total number would be conserved (they are emitted and absorbed
in the black--body material with which they interact). Since in equilibrium $F$ should be
minimum, then $(\partial F/\partial N)_{T,V}=0$. But $(\partial F/\partial
N)_{T,V}=\mu$. and so, $\mu=0$. It follows then that distribution of photons
across the quantum states
obeys BE statistics with $\mu=0$, that is
\begin{equation}
\bar{N}_\omega=\frac{1}{e^{\hbar\omega/kT}-1}.
\end{equation}

The calculation of the density of states here is somewhat different from
the one in Subsection 4.3. Earlier, we considered 
a particle with positive mass $m$, whose
kinetic energy is $\|\vec{\bp}\|^2/2m=\hbar^2\|\vec{\bk}\|^2/2m$, whereas
now we are talking about a photon whose rest mass is zero and whose energy
is $\hbar\omega=\hbar\|\vec{k}\|c=\|\vec{\bp}\|c$ 
($c$ being the speed of light), so the dependence on
$\|\vec{\bk}\|$ is now linear rather than quadratic. This is a relativistic
effect.

Assuming that $V$ is large enough, we can pass to the continuous approximation.
The number of oscillatory modes for which a wave--vector $\vec{\bk}$ resides
in an infinitesimal 
cube $\mbox{d}^3\vec{\bk}=\mbox{d}k_x\mbox{d}k_y\mbox{d}k_z$ 
is $V\cdot 4\pi\|\vec{\bk}\|^2\mbox{d}\|\vec{\bk}\|/(2\pi)^3$, as in phase
space of position and momentum, it is
\begin{equation}
V4\pi\|\vec{\bp}\|^2\mbox{d}\|\vec{\bp}\|/h^3=
V4\pi\|\hbar\vec{\bk}\|^2\mbox{d}\|\vec{\hbar\bk}\|/(2\pi\hbar)^3=
V\cdot 4\pi\|\vec{\bk}\|^2\mbox{d}\|\vec{\bk}\|/(2\pi)^3.
\end{equation}
Using the relation $\omega=c\|\vec{\bk}\|$
and doubling the above
expression (for two directions of polarization), we have that the total number
of quantum states of a photon in the range $[\omega,\omega+\mbox{d}\omega]$
is $V\omega^2\mbox{d}\omega/\pi^2c^3$. Thus, the number of photons in this
frequency range is 
\begin{equation}
\mbox{d}N_\omega=\frac{V}{\pi^2c^3}\cdot\frac{\omega^2\mbox{d}\omega}{e^{\hbar\omega/kT}-1}.
\end{equation}
The contribution of this to the energy is
\begin{equation}
\mbox{d}E_\omega=\hbar\omega\mbox{d}N_\omega=
\frac{\hbar V}{\pi^2c^3}\cdot\frac{\omega^3\mbox{d}\omega}{e^{\hbar\omega/kT}-1}
\end{equation}
This expression for the spectrum of black--body radiation is known as {\it
Planck's law}. 

\begin{figure}[h!t!b!]
\centering
\includegraphics[width=8.5cm, height=8.5cm]{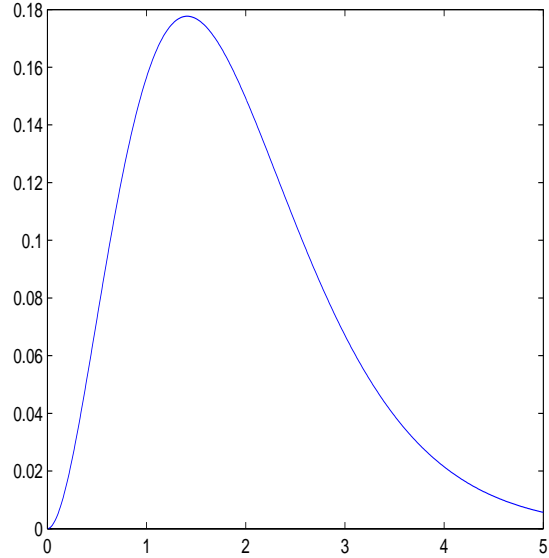}
\caption{\small Illustration of Planck's law. The energy density per unit
frequency as a function of frequency.}
\label{planck}
\end{figure}

\vspace{0.2cm}

\noindent
{\small {\it Exercise 5.6:} Write Planck's law in terms of the wavelength
$\mbox{d}E_\lambda$.}

At low frequencies ($\hbar\omega \ll kT$), this gives
\begin{equation}
\mbox{d}E_\omega\approx \frac{kTV}{\pi^2c^3}\omega^2\mbox{d}\omega
\end{equation}
which is the {\it Rayleigh--Jeans law}. This is actually the classic
limit (see footnote at the Introduction to Chap.\ 4), obtained from
multiplying $kT/2$ by the ``number of waves.''
In the other extreme of $\hbar\omega \gg kT$, we have
\begin{equation}
\mbox{d}E_\omega\approx\hbar\omega\mbox{d}N_\omega=
\frac{\hbar V}{\pi^2c^3}\cdot\omega^3
e^{-\hbar\omega/kT}\mbox{d}\omega, 
\end{equation}
which is {\it Wien's law}. At low temperatures, this is an excellent
approximation over a very wide range of frequencies.
The frequency of maximum radiation is
\begin{equation}
\label{omegamax}
\omega_{\max}=2.822\cdot\frac{kT}{\hbar},
\end{equation}
namely, linear in temperature. 
This relation has immediate applications. For example, the sun is known to be a source of
radiation, which with a good level of approximation, can be considered
a black body. Using a spectrometer, one
can measure the frequency $\omega_{\max}$ of maximum radiation, and
estimate the sun's surface temperature (from eq.\ (\ref{omegamax})),
to be $T\approx 5800^{\mbox{o}}K$. 

Now, the energy density is
\begin{equation}
\frac{E}{V}=\frac{\hbar}{\pi^2c^3}\int_0^\infty
\frac{\omega^3\mbox{d}\omega}{e^{\hbar\omega/kT}-1}=aT^4
\end{equation}
where the second equality is obtained by changing the integration variable to
$x=\hbar\omega/kT$ and then
\begin{equation}
a=\frac{\hbar}{\pi^2c^3}\left(\frac{k}{\hbar}\right)^4\int_0^\infty\frac{x^3\mbox{d}x}{e^x-1}
=\frac{\pi^2k^4}{15\hbar^3c^3}.
\end{equation}
The relation $E/V=aT^4$ is called the {\it Stefan--Boltzmann law}.
The heat capacity in constant volume, $C_V=(\partial E/\partial T)_V$, is therefore
proportional to $T^3$.

\vspace{0.2cm}

\noindent
{\small {\it Exercise 5.7:} Calculate $\rho$, the density of photons.}

\vspace{0.2cm}

Additional thermodynamic quantities can now be calculated
from the logarithm of the grand--canonical partition function
\begin{equation}
\ln\Xi= -\sum_r\ln[1-e^{-\hbar\omega_r/kT}]=
-\frac{V}{\pi^2c^3}\int_0^{\infty}\mbox{d}\omega\cdot
\omega^2\ln[1-e^{-\hbar\omega/kT}]
\end{equation}
For example, the pressure of the photon gas can be calculated from
\begin{eqnarray}
P&=&\frac{kT\ln\Xi}{V}\nonumber\\
&=&-\frac{kT}{\pi^2c^3}\int_0^\infty\mbox{d}\omega\cdot
\omega^2\ln[1-e^{-\hbar\omega/kT}]\nonumber\\
&=&-\frac{(kT)^4}{\pi^2c^3\hbar^3}\int_0^\infty\mbox{d}x\cdot
x^2\ln(1-e^{-x})\nonumber\\
&=&\frac{1}{3}aT^4
=\frac{E}{3V},
\end{eqnarray}
where the integral is calculated using integration by parts.{\footnote{Exercise
5.8: Fill in the details.} Note that while in the ideal gas $P$ was only
linear in $T$, here it is proportional to the fourth power of $T$.
Note also that here, $PV=E/3$, which is different from the ideal gas, where
$PV=2E/3$. 

\newpage
\section{Interacting Particle Systems and Phase Transitions}

In this chapter, we discuss
systems with interacting particles. As we shall see, when the interactions among the
particles are sufficiently significant, the system exhibits
a certain collective behavior that, in the thermodynamic limit,
may be subjected to {\it phase transitions}, i.e., abrupt changes
in the behavior and the properties of the system in the presence of
a gradual change in an external control parameter, like temperature,
pressure, or magnetic field. 

\subsection{Introduction -- Sources of Interaction}

So far, we have dealt almost exclusively with
systems that have additive Hamiltonians, $\calE(\bx)=\sum_i\calE(x_i)$,
which means, under the canonical ensemble,
that the particles are statistically independent and there are no interactions
among them:
each particle behaves as if it was alone in the world. In Nature, of course,
this is seldom really the case. Sometimes this is still a reasonably good
approximation, but in many other cases, the interactions 
are appreciably strong and cannot be neglected.
Among the different particles there could be
many sorts of mutual forces, such as
mechanical, electrical, or magnetic forces, etc.
There could also be interactions that stem from quantum--mechanical effects:
As described earlier, fermions must obey
Pauli's exclusion principle, which asserts that
no quantum state can be populated by more
than one particle. This gives rise to a certain mutual influence between
particles. Another type of interaction stems from the fact that the particles
are indistinguishable, so permutations between them are not considered as
distinct states. For example, referring again to 
BE statistics of Chapter 5, had the $N$
particles been statistically independent, the resulting partition function
would be
\begin{align}
Z_N(\beta)&=\left[\sum_r e^{-\beta\epsilon_r}\right]^N\nonumber\\
&=\sum_{N_1,N_2,\ldots}\delta\left(\sum_rN_r=N\right)\frac{N!}{\prod_r N_r!}\cdot
\exp\left\{-\beta\sum_rN_r\epsilon_r\right\}
\end{align}
whereas in eq.\ (\ref{qg}) (but with summations extending over all integers $\{N_r\}$),
the combinatorial factor, $N!/\prod_rN_r!$, that distinguishes
between the various permutations among the particles, is absent. This introduces
dependency, which physically means interaction. Indeed, in the case
of the ideal boson gas, we have encountered the effect of
Bose--Einstein condensation, which is actually a phase transition, and phase
transitions can occur only in systems of interacting particles, as will be
discussed in this chapter.\footnote{Another way to understand the dependence
is to observe that occupation numbers $\{N_r\}$ are dependent via the
constraint on their sum. This is different from the grand--canonical ensemble,
where they are independent.}

\subsection{Models of Interacting Particles}

The simplest forms of deviation from the purely 
additive Hamiltonian structure 
are those that consists, in addition to the individual energy
terms, $\{\calE(x_i)\}$, also terms that depend on pairs, and/or triples,
and/or even larger cliques of particles. 
In the case of purely pairwise interactions, this
means a structure like the following:
\begin{equation}
\label{pairwiseinteractions}
\calE(\bx)=\sum_{i=1}^N\calE(x_i)+\sum_{(i,j)}\varepsilon(x_i,x_j)
\end{equation}
where the summation over pairs can be defined over all pairs $i\ne j$, 
or over some of the pairs, according to a given rule, e.g., depending
on the distance between particle $i$ and particle $j$, and according to
the geometry of the system, or according to a certain graph whose edges
connect the relevant pairs of variables (that in turn, are designated as nodes).

For example, in a one--dimensional array 
(a lattice) of
particles, a customary model accounts for interactions between neighboring
pairs only, neglecting more remote ones, thus the second term above would be
$\sum_{i}\varepsilon(x_i,x_{i+1})$. A well known special case of this is
that of a polymer or a solid with crystal lattice structure, where 
in the one--dimensional version of the model, atoms are thought of as
a chain of masses connected by springs (see left part of Fig.\ \ref{springs2d}),
i.e., an array of coupled harmonic oscillators. In
this case,
$\varepsilon(x_i,x_{i+1})=\frac{1}{2}K(x_{i+1}-x_i)^2$, where $K$ is a
constant and $x_i$ is the displacement of the $i$-th atom from its
equilibrium location, i.e., the potential energies of the springs.
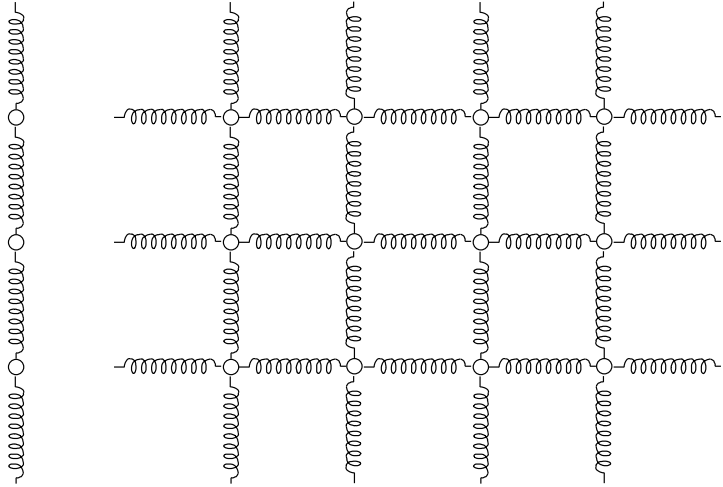
\begin{figure}[ht]
\hspace*{4cm}\input{springs2d.pstex_t}
\caption{\small Elastic interaction forces between adjacent atoms in a
one--dimensional lattice (left part of the figure) and in a two--dimensional
lattice (right part).}
\label{springs2d}
\end{figure}
In higher dimensional arrays (or lattices), similar interactions apply,
there are just more neighbors to each site,
from the various directions 
(see right part of Fig.\ \ref{springs2d}). 
These kinds of models will be discussed in the next chapter in some depth.

In a system where the
particles are mobile and hence their locations vary 
and have no geometrical structure,
like in a gas, the interaction terms are also
potential energies pertaining to 
the mutual forces (see Fig.\ \ref{spheres}), and these normally
depend solely on the distances $\|\vec{r}_i-\vec{r}_j\|$.
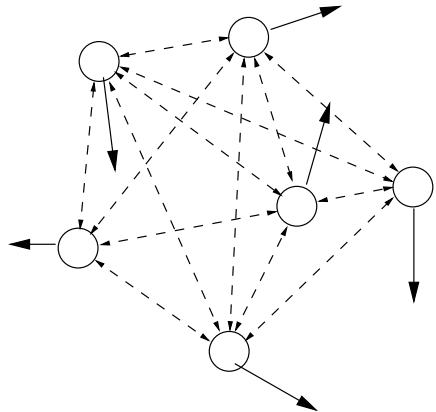
\begin{figure}[ht]
\hspace*{5cm}\input{spheres.pstex_t}
\caption{\small Mobile particles and mutual forces between them.}
\label{spheres}
\end{figure}
For example, in a non--ideal gas,
\begin{equation}
\calE(\bx)=\sum_{i=1}^N\frac{\|\vec{p}_i\|^2}{2m}+
\sum_{i\ne j}\phi(\|\vec{r}_i-\vec{r}_j\|).
\end{equation}
A very simple special case is that of 
hard spheres (Billiard balls), without any
forces, where
\begin{equation}
\phi(\|\vec{r}_i-\vec{r}_j\|)=\left\{\begin{array}{ll}
\infty & \|\vec{r}_i-\vec{r}_j\| < 2R\\
0 & \|\vec{r}_i-\vec{r}_j\| \ge 2R
\end{array}\right.
\end{equation}
which expresses the simple fact that balls cannot physically overlap.
The analysis of this model can be carried out using diagrammatic techniques (the
cluster expansion, etc.), but we will not get into the details of it in the framework of this
course.\footnote{The reader can find the derivations in any textbook
on elementary statistical mechanics, for example, \cite[Chap.\ 9]{Pathria96}.} To
demonstrate, however, the effect of interactions on the deviation from the
equation of state of the ideal gas, we consider next a simple
one--dimensional example.

\vspace{0.2cm}

\noindent
{\small {\it Example 6.1 -- Non--ideal gas in one dimension.}
Consider a one--dimensional object of length $L$ that contains $N+1$
particles, whose locations are $0\equiv r_0 \le r_1 \le \ldots \le r_{N-1}\le
r_N\equiv L$, namely, the first and the last particles are fixed at the edges.
The order of the particles is fixed, namely, they cannot be swapped.
Let the Hamiltonian be given by
\begin{equation}
\calE(\bx)=\sum_{i=1}^N\phi(r_i-r_{i-1})+\sum_{i=1}^n \frac{p_i^2}{2m}
\end{equation}
where $\phi$ is a given potential function designating the interaction between
two neighboring particles along the line. The partition function, which is an
integral of the Boltzmann factor pertaining to this Hamiltonian, should
incorporate the fact that the positions $\{r_i\}$ are not independent.
It is convenient to change variables to $\xi_i=r_i-r_{i-1}$, $i=1,2,\ldots,N$, where it should
be kept in mind that $\xi\ge 0$ for all $i$ and $\sum_{i=1}^N\xi_i=L$.
Let us assume that $L$ is an extensive variable, i.e., $L=N\xi_0$ for some
constant $\xi_0> 0$. Thus, the partition function is 
\begin{eqnarray}
Z_N(\beta,L)&=&\frac{1}{h^N}\int\mbox{d}p_1\cdot\cdot\cdot\mbox{d}p_N
\int_{\reals_N^+}\mbox{d}\xi_1\cdot\cdot\cdot\mbox{d}\xi_N
e^{-\beta\sum_{i=1}^N[\phi(\xi_i)+p_i^2/2m]}\cdot\delta\left(L-\sum_{i=1}^N\xi_i\right)\\
&=&\frac{1}{\lambda^N}
\int_{\reals_N^+}\mbox{d}\xi_1\cdot\cdot\cdot\mbox{d}\xi_N
e^{-\beta\sum_{i=1}^N\phi(\xi_i)}\cdot\delta\left(L-\sum_{i=1}^N\xi_i\right),
\end{eqnarray}
where $\lambda=h/\sqrt{2\pi mkT}$.
The constraint $\sum_{i=1}^N\xi_i=L$ makes the analysis of the configurational partition
function difficult. 
Let us pass to the corresponding Gibbs ensemble where instead of fixing the
length $L$, we control it by applying a 
force $f$.\footnote{Here we use the principle of ensemble equivalence.}
The corresponding partition function now reads
\begin{eqnarray}
Y_N(\beta,f)&=&\lambda^{-N}\int_0^\infty\mbox{d}L e^{-\beta fL}Z_N(\beta,L)\nonumber\\
&=&\lambda^{-N}\int_0^\infty\mbox{d}L e^{-\beta fL}
\int_{\reals_N^+}\mbox{d}\xi_1\cdot\cdot\cdot\mbox{d}\xi_N
e^{-\beta\sum_{i=1}^N\phi(\xi_i)}\cdot\delta\left(L-\sum_{i=1}^N\xi_i\right)
\nonumber\\
&=&\lambda^{-N}\int_{\reals_N^+}\mbox{d}\xi_1\cdot\cdot\cdot\mbox{d}\xi_N
\left[\int_0^\infty\mbox{d}L e^{-\beta fL}
\delta\left(L-\sum_{i=1}^N\xi_i\right)\right]
e^{-\beta\sum_{i=1}^N\phi(\xi_i)}\nonumber\\
\nonumber\\
&=&\lambda^{-N}\int_{\reals_N^+}\mbox{d}\xi_1
\cdot\cdot\cdot\mbox{d}\xi_N\exp\left\{-\beta\left[f\sum_{i=1}^N\xi_i+\sum_{i=1}^N\phi(\xi_i)
\right]\right\}\nonumber\\
&\dfn&\lambda^{-N}\int_{\reals_N^+}\mbox{d}\xi_1
\cdot\cdot\cdot\mbox{d}\xi_N\exp\left[-s\sum_{i=1}^N\xi_i-\beta\sum_{i=1}^N\phi(\xi_i)
\right]\nonumber\\
&=&\left\{\frac{1}{\lambda}\int_0^\infty\mbox{d}\xi\cdot
e^{-[s\xi+\beta\phi(\xi)]}
\right\}^N
\end{eqnarray}
With a slight abuse of notation, from now on, we will denote the last
expression by $Y_N(\beta,s)$.
Consider now the following potential function
\begin{equation}
\phi(\xi)=\left\{\begin{array}{ll}
\infty & 0\le\xi\le d\\
-\epsilon & d < \xi\le d+\delta\\
0 & \xi > d+\delta\end{array}\right.
\end{equation}
In words, distances below $d$ are strictly forbidden (e.g., because of the size of the
particles), in the range between $d$ and $d+\delta$ there is a negative
potential $-\epsilon$, and beyond $d+\delta$ the potential is
zero.\footnote{This is a caricature of
the Lennard--Jones potential function
$\phi(\xi)\propto[(d/\xi)^{12}-(d/\xi)^6]$, which begins from $+\infty$,
decreases down to a negative minimum, and finally increases and tends to zero.}
Now, for this potential function, the one--dimensional integral above is given by
\begin{equation}
I=\int_0^\infty\mbox{d}\xi
e^{-[s\xi+\beta\phi(\xi)]}=\frac{e^{-sd}}{s}[e^{-s\delta}(1-e^{\beta\epsilon})+e^{\beta\epsilon}],
\end{equation}
and so,
\begin{eqnarray}
Y_N(\beta,s)&=&
\frac{e^{-sdN}}{\lambda^Ns^N}[e^{-s\delta}(1-e^{\beta\epsilon})+e^{\beta\epsilon}]^N\nonumber\\
&=&\exp\left\{N\left[
\ln[e^{-s\delta}(1-e^{\beta\epsilon})+e^{\beta\epsilon}]-
sd-\ln (\lambda s)\right]\right\}
\end{eqnarray}
Now, the average length of the system is given by
\begin{eqnarray}
\left<L\right>&=&-\frac{\partial \ln Y_N(\beta,s)}{\partial
s}\nonumber\\
&=&\frac{N\delta
e^{-s\delta}(1-e^{\beta\epsilon})}{e^{-s\delta}(1-e^{\beta\epsilon})+e^{\beta\epsilon}}
+Nd+\frac{N}{s}
\end{eqnarray}
Or, equivalently, 
$\left<\Delta L\right>=\left<L\right>-Nd$, which is the excess length beyond the
possible minimum, is given by
\begin{eqnarray}
\left<\Delta L\right>&=&
\frac{N\delta
e^{-f\delta/kT}(1-e^{\epsilon/kT})}{e^{-f\delta/kT}(1-e^{\epsilon/kT})+e^{\epsilon/kT}}
+\frac{NkT}{f}.
\end{eqnarray}
Thus,
\begin{eqnarray}
f\cdot\left<\Delta L\right>&=&NkT+
\frac{Nf\delta
e^{-f\delta/kT}(1-e^{\epsilon/kT})}{e^{-f\delta/kT}(1-e^{\epsilon/kT})+e^{\epsilon/kT}}\nonumber\\
&=&N\left[kT-\frac{f\delta}{e^{(\epsilon+f\delta)/kT}/(e^{\epsilon/kT}-1)-1}\right]
\end{eqnarray}
where the last line is obtained after some standard algebraic manipulation.
Note that without the potential well of the intermediate
range of distances ($\epsilon=0$ or $\delta=0$), the second term in the square
brackets disappears and we get a one dimensional version of the equation of
state of the ideal gas (with the volume being replaced by length and the pressure -- replaced
by force). The second term is then a correction term due to the interaction. The
attractive potential reduces the product $f\cdot\Delta L$.} $\Box$

Yet another example of a model, or more precisely, a very large class of
models with interactions, are those of magnetic materials. These models
will closely accompany our discussions from this point onward in this chapter.
Few of these models are
solvable, most of them are not. For the purpose of our discussion, a
magnetic material is one for which the relevant  property of each particle
is its {\it magnetic moment}. As a reminder, the magnetic moment is a vector proportional to
the angular momentum of a revolving charged particle (like a rotating electron,
or a current loop), or the {\it spin}, and it designates the intensity of its
response to the net magnetic field that this particle `feels'. This magnetic
field may be 
the superposition of an externally applied magnetic field and the magnetic
fields generated by the neighboring spins. 

Quantum mechanical considerations
dictate that each spin, which will be denoted by $s_i$,
is quantized, that is, it may take only one out of finitely many 
values. In the simplest case to be adopted in our study -- two values only.
These will be designated by $s_i=+1$ (``spin up'') and $s_i=-1$ (``spin
down''), corresponding to the same intensity, but in two opposite directions,
one parallel to the magnetic field, and the other -- anti-parallel (see Fig.\
\ref{spinglass}).
\begin{figure}[ht]
\hspace*{5cm}\input{spinglass1.pstex_t}
\caption{\small Illustration of a spin array on a square lattice.}
\label{spinglass}
\end{figure}
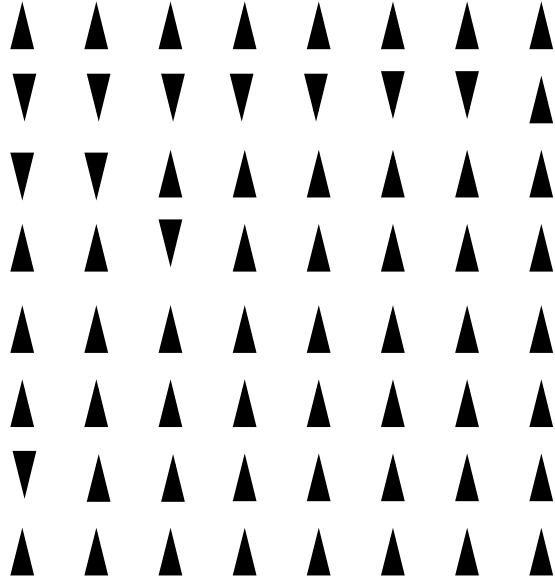
The Hamiltonian associated with an array of spins $\bs=(s_1,\ldots,s_N)$
is customarily modeled (up to certain constants that, 
among other things, accommodate for the
physical units) with a structure like this:
\begin{equation}
\calE(\bs)=-B\cdot\sum_{i=1}^Ns_i-\sum_{(i,j)} J_{ij}s_is_j,
\end{equation}
where $B$ is the externally applied magnetic field 
and $\{J_{ij}\}$
are the coupling constants that designate the levels of interaction between
spin pairs, and they depend on properties of the magnetic material and on the
geometry of the system. The first term accounts for the 
contributions of potential energies 
of all spins due to the magnetic field, which in general, are 
given by the inner
product $\vec{B}\cdot\vec{s}_i$, but since each $\vec{s}_i$ is either
parallel or anti-parallel to $\vec{B}$, as said, these boil down to simple
products, where only the sign of each $s_i$ counts. Since $P(\bs)$ is
proportional to $e^{-\beta\calE(\bs)}$, the spins `prefer' to be parallel,
rather than anti-parallel to the magnetic field. 
The second term in the above Hamiltonian accounts for the interaction energy.
If $J_{ij}$ are all positive,
they also prefer to be parallel to one another (the probability for this is
larger), which is the case where
the material is called {\it ferromagnetic} 
(like iron and nickel). If they are all
negative, the material is {\it antiferromagnetic}. In the mixed case, it is
called a {\it spin glass}. In the latter, the behavior is rather complicated.

The case where all $J_{ij}$ are equal and the double summation over
$\{(i,j)\}$ is over nearest neighbors only is called the {\it Ising model}.
A more general version of it is called the $O(n)$ model. according to which
each spin is an $n$--dimensional unit vector $\vec{s}_i=(s_i^1,\ldots,s_i^n)$
(and so is the magnetic field),
where $n$ is not necessarily related to the dimension $d$ of the lattice in
which the spins reside. The case $n=1$ is then the Ising model.
The case $n=2$ is called the {\it XY model}, and the case $n=3$ is called the
{\it Heisenberg model}.

Of course, the above models for the Hamiltonian can (and, in fact, is being)
generalized to include interactions formed also, by triples, quadruples, or
any fixed size $p$ (that does not grow with $N$) of spin--cliques.

We next discuss a very important effect that exists in some
systems with strong interactions (both in magnetic
materials and in other models): the effect of {\it phase transitions}.

\subsection{A Qualitative Discussion on Phase Transitions}

As was mentioned in the introductory paragraph of this chapter,
a phase transition means an abrupt change in the
collective behavior of a physical system, as we change gradually
one of the externally controlled parameters, like the temperature, pressure,
or magnetic field. The most common 
example of a phase transition in our everyday
life is the water that we boil in the kettle when we make coffee, or when it
turns into ice as we put it in the freezer. 

What exactly these phase transitions are?
In physics, phase transitions 
can occur only if the system has interactions. Consider,
the above example of an array of spins with $B=0$, and let us
suppose that all $J_{ij}> 0$ are equal, 
and thus will be denoted commonly by $J$ (like in the $O(n)$ model). Then,
\begin{equation}
P(\bs)=\frac{\exp\left\{\beta J\sum_{(i,j)}s_is_j\right\}}{Z(\beta)}
\end{equation}
and, as mentioned earlier, this is a ferromagnetic model, where all spins
`like' to be in the same direction, especially when $\beta$ and/or $J$ is
large. In other words, the interactions, in this case, tend to introduce {\it
order} into the system. On the other hand, the second law talks about maximum
entropy, which tends to increase the {\it disorder}. So there are two
conflicting effects here. Which one of them prevails? 

The answer turns out to depend on
temperature. Recall that in the canonical ensemble, equilibrium is attained at
the point of minimum free energy $f=\epsilon-Ts(\epsilon)$. Now, $T$ plays
the role of a weighting factor for the entropy. At low temperatures, the
weight of the second term of $f$ is small, and minimizing $f$ is approximately
equivalent to minimizing $\epsilon$, which is obtained by states with a high
level of order, as $\calE(\bs)=-J\sum_{(i,j)}s_is_j$, in this example. 
As $T$ grows, however, the weight of the term $-Ts(\epsilon)$ increases, and 
$\min f$, becomes more 
and more equivalent to $\max s(\epsilon)$, which is achieved by
states with a high level of disorder (see Fig.\ \ref{fgraph}). 
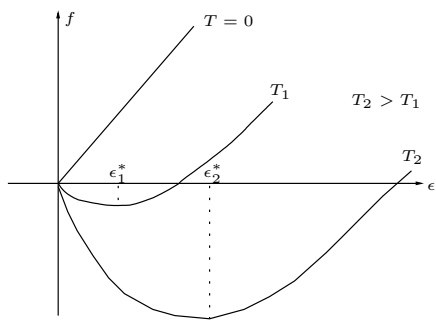
\begin{figure}[ht]
\hspace*{5cm}\input{fgraph.pstex_t}
\caption{\small Qualitative graphs of $f(\epsilon)$ at various temperatures.
The minimizing $\epsilon$ increases with $T$.}
\label{fgraph}
\end{figure}
Thus, the order--disorder characteristics depend
primarily on temperature. 
It turns out that for 
some magnetic systems of this kind, this transition between
order and disorder may be abrupt, in which case, we call it a {\it phase
transition}.
At a certain critical temperature, called
the {\it Curie temperature}, there
is a sudden transition between order and disorder. In the ordered phase,
a considerable fraction of the spins align in the same
direction, which means that the system is spontaneously magnetized (even
without an external magnetic field), whereas in
the disordered phase, about half of the spins are in either direction, and
then the
net magnetization vanishes.
This happens if the
interactions, or more precisely, their dimension in some sense, is strong
enough. 

What is the mathematical significance of a phase transition? If we look at
the partition function, $Z_N(\beta)$, which is the key to all physical
quantities of interest, then
for every finite $N$, this is simply the sum of finitely many exponentials in
$\beta$ and therefore it is continuous and differentiable infinitely 
many times.
So what kind of abrupt changes could there possibly be in the behavior
of this function?
It turns out that while this 
is true for all finite $N$, it is no longer necessarily true if we
look at the thermodynamical limit, i.e., if we look at the
behavior of 
\begin{equation}
\phi(\beta)=\lim_{N\to\infty}\frac{\ln
Z_N(\beta)}{N}. 
\end{equation}
While $\phi(\beta)$ must be continuous for all $\beta >0$
(since it is convex), it need not necessarily have continuous derivatives.
Thus, a phase transition, if exists, 
is fundamentally an asymptotic property, it
may exist in the thermodynamical limit only. While a physical system is, after
all finite, it is nevertheless well approximated by the thermodynamical limit
when it is very large.

The above discussion explains also why a system without interactions,
where all $\{x_i\}$ are i.i.d., cannot have phase transitions. In this case,
$Z_N(\beta)=[Z_1(\beta)]^N$, and so, $\phi(\beta)=\ln Z_1(\beta)$, which is
always a smooth function without any irregularities. For a phase transition to
occur, the particles must behave in some collective manner, which is the case
only if interactions take place.

There is a distinction between two types of phase transitions:
\begin{itemize}
\item If $\phi(\beta)$ has a discontinuous first order derivative, then this
is called a {\it first order phase transition}. 
\item If $\phi(\beta)$ has a continuous first order derivative, 
but a discontinuous second order derivative then this
is called a {\it second order phase transition}, or a {\it continuous phase
transition}.
\end{itemize}

We can talk, of course, 
about phase transitions w.r.t.\ additional parameters other than
temperature. In the above magnetic example, if we introduce back the magnetic
field $B$ into the picture, 
then $Z$, and hence also $\phi$, become functions of $B$
too. If we then look at derivative of
\begin{align}
\phi(\beta,B)&=\lim_{N\to\infty}\frac{\ln Z_N(\beta,B)}{N}\nonumber\\
&=
\lim_{N\to\infty}\frac{1}{N}\ln\left[\sum_{\bs}\exp\left\{\beta
B\sum_{i=1}^Ns_i+\beta J\sum_{(i,j)}s_is_j\right\}\right]
\end{align}
w.r.t.\ the product $(\beta B)$, which multiplies the magnetization,
$\sum_is_i$, at the exponent,
this would give exactly the average magnetization per spin
\begin{equation}
m(\beta,B)=\left<\frac{1}{N}\sum_{i=1}^N S_i\right>,
\end{equation}
and this quantity might not always be continuous. 
Indeed, as mentioned earlier, below the Curie
temperature there might be a spontaneous magnetization. If $B\downarrow
0$, then this magnetization is positive, and if $B\uparrow 0$, it is negative,
so there is a discontinuity at $B=0$.
We shall see this more concretely later on.  

\subsection{The One--Dimensional Ising Model}

The most familiar model of a magnetic system with interactions is 
the one--dimensional Ising model,
according to which
\begin{equation}
\calE(\bs)=-B\sum_{i=1}^Ns_i -J\sum_{i=1}^N s_is_{i+1}
\end{equation}
with the periodic boundary condition $s_{N+1}=s_1$. Thus, 
\begin{align}
Z_N(\beta,B)&=\sum_{\bs}\exp\left\{\beta B\sum_{i=1}^Ns_i+\beta J\sum_{i=1}^N
s_is_{i+1}\right\}
\nonumber\\
&=\sum_{\bs}\exp\left\{h\sum_{i=1}^Ns_i+K\sum_{i=1}^N
s_is_{i+1}\right\}~~~~~h\dfn\beta B,~~K\dfn \beta J\nonumber\\
&=\sum_{\bs}\exp\left\{\frac{h}{2}\sum_{i=1}^N(s_i+s_{i+1})+K\sum_{i=1}^N
s_is_{i+1}\right\}.
\end{align}
Consider now the $2\times 2$ matrix $P$ whose entries are
$\exp\{\frac{h}{2}(s+s')+Kss'\}$, $s,s'\in\{-1,+1\}$, i.e., 
\begin{equation}
P=\left(\begin{array}{cc}
e^{K+h} & e^{-K} \\
e^{-K} & e^{K-h}\end{array}\right).
\end{equation}
Also, $s_i=+1$ will be represented by the column vector $\sigma_i=(1,0)^T$ and
$s_i=-1$ will be represented by $\sigma_i=(0,1)^T$. Thus,
\begin{align}
Z(\beta,B)&=\sum_{\sigma_1}\cdot\cdot\cdot\sum_{\sigma_N} 
(\sigma_1^T
P\sigma_2)\cdot(\sigma_2^TP\sigma_2)\cdot\cdot\cdot(\sigma_N^TP\sigma_1)\nonumber\\
&=\sum_{\sigma_1}\sigma_1^TP\left(\sum_{\sigma_2}\sigma_2\sigma_2^T\right)P
\left(\sum_{\sigma_3}\sigma_3\sigma_3^T\right)P\cdot\cdot\cdot P
\left(\sum_{\sigma_N}\sigma_N\sigma_N^T\right)P\sigma_1\nonumber\\
&=\sum_{\sigma_1}\sigma_1^TP\cdot I\cdot P\cdot I
\cdot\cdot\cdot I\cdot P\sigma_1\nonumber\\
&=\sum_{\sigma_1}\sigma_1^TP^N\sigma_1\nonumber\\
&=\mbox{tr}\{P^N\}\nonumber\\
&=\lambda_1^N+\lambda_2^N
\end{align}
where $\lambda_1$ and $\lambda_2$ are the eigenvalues of $P$, which are
\begin{equation}
\lambda_{1,2}=e^K\cosh(h)\pm\sqrt{e^{-2K}+e^{2K}\sinh^2(h)}.
\end{equation}
Letting $\lambda_1$ denote the larger (the dominant) eigenvalue, i.e.,
\begin{equation}
\lambda_{1}=e^K\cosh(h)+\sqrt{e^{-2K}+e^{2K}\sinh^2(h)},
\end{equation}
then clearly,
\begin{equation}
\phi(h,K)=\lim_{N\to\infty}\frac{\ln Z_N(h,K)}{N}=\ln\lambda_1.
\end{equation}
The average magnetization is
\begin{align}
M(h,K)&=\left< \sum_{i=1}^N S_i\right>\nonumber\\
&=\frac{\sum_{\bs}(\sum_{i=1}^Ns_i)\exp\{h\sum_{i=1}^Ns_i+K\sum_{i=1}^Ns_is_{i+1}\}}
{\sum_{\bs}\exp\{h\sum_{i=1}^Ns_i+K\sum_{i=1}^Ns_is_{i+1}\}}\nonumber\\
&=\frac{\partial \ln Z(h,K)}{\partial h}
\end{align}
and so, the per--spin magnetization is:
\begin{equation}
m(h,K)\dfn\lim_{N\to\infty}\frac{M(h,K)}{N}=\frac{\partial \phi(h,K)}{\partial
h}= \frac{\sinh(h)}{\sqrt{e^{-4K}+\sinh^2(h)}}
\end{equation}
or, returning to the original parametrization:
\begin{equation}
m(\beta,B)=
\frac{\sinh(\beta B)}{\sqrt{e^{-4\beta J}+\sinh^2(\beta B)}}.
\end{equation}
For $\beta > 0$ and $B > 0$ this is a smooth 
function, and so, there is are no
phase transitions and no spontaneous magnetization 
at any finite temperature.\footnote{Note, in particular, that for
$J=0$ (i.i.d.\ spins) we get paramagnetic characteristics
$m(\beta,B)=\tanh(\beta B)$, in agreement with
the result pointed out in the example of two--level systems, in one of our
earlier discussions.} However, at the absolute zero ($\beta\to\infty$), we get
\begin{equation}
\lim_{B\downarrow 0}\lim_{\beta\to\infty}m(\beta,B)=+1;~~
\lim_{B\uparrow 0}\lim_{\beta\to\infty}m(\beta,B)=-1,
\end{equation}
thus $m$ is discontinuous w.r.t.\ $B$ at $\beta\to\infty$, which means that
there is a phase transition at $T=0$. In other words, the Curie temperature is
$T_c=0$ independent of $J$.

We see then that one--dimensional Ising model is easy to handle, but it is not
very interesting in the sense that there is actually no phase transition. The
extension to the two--dimensional Ising model on the square lattice is
surprisingly more difficult, but it is 
still solvable, albeit without a magnetic
field. It was first solved in 1944 by
Onsager \cite{Onsager44}, who has shown that it exhibits a phase transition with
Curie temperature given by 
\begin{equation}
T_c=\frac{2J}{k\ln(\sqrt{2}+1)}.
\end{equation}
For lattice dimension larger than two, the problem is still open.

It turns out then that whatever 
counts for the existence of phase transitions, is
not only the intensity of the interactions (designated by the magnitude of $J$),
but more importantly, the ``dimensionality'' of the 
structure of the pairwise interactions. If we
denote by
$n_\ell$ the number of $\ell$--th order 
neighbors of every given site, namely, the
number of sites that can be 
reached within $\ell$ steps from the given site, then
whatever counts is how fast does the sequence $\{n_\ell\}$ grow, or more
precisely, what is the value of $d\dfn \lim_{\ell\to\infty}\frac{\ln
n_\ell}{\ln\ell}$, which is exactly the ordinary dimensionality for hyper-cubic lattices. 
Loosely speaking, this dimension must be sufficiently large for a phase transition to exist.

To demonstrate this point, we next discuss an extreme case of a model where
this dimensionality
is actually infinite. In this model ``everybody is a neighbor of everybody
else'' and to the same extent, so it definitely has the highest connectivity
possible. This is not quite a physically realistic model, but
the nice thing about it is that it is easy to solve and that it exhibits a
phase transition that is 
fairly similar to those that exist in real systems. It is also
intimately related to a very popular approximation method in statistical
mechanics, called the {\it mean field approximation}. Hence it is sometimes
called the {\it mean field model}. It is also known as the {\it Curie--Weiss
model} or the {\it infinite range model}. 

Finally, we should comment that there are
other ``infinite--dimensional'' Ising models, like the one defined on the
Bethe lattice (an infinite tree without a root and without leaves), 
which is also easily solvable (by recursion) and it also
exhibits phase transitions \cite{Baxter82},
but we will not discuss it here.

\subsection{The Curie--Weiss Model}

According to the Curie--Weiss (C--W) model,
\begin{equation}
\calE(\bs)=-B\sum_{i=1}^N s_i-\frac{J}{2N}\sum_{i\ne j}s_is_j.
\end{equation}
Here, all pairs $\{(s_i,s_j)\}$ communicate to the same extent,
and without any geometry. The $1/N$ factor here
is responsible for
keeping the energy of the system extensive (linear in $N$), as the number of
interaction terms is quadratic in $N$. The factor $1/2$ compensates for the
fact that the summation over $i\ne j$ counts each pair twice. The first
observation is the trivial fact that
\begin{equation}
\left(\sum_is_i\right)^2=\sum_is_i^2+\sum_{i\ne j}s_is_j=N+\sum_{i\ne j}s_is_j
\end{equation}
where the second equality holds since $s_i^2\equiv 1$. It follows then, that
our Hamiltonian is, up to a(n immaterial) constant, equivalent to
\begin{align}
\calE(\bs)&=-B\sum_{i=1}^N s_i-\frac{J}{2N}
\left(\sum_{i=1}^Ns_i\right)^2\nonumber\\
&=
-N\left[B\cdot\left(\frac{1}{N}\sum_{i=1}^N s_i\right)+
\frac{J}{2}\left(\frac{1}{N}\sum_{i=1}^Ns_i\right)^2\right],
\end{align}
thus $\calE(\bs)$ depends on $\bs$ only via the magnetization
$m(\bs)=\frac{1}{N}\sum_is_i$. This fact makes the C--W model very easy to
handle:
\begin{align}
Z_N(\beta,B)&=\sum_{\bs}\exp\left\{N\beta\left[B\cdot
m(\bs)+\frac{J}{2}m^2(\bs)\right]\right\}\nonumber\\
&=\sum_{m=-1}^{+1}\Omega(m)\cdot e^{N\beta(Bm+Jm^2/2)}\nonumber\\
&\exe\sum_{m=-1}^{+1}e^{Nh_2((1+m)/2)}\cdot e^{N\beta(Bm+Jm^2/2)}\nonumber\\
&\exe\exp\left\{N\cdot\max_{|m|\le 1}\left[h_2\left(\frac{1+m}{2}\right)+
\beta Bm+\frac{\beta m^2J}{2}\right]\right\}
\end{align}
and so,
\begin{equation}
\phi(\beta,B)=\max_{|m|\le 1}\left[h_2\left(\frac{1+m}{2}\right)+
\beta Bm+\frac{\beta m^2J}{2}\right].
\end{equation}
The maximum is found by equating the derivative to zero, i.e.,
\begin{equation}
0=\frac{1}{2}\ln\left(\frac{1-m}{1+m}\right)+\beta B+\beta Jm
\equiv -\tanh^{-1}(m)+\beta B+\beta Jm
\end{equation}
or equivalently, the maximizing (and hence the dominant) 
$m$ is a solution $m^*$ to the equation\footnote{Once again, for $J=0$, we are
back to non--interacting spins and then this equation gives the
paramagnetic behavior $m=\tanh(\beta B)$.}
$$m=\tanh(\beta B+\beta Jm).$$
Consider first the case
$B=0$, where the equation boils down to
\begin{equation}
m=\tanh(\beta Jm).
\end{equation}
It is instructive to look at this equation graphically. 
Referring to Fig.\ \ref{cw1}, we have to make a distinction between
two cases: If $\beta J <1$, namely, $T > T_c\dfn J/k$, the slope of the
function $y=\tanh(\beta Jm)$ at the origin, 
$\beta J$, is smaller than the slope of the
linear function $y=m$, which is $1$, thus these two graphs intersect only at
the origin. It is easy to check that in this case, the second derivative of
\begin{equation}
\psi(m)\dfn h_2\left(\frac{1+m}{2}\right)+\frac{\beta 
Jm^2}{2} 
\end{equation}
at $m=0$ is negative, and therefore it is indeed the
maximum (see Fig.\ \ref{cw2}, left part). Thus,
the dominant magnetization is $m^*=0$, which means
disorder and hence no spontaneous
magnetization for $T > T_c$.
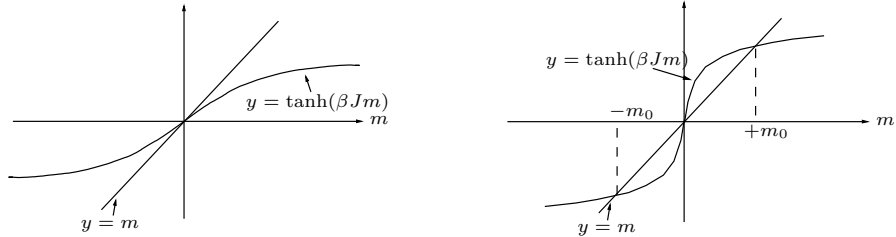
\begin{figure}[ht]
\hspace*{3cm}\input{cw1.pstex_t}
\caption{\small Graphical solutions of equation $m=\tanh(\beta Jm)$: The
left part corresponds to the case $\beta J < 1$, where there is one
solution only, $m^*=0$. The right part corresponds to the case $\beta J > 1$, where
in addition to the zero solution, there are two non--zero solutions $m^*=\pm
m_0$.}
\label{cw1}
\end{figure}
On the other hand, when $\beta J > 1$, which means temperatures lower than
$T_c$, the initial slope of the $\tanh$ function is larger than that of the
linear function, but since the $\tanh$ function cannot take values outside the interval
$(-1,+1)$, the two functions must intersect also at two additional, symmetric, non--zero
points, which we denote by $+m_0$ and $-m_0$ (see Fig.\ \ref{cw1}, right part). 
In this case, it can readily be
shown that the second derivative of 
$\psi(m)$ is positive at the origin (i.e., there is a local
minimum at $m=0$) and negative at $m=\pm m_0$, which means that 
there are maxima at these two points (see Fig.\ \ref{cw2}, right part). 
Thus, the dominant magnetizations are
$\pm m_0$, each capturing about half of the probability.
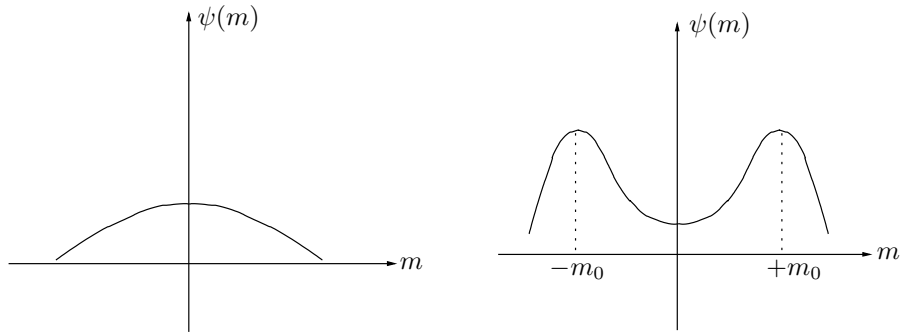
\begin{figure}[ht]
\hspace*{3cm}\input{cw2.pstex_t}
\caption{\small The function $\psi(m)=  h_2((1+m)/2)+\beta Jm^2/2$ has a
unique maximum at $m=0$ when $\beta J < 1$ (left graph) and two local maxima
at $\pm m_0$,
in addition to a local minimum at $m=0$, when $\beta J > 1$ (right graph).}
\label{cw2}
\end{figure}

Consider now the case $\beta J > 1$, where the magnetic field $B$ is brought
back into the picture. This will break the symmetry of the right graph of
Fig.\ \ref{cw2} and the corresponding graphs of $\psi(m)$ would be as in Fig.\
\ref{cw3}, where now the higher local maximum (which is also the global one)
is at $m_0(B)$ whose sign is as that of $B$. But as $B\to 0$, $m_0(B)\to m_0$
of Fig.\ \ref{cw2}.
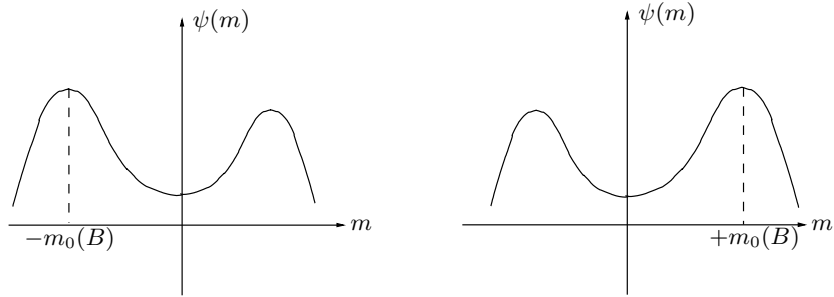
\begin{figure}[ht]
\hspace*{3cm}\input{cw3.pstex_t}
\caption{\small The case $\beta J > 1$ with a magnetic field $B$. The left
graph corresponds to $B < 0$ and the right graph -- to $B > 0$.}
\label{cw3}
\end{figure}
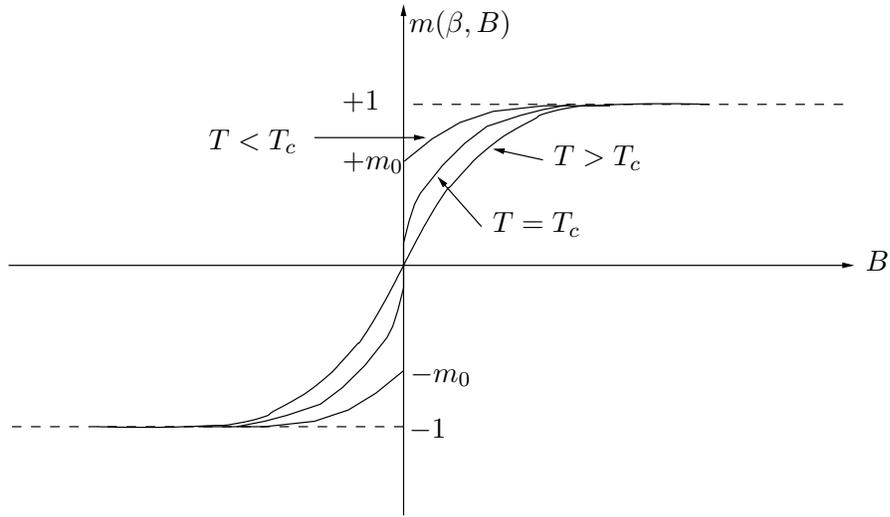
\begin{figure}[ht]
\hspace*{3cm}\input{cw4.pstex_t}
\caption{\small
Magnetization vs.\ magnetic field: For $T < T_c$ there is spontaneous
magnetization: $\lim_{B\downarrow 0}m(\beta,B)=+m_0$ and
$\lim_{B\uparrow 0}m(\beta,B)=-m_0$, and so there is a discontinuity at
$B=0$.}
\label{cw4}
\end{figure}
Thus, we see the spontaneous magnetization here. Even after removing the
magnetic field, the system remains magnetized to the level of $m_0$, depending
on the direction (the sign) of $B$ before its removal. Obviously, the
magnetization $m(\beta,B)$ has a 
discontinuity at $B=0$ for $T < T_c$, which is a first
order phase transition w.r.t.\ 
$B$ (see Fig.\ \ref{cw4}). We note that the point $T=T_c$ is
the boundary between the region of existence and the 
region of non--existence of a phase transition
w.r.t. $B$. Such a point is called a {\it critical point}. 
The phase transition w.r.t.\ $\beta$ is of the second order.

Finally, we should mention here an alternative technique that can be used
to analyze this model, which
is based on the 
Hubbard--Stratonovich transform.
Specifically, we have the following chain of equalities:
\begin{align}
Z(h,K)&=\sum_{\bs} \exp\left\{h\sum_{i=1}^N
s_i+\frac{K}{2N}\left(\sum_{i=1}^Ns_i\right)^2\right\}~~~~h\dfn \beta
B,~K\dfn \beta J\nonumber\\
&=\sum_{\bs} \exp\left\{h\sum_{i=1}^N
s_i\right\}\cdot\exp\left\{\frac{K}{2N}\left(\sum_{i=1}^Ns_i\right)^2\right\}\nonumber\\
&=\sum_{\bs} \exp\left\{h\sum_{i=1}^N
s_i\right\}\cdot\sqrt{\frac{N}{2\pi K}}\int_\reals \dd z
\exp\left\{-\frac{Nz^2}{2K}+z\cdot\sum_{i=1}^Ns_i\right\}\nonumber\\
&=\sqrt{\frac{N}{2\pi K}}\int_\reals \dd z e^{-Nz^2/(2K)}
\sum_{\bs} \exp\left\{(h+z)\sum_{i=1}^N
s_i\right\}\nonumber\\
&=\sqrt{\frac{N}{2\pi K}}\int_\reals \dd z e^{-Nz^2/(2K)}
\left[\sum_{s=-1}^1 e^{(h+z)s}\right]^N\nonumber\\
&=\sqrt{\frac{N}{2\pi K}}\int_\reals \dd z e^{-Nz^2/(2K)}
[2\cosh(h+z)]^N\nonumber\\
&=2^N\cdot\sqrt{\frac{N}{2\pi K}}\int_\reals \dd z
\exp\{N[\ln\cosh(h+z)-z^2/(2K)]\}
\end{align}
This integral can be shown\footnote{The basics of saddle--point integration, or
at least Laplace integration, should be taught in a recitation.}
dominated by $e$ to the $N$ times the maximum of the function in the square brackets at the
exponent of the integrand, or equivalently, the minimum of the function
\begin{equation}
\gamma(z)=\frac{z^2}{2K}-\ln\cosh(h+z).
\end{equation}
by equating its derivative to zero, we get the very same equation as
$m=\tanh(\beta B+\beta Jm)$ by setting $z=\beta Jm$.
The function $\gamma(z)$ is different from the function
$\psi$ that we maximized earlier, but the extremum is the same. This function
is called the {\it Landau free energy}.

\subsection{Spin Glasses$^*$}

So far we discussed only models where the non--zero coupling coefficients,
$\bJ=\{J_{ij}\}$ are equal, 
thus they are either all positive (ferromagnetic models) or
all negative (antiferromagnetic models). As mentioned earlier, there are also
models where the signs of these coefficients are mixed, which are called {\it
spin glass} models. 

Spin glass models have a much more complicated and more interesting behavior
than ferromagnets, because there might be metastable states due to the fact
that not necessarily all spin pairs $\{(s_i,s_j)\}$ can be in their preferred
mutual polarization. It might be the case that some of these pairs are
``frustrated.'' In order to model situations of amorphism and disorder in
such systems, it is customary to model the coupling coefficients as random
variables. This model with random
parameters means that there are now 
two levels of randomness:
\begin{itemize}
\item Randomness of the coupling coefficients $\bJ$.
\item Randomness of the spin configuration $\bs$ given $\bJ$, according
to the Boltzmann distribution, i.e.,
\begin{equation}
P(\bs|\bJ)=\frac{\exp\left\{\beta\left[B\sum_{i=1}^Ns_i+\sum_{(i,j)}J_{ij}s_is_j\right]\right\}}
{Z(\beta,B|\bJ)}.
\end{equation}
\end{itemize}
However, these two sets of random variables have a rather different stature.
The underlying setting is normally such that $\bJ$ is considered to be
randomly drawn once and for all, and then remain fixed, whereas $\bs$ keeps
varying all the time (according to the dynamics of the system).
At any rate, the time scale along which $\bs$ varies is much smaller than that
of $\bJ$. Another difference is that $\bJ$ is normally not
assumed to depend on temperature, whereas $\bs$, of course, does. In the
terminology of physicists, $\bs$ is considered 
an {\it annealed} random variable, whereas
$\bJ$ is considered a {\it quenched} random variable. Accordingly, there is a
corresponding distinction between {\it annealed averages} and {\it quenched
averages}.

Let us see what is exactly the difference
between the quenched averaging and the annealed one. If we examine, for
instance, the free energy, or the log--partition function, $\ln
Z(\beta|\bJ)$, this is now a random variable, of course, because it depends on the random
$\bJ$. If we denote by $\langle\cdot\rangle_{\bJ}$ the
expectation w.r.t.\ the randomness of $\bJ$, then quenched averaging means
$\langle\ln Z(\beta|\bJ)\rangle_{\bJ}$ (with the motivation of
the self--averaging property of the random variable $\ln Z(\beta|\bJ)$ in many cases),
whereas annealed averaging means $\ln\langle Z(\beta|\bJ)\rangle_{\bJ}$.
Normally, the relevant average is the quenched one, but it is typically
also much harder to calculate.
Clearly, the annealed average is never smaller than the quenched one because of Jensen's
inequality, but they sometimes coincide at high temperatures.
The difference between them is that in quenched averaging, the
dominant realizations of $\bJ$ are the typical ones, whereas in annealed
averaging, this is not necessarily the case. This follows from the
following sketchy consideration. As for the annealed average, we have:
\begin{align}
\left<Z(\beta|\bJ)\right>_{\bJ}&=\sum_{\bJ}P(\bJ)Z(\beta|\bJ)\nonumber\\
&\approx\sum_{\alpha}\mbox{Pr}\{\bJ:~Z(\beta|\bJ)\exe e^{N\alpha}\}\cdot
e^{N\alpha}\nonumber\\
&\approx\sum_{\alpha}e^{-NE(\alpha)}\cdot
e^{N\alpha}~~~~~~\mbox{(assuming exponential probabilities)}\nonumber\\
&\exe e^{N\max_\alpha[\alpha-E(\alpha)]}
\end{align}
which means that the annealed average is dominated by realizations
of the system with 
\begin{equation}
\frac{\ln Z(\beta|\bJ)}{N}\approx \alpha^*\dfn
\mbox{arg}\max_\alpha[\alpha-E(\alpha)],
\end{equation}
which may differ from the typical value of $\alpha$,
which is 
\begin{equation}
\alpha=\phi(\beta)
\equiv\lim_{N\to\infty}\frac{1}{N}\left<\ln
Z(\beta|\bJ)\right>.
\end{equation}
On the other hand, when it comes to quenched averaging, 
the random variable $\ln Z(\beta|\bJ)$ behaves linearly in $N$, 
and concentrates strongly around 
the typical value $N\phi(\beta)$,
whereas other values are weighted by (exponentially) decaying probabilities.

The literature on spin glasses includes many models for the randomness of
the coupling coefficients. We end this part by listing just a few.
\begin{itemize}
\item The {\it Edwards--Anderson} (E--A) model, where $\{J_{ij}\}$ are non--zero
for nearest--neighbor pairs only (e.g., $j=i\pm 1$ in one--dimensional model).
According to this model, these $J_{ij}$'s are i.i.d.\ random variables, which are normally
modeled to have a zero--mean Gaussian pdf, or binary symmetric with levels $\pm J_0$.
It is customary to work with a zero--mean distribution if we have a pure spin
glass in mind. If the mean is nonzero, the model has either a ferromagnetic
or an anti-ferromagnetic bias, according to the sign of the mean.
\item The {\it Sherrington--Kirkpatrick} (S--K) model, which is similar to the
E--A model, except that the support of $\{J_{ij}\}$ is extended to include all
$N(N-1)/2$ pairs, and not only nearest--neighbor pairs. This can be thought of
as a stochastic version of the C--W model in the sense that here too, there is
no geometry, and every spin `talks' to every other spin to the same extent,
but here the coefficients are random, as said. 
\item The {\it $p$--spin} model, which is similar to the S--K model, but now
the interaction term consists, not only of pairs, but also triples,
quadruples, and so on, up to cliques of size $p$, i.e., products 
$s_{i_1}s_{i_2}\cdot\cdot\cdot s_{i_p}$, where $(i_1,\ldots,i_p)$ exhaust all
possible subsets of $p$ spins out of $N$. Each such term has a Gaussian
coefficient $J_{i_1,\ldots,i_p}$ with an appropriate variance.
\end{itemize}
Considering the $p$--spin model, it turns out that if we look at the extreme
case of $p\to\infty$ (taken after the thermodynamic limit $N\to\infty$), the
resulting behavior turns out to be extremely erratic: all energy levels
$\{\calE(\bs)\}_{\bs\in\{-1,+1\}^N}$ become i.i.d.\ 
Gaussian random variables. This is,
of course, a toy model, which has very little to do with reality (if any),
but it is surprisingly interesting and easy to work with. It is called the
{\it random energy model} (REM). 

\newpage
\section{Vibrations in a Solid -- Phonons and Heat Capacity$^*$}

\subsection{Introduction}

A problem mathematically similar to that of black--body radiation, discussed
earlier, arises from vibrational modes of a solid. As in the case of
black--body radiation, the analysis of vibrational modes in a solid can be
viewed either by regarding the system as a bunch of {\it interacting} harmonic oscillators or as
a gas of particles, called {\it phonons} -- the analogue of photons, but in
the context of sound waves, rather than electromagnetic waves.

In this chapter, we shall use this point of view and 
apply statistical mechanical methods to calculate
the heat capacity of solids, or more precisely, the heat capacity
pertaining to the lattice vibrations of crystalline solids.\footnote{
In general, there are additional contributions to the heat capacity (e.g.,
from orientational ordering in paramagnetic salts, or from conduction
electrons in metals, etc.), but here we shall consider only the vibrational
heat capacity.}

There are two basic experimental facts about heat capacity in solids which any
good theory must explain. The first is that in room temperature the heat
capacity of most solids is about $3k$ per atom\footnote{Each atom has 6
degrees of freedom (3 of position + 3 of momentum). Classically, each one
of them contributes one quadratic term to the Hamiltonian, whose mean is
$kT/2$, thus a total mean energy of $3kT$, which means specific heat
of $3k$ per atom.}
ideal gas). This is essentially the law of Dulong and Petit (1819), but this
is only an approximation, and it might be quite wrong. The second fact is
that at low temperatures, the heat capacity at constant volume, $C_V$, decreases, and actually vanishes
at $T=0$. Experimentally, it was observed that the low--temperature dependence is of the
form
\begin{equation}
C_V=\alpha T^3+\gamma T,
\end{equation}
where $\alpha$ and $\gamma$ are constants that depend on the material and the
volume. For certain insulators, like potassium chloride, $\gamma=0$, namely,
$C_V$ is proportional to $T^3$. For metals (like copper), the linear term is
present, but it is contributed by the conduction electrons. A good theory
of the vibrational contribution to heat capacity should therefore predict
$T^3$ behavior at low temperatures.

In classical statistical mechanics, the equipartition theorem leads to a
constant heat capacity {\it at all temperatures}, in contradiction with both
experiments and with the third law of thermodynamics that asserts that as
$T\to 0$, the entropy $S$ tends to zero (whereas a constant heat capacity
would yield $S\propto \ln T$ for small $T$).

A fundamental step to resolve this discrepancy between classical 
theory and experiment was taken by Einstein in 1907, who treated the lattice
vibrations quantum mechanically. Einstein's theory qualitatively reproduces
the observed features. However, he used a simplified model and didn't expect
full agreement with experiment, but he pointed out the kind of modifications
which the model requires. Einstein's theory was later (1912) improved by Debye who 
considered a more realistic model.

\subsection{Formulation}

Consider a Hamiltonian of a classical solid composed
of $N$ atoms whose positions in space are specified by the coordinates
$\bx=(x_1,\ldots,x_{3N})$. In the state of lowest energy (the ground state),
these coordinates are denoted by 
$\bar{\bx}=(\bar{x}_1,\ldots,\bar{x}_{3N})$, which are normally points of a lattice
in the three--dimensional space, if
the solid in question is a crystal. Let $\xi_i=x_i-\bar{x}_i$,
$i=1,\ldots,3N$, denote the displacements. The kinetic energy of the
system is is clearly
\begin{equation}
K=\frac{1}{2}m\sum_{i=1}^{3N}\dot{x}_i^2=
\frac{1}{2}m\sum_{i=1}^{3N}\dot{\xi}_i^2
\end{equation}
and the potential energy is
\begin{equation}
\Phi(\bx)=\Phi(\bar{\bx})+\sum_i\left(\frac{\partial\Phi}{\partial
x_i}\right)_{\bx=\bar{\bx}}\xi_i+\sum_{i,j}
\frac{1}{2}\left(\frac{\partial^2\Phi}{\partial x_i\partial
x_j}\right)_{\bx=\bar{\bx}}\xi_i\xi_j+\ldots
\end{equation}
The first term in this expansion represents 
the minimum energy when all atoms are at rest in their mean positions
$\bar{x}_i$. We henceforth denote this energy by $\Phi_0$. 
The second term is identically zero because $\Phi(\bx)$ is minimized at
$\bx=\bar{\bx}$. The second order terms of this expansion represent the 
{\it harmonic component} of the vibrations. If we assume that the overall
amplitude of the vibrations is reasonably small, we can safely neglect all
successive terms and then we are working with
the so called {\it harmonic approximation}. Thus, we may write
\begin{equation}
\calE(\bx)=\Phi_0+\frac{1}{2}m\sum_{i=1}^{3N}\dot{\xi}_i^2+\sum_{i,j}\alpha_{ij}\xi_i\xi_j
\end{equation}
where we have denoted
\begin{equation}
\alpha_{ij}=\frac{1}{2}\cdot\frac{\partial^2\Phi}{\partial x_i\partial
x_j}\bigg|_{\bx=\bar{\bx}}.
\end{equation}
This Hamiltonian corresponds to harmonic oscillators that are coupled to one
another, as discussed in Subsections 3.2.2 and 6.2, where the off--diagonal
terms of the matrix $A=\{\alpha_{ij}\}$ designate the pairwise interactions. 
This Hamiltonian obeys the general form of eq.\ (\ref{pairwiseinteractions}).

While Einstein neglected the off--diagonal terms of $A$ in the first place,
Debye did not. In the following, we present the latter approach, which is
more general (and more realistic), whereas the former
will essentially be a special case.

\subsection{Heat Capacity Analysis}

The first idea of the analysis is to transform the coordinates
into a new domain where the components 
are all decoupled. This means diagonalizing
the matrix $A$. Since $A$ is a symmetric non--negative definite matrix, it is
clearly diagonalizable by a unitary matrix formed by its eigenvectors,
and the diagonal elements of the diagonalized matrix (which are the eigenvalues of $A$) 
are non--negative. Let us
denote the new coordinates of the system by $q_i$, $i=1,\ldots,3N$, and the
eigenvalues -- by $\frac{1}{2}m\omega_i^2$. By linearity of the
differentiation operation, the same transformation take us from
the vector of velocities $\{\dot{\xi}_i\}$ (of the kinetic component of the
Hamiltonian) to the vector of derivatives of $\{q_i\}$, which will be denoted
$\{\dot{q}_i\}$. Fortunately enough, since the transformation is unitary it
leaves the components $\{\dot{q}_i\}$ decoupled. In other words, by the Parseval theorem,
the norm of $\{\dot{\xi}_i\}$ is equal to the norm of $\{\dot{q}_i\}$. Thus,
in the transformed domain, the Hamiltonian reads
\begin{equation}
\calE(\bq)=\Phi_0+\frac{1}{2}m\sum_i(\dot{q}_i^2+\omega_i^2q_i^2).
\end{equation}
which can be viewed as $3N$ {\it decoupled} harmonic
oscillators, each one oscillating
in its individual normal mode $\omega_i$.
The parameters $\{\omega_i\}$ are
called {\it characteristic frequencies} or {\it normal modes}.

\vspace{0.2cm}

\noindent
{\small {\it Example 7.1 -- One--dimensional ring of springs.}
If the system has translational symmetry and if, in addition, there are periodic
boundary conditions, then the matrix $A$ is circulant, which means that it is always
diagonalized by the discrete Fourier transform (DFT). In this case, $q_i$ are
the corresponding spatial frequency variables, conjugate to the 
location displacement variables
$\xi_i$. The simplest example of this is a ring of $N$ one--dimensional springs, as
discussed in Subsection 6.2 (see left part of Fig.\ \ref{springs2d}), where
the Hamiltonian (in the current notation) is
\begin{equation}
\calE(\bx)=\Phi_0+\frac{1}{2}m\sum_i\dot{\xi}_i^2+\frac{1}{2}K\sum_i(\xi_{i+1}-\xi_i)^2
\end{equation}
In this case, the matrix $A$ is given by
\begin{equation}
A=K\cdot \left(\begin{array}{cccccc}
1 & -\frac{1}{2} & 0 & \ldots & 0 & -\frac{1}{2}\\
-\frac{1}{2} & 1 & -\frac{1}{2} & 0 & \ldots & 0 \\
0 & -\frac{1}{2} & 1 & -\frac{1}{2} & 0 & \ldots \\
\cdot & \cdot & \cdot & \cdot & \cdot & \cdot\\
0 & 0 & \ldots & -\frac{1}{2} & 1 & -\frac{1}{2}\\
-\frac{1}{2} & 0 & \ldots & 0 & -\frac{1}{2} & 1 \\
\end{array}\right)
\end{equation}
The eigenvalues of $A$ are $\lambda_i=K[1-\cos(2\pi i/N)]$, which are simply
the DFT coefficients of the $N$--sequence formed by any row of $A$ (removing the
complex exponential of the phase factor). This means
that the normal modes are
$\omega_i=\sqrt{2K[1-\cos(2\pi i/N)]/m}$.} $\Box$

\vspace{0.2cm}

Classically, each of the $3N$
normal modes of vibration corresponds to a wave of distortion of the lattice.
Quantum--mechanically, these modes give rise to quanta called {\it phonons},
in analogy to the fact that vibrational modes of electromagnetic waves give
rise to photons. There is one important difference, however: While the number
of normal modes in the case of an electromagnetic wave is infinite, here the
number of modes (or the number of phonon energy levels)
is finite -- there are exactly $3N$ of them. This gives rise to a few
differences in the physical behavior, but at low temperatures, where the
high--frequency modes of the solid become unlikely to be excited, these
differences become insignificant.

The eigenvalues of the Hamiltonian are then
\begin{equation}
E(n_1,n_2,\ldots)=\Phi_0+\sum_i\left(n_i+\frac{1}{2}\right)\hbar\omega_i.
\end{equation}
where the non-negative integers $\{n_i\}$ denote the `states of excitation' of
the various oscillators, or equally well, the occupation numbers of the
various phonon levels in the system.
The internal energy is then
\begin{eqnarray}
\left<E\right>&=&-\frac{\partial}{\partial\beta}\ln Z_{3N}(\beta)\nonumber\\
&=&-\frac{\partial}{\partial\beta}\ln\left(\sum_{n_1}\sum_{n_2}\ldots
\exp\left\{-\beta\left[\Phi_0+\sum_i\left(n_i+\frac{1}{2}\right)
\hbar\omega_i\right]\right\}\right)\nonumber\\
&=&-\frac{\partial}{\partial\beta}\ln\left[e^{-\beta\Phi_0}\prod_i\frac{e^{-\beta\hbar\omega_i/2}}{1-
e^{-\beta\hbar\omega_i}}\right]\nonumber\\
&=&\Phi_0+\sum_i\frac{1}{2}\hbar\omega_i+\sum_i\frac{\partial}{\partial\beta}
\ln(1-e^{-\beta\hbar\omega_i})\nonumber\\
&=&\Phi_0+\sum_i\frac{1}{2}\hbar\omega_i+\sum_i\frac{\hbar\omega_i}{1-e^{-\beta\hbar\omega_i}}.
\end{eqnarray}
Only the last term of the last expression depends on $T$. Thus, the
heat capacity at constant volume\footnote{Exercise 7.1: Why is this the heat
capacity at constant volume? Where is the assumption of constant volume being
used here?} is:
\begin{equation}
C_V=\frac{\partial \left<E\right>}{\partial T}=k\sum_i
\frac{(\hbar\omega_i/kT)^2e^{h\omega_i/kT}}{(e^{h\omega_i/kT}-1)^2}.
\end{equation}
To proceed from here, one has 
to know (or assume something) about the form of the
density $g(\omega)$ of $\{\omega_i\}$ and
then pass from summation to integration. 
At this point begins the difference
between Einstein's approach and Debye's approach.

\subsubsection{Einstein's Theory}

For Einstein, who assumed that the
oscillators do not interact in the original, $\xi$--domain,
all the normal modes are equal $\omega_i=\omega_E$ for all $i$, because
then (assuming translational symmetry) $A$ is proportional to the identity
matrix and then all its eigenvalues are the same. Thus, in Einstein's model
$g(\omega)=3N\delta(\omega-\omega_E)$, and the result is
\begin{equation}
C_V=3Nk\bE(x)
\end{equation}
where $\bE(x)$ is the so--called the {\it Einstein function}:
\begin{equation}
\bE(x)=\frac{x^2e^x}{(e^x-1)^2}
\end{equation}
with
\begin{equation}
x=\frac{\hbar\omega_E}{kT}\dfn \frac{\Theta_E}{T}.
\end{equation}
where $\Theta_E=\hbar\omega_E/k$ is called the {\it Einstein temperature}.
At high temperatures $T \gg \Theta_E$, where $x\ll 1$ and then $\bE(x)\approx
1$, we readily see that $C_V(T)\approx 3Nk$, 
in agreement with classical physics.
For low temperatures, $C_V(T)$ falls exponentially fast as $T\to 0$. 
This theoretical rate of decay, however, is way too fast compared to the
observed rate, which is cubic, as described earlier. But at least, Einstein's
theory predicts the qualitative behavior correctly.

\subsubsection{Debye's Theory}

Debye (1912), on the other hand, assumed a continuous density $g(\omega)$.
He assumed some cutoff frequency $\omega_D$, so that
\begin{equation}
\int_0^{\omega_D}g(\omega)\mbox{d}\omega=3N.
\end{equation}
For $g(\omega)$ in the range $0\le\omega\le\omega_D$, Debye adopted a Rayleigh
expression in the spirit of the one we saw in black--body radiation, but
with a distinction between the longitudinal mode and the two independent transverse modes
associated with the propagation of each wave at a given frequency.
Letting $v_L$ and $v_T$ denote the corresponding velocities of these modes,
this amounts to
\begin{equation}
g(\omega)\mbox{d}\omega=V\left(\frac{\omega^2\mbox{d}\omega}{2\pi^2v_L^3}+
\frac{\omega^2\mbox{d}\omega}{\pi^2v_T^3}\right).
\end{equation}
This, together with the previous equation, determines the cutoff frequency to
be
\begin{equation}
\omega_D=\left[\frac{18\pi^2\rho}{1/v_L^3+2/v_T^3}\right]^{1/3}
\end{equation}
where $\rho=N/V$ is the density of the atoms.
Accordingly,
\begin{equation}
g(\omega)=\left\{\begin{array}{cc}
\frac{9N}{\omega_D^3}\omega^2 & \omega\le\omega_D\\
0 & \mbox{elsewhere}\end{array}\right.
\end{equation}
The Debye formula for the heat capacity is now
\begin{equation}
C_V=3Nk\bD(x_0)
\end{equation}
where $\bD(\cdot)$ is called the {\it Debye function}
\begin{equation}
\bD(x_0)=\frac{3}{x_0^3}\int_0^{x_0}\frac{x^4e^x\mbox{d}x}{(e^x-1)^2}
\end{equation}
with
\begin{equation}
x_0=\frac{\hbar\omega_D}{kT}\dfn \frac{\Theta_D}{T},
\end{equation}
where $\Theta_D=\hbar\omega_D/k$ is called the {\it Debye temperature}.
Integrating by parts, the Debye function can also be written as
\begin{equation}
\bD(x_0)=-\frac{3x_0}{e^{x_0}-1}+\frac{12}{x_0^3}\int_0^{x_0}\frac{x^3\mbox{d}x}{e^x-1}.
\end{equation}
Now, for $T\gg \Theta_D$, which means $x_0\ll 1$, $D(x_0)$ can be approximated
by a Taylor series expansion:
\begin{equation}
\bD(x_0)=1-\frac{x_0^2}{20}+\ldots
\end{equation}
Thus, for high temperatures, we again recover the classical result $C_V=3Nk$.
On the other hand, for $T\ll \Theta_D$, which is $x_0\gg 1$, the dominant term
in the integration by parts is the second one, which gives the approximation
\begin{equation}
\bD(x_0)\approx \frac{12}{x_0^3}\int_0^\infty\frac{x^3\mbox{d}x}{e^x-1}=
\frac{4\pi^4}{5x_0^3}=\frac{4\pi^4}{5}\left(\frac{T}{\Theta_D}\right)^3.
\end{equation}
Therefore, at low temperatures, the heat capacity is
\begin{equation}
C_V\approx \frac{12\pi^4}{5}Nk\left(\frac{T}{\Theta_D}\right)^3.
\end{equation}
In other words, Debye's theory indeed recovers the $T^3$ behavior at low
temperatures, in agreement with experimental evidence. Moreover, the
match to experimental results is very good, not only
near $T=0$, but across a rather wide range of temperatures. In some textbooks,
like \cite[p. 164, Fig.\ 6.7]{Mandl71}, or \cite[p.\ 177, Fig.\
7.10]{Pathria96}, there are plots of $C_V$ as a function of $T$ for certain
materials, which show impressive proximity between theory and measurements.

\vspace{0.2cm}

\noindent
{\small {\it Exercise 7.2.} Extend Debye's analysis to allow two different
cutoff frequencies, $\omega_L$ and $\omega_T$ -- for the longitudinal and the
transverse modes, respectively.}

\vspace{0.2cm}

\noindent
{\small {\it Exercise 7.3.} Calculate the density $g(\omega)$ for a ring of
springs as described in Example 7.1. Write an expression for $C_V$ as an
integral and try to simplify it as much as you can.}

\newpage
\section{Fluctuations, Stochastic Dynamics and Noise}

So far we have discussed mostly systems in equilibrium. Extensive quantities
like volume, energy, occupations numbers, etc., have been calculated as means
of certain ensembles, with the justification that these are not only just means but
moreover, the values around which most of the probability concentrates in the
thermodynamic limit, namely, the limit of a very large system.
In this chapter, we will investigate the statistical fluctuations around
these means, as well as dynamical issues, like the rate of approach to equilibrium
when a system is initially away from equilibrium. We will also discuss noise
phenomena and noise generation mechanisms as well as their implications and impact on
electric circuits and other systems.

Historically, the theory of fluctuations has been interesting and useful because it
made explicable several experimental effects, which late--nineteenth--century
physicists, firmly adhering to thermostatics and classical mechanics, were
not able to explain rigorously.
One such phenomenon is Brownian motion -- the irregular, random motion 
of light particles suspended 
in a drop of liquid, which is observable with a microscope.
Another phenomenon is electrical noise, such as thermal noise and shot noise, 
as mentioned in the previous paragraph. 

Classical thermodynamics cannot explain fluctuations and, in fact,
even denies their existence, because as we will see, a fluctuation into a less
likely state involves a decrease of entropy, which is seemingly contradictory
to the nineteenth--century ideas of the steady increase of entropy. This
contradiction is resolved by the statistical--mechanical viewpoint, according
to which the increase of entropy holds true only on the average (or with high
probability), not deterministically.

Apart from their theoretical interest, fluctuations are important to
understand in order to make accurate measurements
of physical properties and at the same time, to realize that the precision is limited by the
fluctuations. 

\subsection{Elements of Fluctuation Theory}

So far, we have established probability distributions for various physical
situations and have taken for granted the most likely value (or the mean value) as {\it
the value} of the physical quantity of interest. For example, the internal
energy in the canonical ensemble was taken to be $\left<E\right>$, which is
also the most likely value, with a very sharp peak as $N$ grows.

The first question that we ask now is what is the probabilistic
characterization of the departure from the mean. One of the most natural measures
of this departure is the variance. In the above--mentioned example of the energy,
$\mbox{Var}\{E\}=\left<E^2\right>-\left<E\right>^2$ or the relative standard
deviation $\sqrt{\mbox{Var}\{E\}}/\left<E\right>$. When several physical quantities are
involved, then the covariances between them are also measures of fluctuation.
There are two possible routes to assess fluctuations in this second order
sense. The first is directly from the relevant ensemble, and the second is by
a Gaussian approximation around the mean.
It should be emphasized that when it comes to fluctuations, the principle of
ensemble equivalence no longer holds in general. For example, in the
microcanonical ensemble, $\mbox{Var}\{E\}=0$ (since $E$ is strictly fixed), 
whereas in the canonical ensemble,
it is normally extensive, as we shall see shortly 
(think of the variance of the sum of $N$ i.i.d.\ 
random variables). Only $\sqrt{\mbox{Var}\{E\}}/\left<E\right>$,
which is proportional to $1/\sqrt{N}$ and hence tends to $0$, can be
considered asymptotically equivalent (in a very rough sense) 
to that of the microcanonical ensemble.

Consider first a composite system,
consisting of two subsystems, labeled 1 and 2, 
which together reside in a microcanonical ensemble with common temperature $T$
and common pressure $P$. We think of subsystem no.\ 1 as the subsystem of
interest, whereas subsystem no.\ 2 is the reservoir.
The two subsystems are allowed to exchange volume and
energy, but not
particles. Let $\Delta S$ denote
the deviation in the entropy of the composite system from its equilibrium
value $S_0$, i.e., 
\begin{equation}
\Delta S \equiv S-S_0 =k\ln\Omega_f-k\ln\Omega_0,
\end{equation}
where $\Omega_f$ and $\Omega_0$ are the numbers of accessible microstates
corresponding to the presence and absence of the fluctuation, respectively.
The probability that this fluctuation would occur is obviously
\begin{equation}
p=\frac{\Omega_f}{\Omega_0}=\exp(\Delta S/k).
\end{equation}
Considering the two subsystems, of course, 
\begin{equation}
\Delta S= \Delta S_1+\Delta S_2=\Delta
S_1+\int_0^f\frac{\mbox{d}E_2+P\mbox{d}V_2}{T}
\approx \Delta S_1+\frac{\Delta E_2+P\Delta V_2}{T}
\end{equation}
where the second equality is approximate since we have neglected the resulting perturbations
in pressure and temperature, as these have only second order effects.
At the same time, $\Delta E_1=-\Delta E_2$ and 
$\Delta V_1=-\Delta V_2$, so 
\begin{equation}
\Delta S = \Delta S_1-\frac{\Delta E_1+P\Delta V_1}{T}
\end{equation}
Accordingly,
\begin{equation}
p\propto \exp\{-(\Delta E_1-T\Delta S_1+P\Delta V_1)/kT\}.
\end{equation}
At this point, it is observed that this probability distribution does not
depend on subsystem no.\ 2, it depends only on system no.\ 1. So from now on, we
might as well forget about subsystem no.\ 2 and hence drop the subscript 1.
I.e.,
\begin{equation}
\label{p}
p\propto \exp\{-(\Delta E-T\Delta S+P\Delta V)/kT\}.
\end{equation}
By a Taylor series expansion
\begin{equation}
\Delta E-T\Delta S+P\Delta V=\frac{1}{2}
\left(\frac{\partial^2E}{\partial S^2}\right)_0(\Delta S)^2+
\left(\frac{\partial^2E}{\partial S\partial V}\right)_0\Delta S\Delta V+
\frac{1}{2}\left(\frac{\partial^2E}{\partial V^2}\right)_0(\Delta V)^2 +
\ldots
\end{equation}
Plugging this into eq.\ (\ref{p}) and retaining 
only the second order terms, we end up with
\begin{equation}
p\propto \exp\left\{-\frac{1}{2kT}\left[\left(\frac{\partial^2E}{\partial
S^2}\right)_0(\Delta S)^2+
2\left(\frac{\partial^2E}{\partial S\partial V}\right)_0\Delta S\Delta V+
\frac{1}{2}\left(\frac{\partial^2E}{\partial V^2}\right)_0(\Delta
V)^2\right]\right\},
\end{equation}
which means that the random vector $(\Delta V,\Delta S)$ is approximately
Gaussian\footnote{This finding can be thought of as a variation of the central
limit theorem.}
with a covariance matrix that depends on the partial second order derivatives 
of $E$ w.r.t.\ $V$ and $S$. In particular, this covariance matrix is
proportional to the inverse of the Hessian matrix of $E(S,V)$, which can be
calculated for a given system in the microcanonical ensemble.
We note that the second order partial derivatives of $E$ are inversely
proportional to the system size (i.e., $N$ or $V$), 
and therefore by the aforementioned inverse relationship, the
covariances are proportional to the system size (extensive). This means that
the relative standard deviations are inversely proportional to the
square root of the system size. Of course, one can extend the scope of this
derivation and allow also fluctuations in $N$, finding that the random vector
$(\Delta V,\Delta S,\Delta N)$ is also jointly Gaussian with a well--defined
covariance matrix.\footnote{Exercise 8.1: Carry out this extension.}

\vspace{0.2cm}

\noindent
{\small {\it Example 8.1 -- Ideal gas.} In the case of the ideal gas,
eq.\ (\ref{entropyidealgas}) gives
\begin{equation}
E(S,V)= \frac{3N^{5/3}h^2}{4\pi e^{5/3}mV^{2/3}}e^{2S/(3Nk)},
\end{equation}
whose Hessian is
\begin{equation}
\nabla^2E=\frac{2E}{9}\left(\begin{array}{cc}
2/(Nk)^2 & -2/(NkV)\\
-2/(NkV) & 5/V^2\end{array}\right).
\end{equation}
Thus, the covariance matrix of $(\Delta V,\Delta S)$ is
\begin{equation}
\Lambda=kT\cdot(\nabla^2E)^{-1}=\frac{9kT}{2E}\cdot\frac{(NkV)^2}{6}\cdot
\left(\begin{array}{cc}
5/V^2 & 2/(NkV)\\
2/(NkV) & 2/(Nk)^2\end{array}\right)
\end{equation}
or, using the relation $E=3NkT/2$, 
\begin{equation}
\Lambda=\left(\begin{array}{cc}
5Nk^2/2 & kV\\
kV & V^2/N\end{array}\right).
\end{equation}
Thus, $\mbox{Var}\{\Delta S\}=5Nk^2/2$, $\mbox{Var}\{\Delta V\}=V^2/N$, and
$\left<\Delta S\Delta V\right>=kV$, which are all extensive.} $\Box$

Similar derivations can be carried out for other ensembles, like the
canonical ensemble and the grand--canonical ensemble.

\subsection{Brownian Motion and the Langevin Equation}

We now discuss the topic of Brownian motion, which is of fundamental
importance. The term ``Brownian motion'' is after the botanist Robert Brown,
who in 1828 made careful observations of
tiny pollen grains in a liquid under a microscope and saw that they move in
a rather random fashion, and that this motion was not triggered by any currents or
other processes that take place in the liquid, 
like evaporation, etc. The movement of this pollen grain was
caused by frequent collisions with the particles of the liquid. Einstein (1905) was the
first to provide a sound theoretical analysis 
of the Brownian motion on the basis of the so called ``random walk problem.''
Here, we introduce the topic using a formulation due to the French physicist
Paul Langevin (1872--1946), which makes
the derivation of the dynamics extremely simple.

Langevin focuses on the motion of a relatively large particle of mass $m$
whose center of mass is at $x(t)$ at time $t$ and whose velocity is
$v(t)=\dot{x}(t)$. The particle is subjected to the influence of a force,
which is composed of two components, one is a slowly varying macroscopic force
and the other is varying very rapidly and randomly.
The latter component of the force has zero mean, but it fluctuates.
In the one--dimensional case then, the location of the particle obeys the
simple differential equation
\begin{equation}
\label{langevin}
m\ddot{x}(t)+\gamma\dot{x}(t)=F+F_r(t),
\end{equation}
where $\gamma$ is a frictional (dissipative) coefficient and $F_r(t)$ is the random
component of the force. While this differential equation is nothing but
Newton's law and hence obvious in macroscopic physics, it should not be taken
for granted in the microscopic regime. In elementary Gibbsian statistical
mechanics, all processes are time reversible in the microscopic level, since
energy is conserved in collisions as the effect of dissipation in binary
collisions is traditionally neglected. A reasonable theory, however, should
incorporate the dissipative term.

Clearly, the response of $x(t)$ to $F+F_r(t)$ is the superposition of the
individual responses to $F$ and to $F_r(t)$ separately. The former is the
solution to a simple (deterministic) differential equation, which is not the
center of our interest here. Consider then the response to the random
component $F_r(t)$ alone.
Multiplying  eq.\ (\ref{langevin}) by $x(t)$, we get
\begin{equation}
mx(t)\ddot{x}(t)\equiv
m\left[\frac{\mbox{d}(x(t)\dot{x}(t))}{\mbox{d}t}-\dot{x}^2(t)\right]=-\gamma
x(t)\dot{x}(t)+x(t)F_r(t).
\end{equation}
Now, let's take the expectation, assuming that due to
the randomness of $\{F_r(t)\}$, $x(t)$ and $F_r(t)$ at time $t$, are independent,
and so, $\left<x(t)F_r(t)\right>=\left<x(t)\right>\left<F_r(t)\right>=0$.
Also, note that $m\left<\dot{x}^2(t)\right>=kT$ 
by the energy equipartition theorem
(which applies here since we are assuming the classical regime),
and so, we end up with
\begin{equation}
m\frac{\mbox{d}\left<x(t)\dot{x}(t)\right>}{\mbox{d}t}=kT-\gamma
\left<x(t)\dot{x}(t)\right>,
\end{equation}
a simple first order differential equation, whose solution is
\begin{equation}
\left<x(t)\dot{x}(t)\right>=\frac{kT}{\gamma}+Ce^{-\gamma t/m},
\end{equation}
where $C$ is a constant of integration.
Imposing the condition that $x(0)=0$, this gives $C=-kT/\gamma$, and so
\begin{equation}
\frac{1}{2}\frac{\mbox{d}\left<x^2(t)\right>}{\mbox{d}t}\equiv 
\left<x(t)\dot{x}(t)\right>=\frac{kT}{\gamma}\left(1-e^{-\gamma t/m}\right),
\end{equation}
which yields
\begin{equation}
\label{exactvar}
\left<x^2(t)\right>=\frac{2kT}{\gamma}\left[t-\frac{m}{\gamma}(1-e^{-\gamma
t/m})\right].
\end{equation}
The last equation gives the mean square deviation of a particle 
away from its origin, after time $t$. 
The time constant of the dynamics, a.k.a.\ the {\it relaxation time}, is 
$\theta=m/\gamma$.
For short times ($t\ll \theta$),
$\left<x^2(t)\right>\approx kTt^2/m$, which means
that it looks like
the particle is moving at constant velocity of 
$\sqrt{kT/m}$. For $t\gg \theta$, however,
\begin{equation}
\label{variance}
\left<x^2(t)\right>\approx \frac{2kT}{\gamma}\cdot t.
\end{equation}
It should now be pointed out that
this linear growth rate of $\left<x^2(t)\right>$ is a characeristic of
Brownian motion. Here it only an approximation for $t\gg \theta$, 
as for $m > 0$, $\{x(t)\}$
is not a pure Brownian motion. Pure Brownian motion corresponds to the 
case $m=0$ (hence $\theta=0$), namely,
the term $m\ddot{x}(t)$ in the Langevin equation 
can be neglected,
and then $x(t)$ is simply proportional to 
$\int_0^t F_r(\tau)\mbox{d}\tau$ where $\{F_t(t)\}$ is white noise.
Fig.\ \ref{brown} illustrates a few realizations of a Brownian motion
in one dimenision and in two dimensions.

\begin{figure}[h!t!b!]
\centering
\includegraphics[width=4.5cm, height=4.5cm]{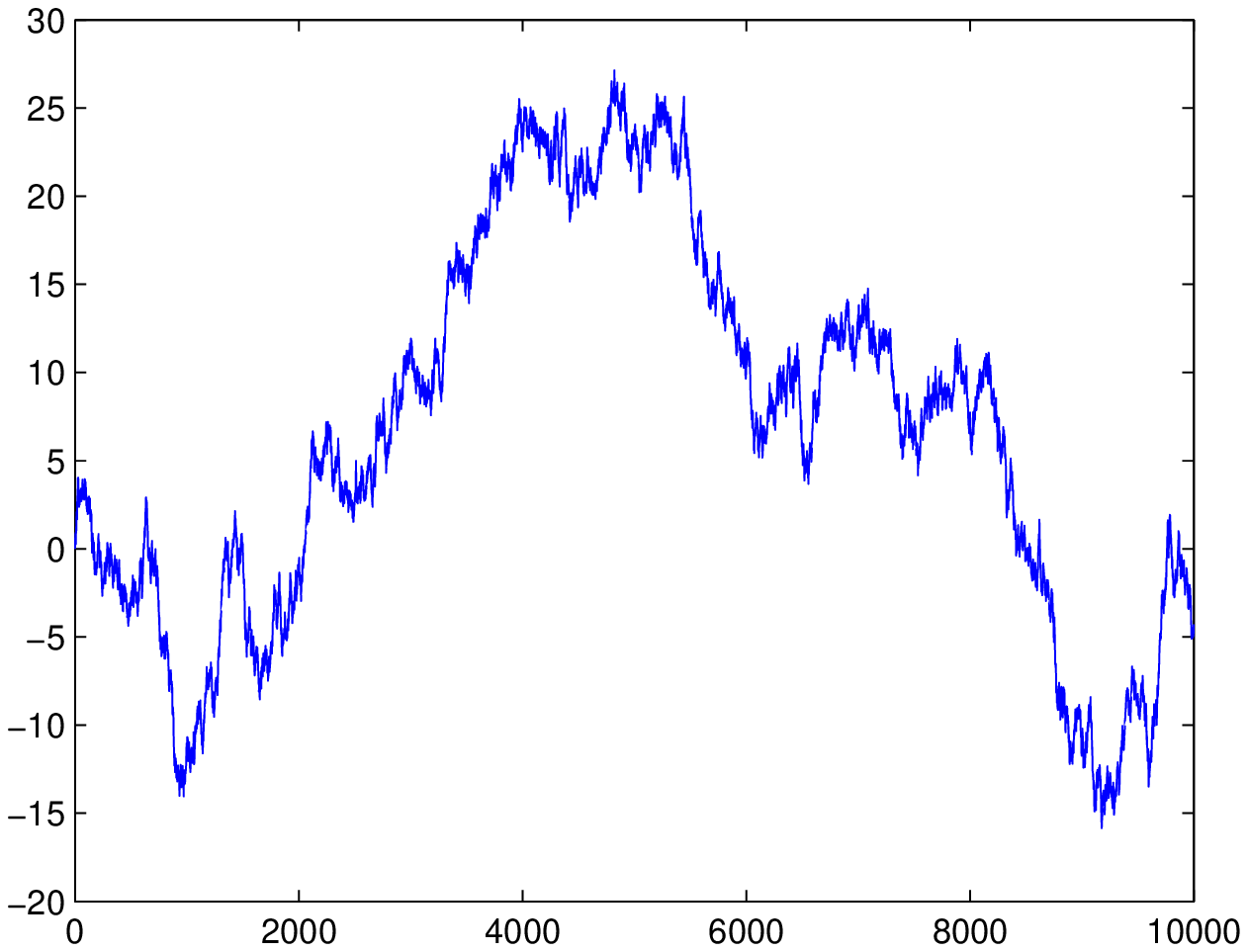}
\includegraphics[width=4.5cm, height=4.5cm]{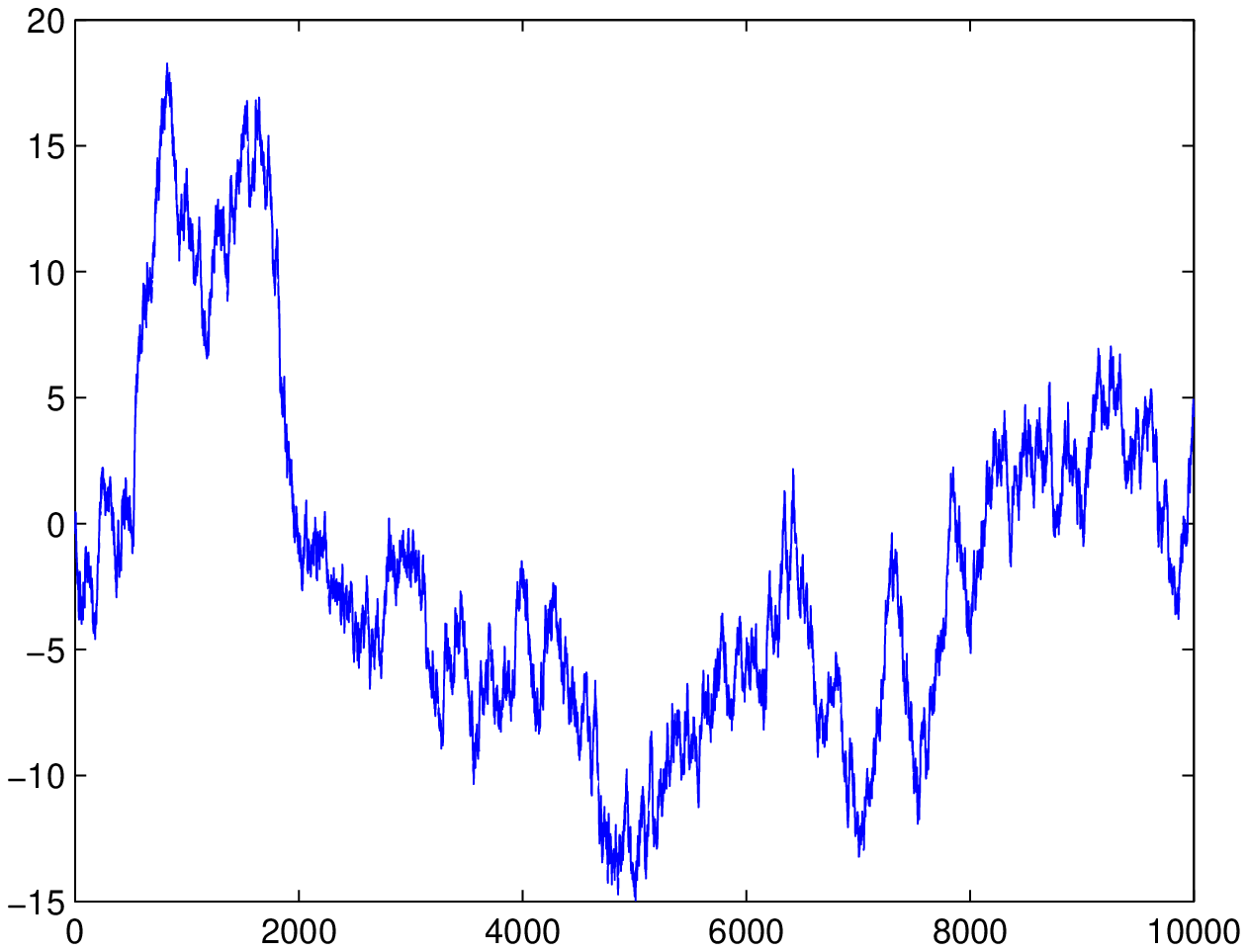}
\includegraphics[width=4.5cm, height=4.5cm]{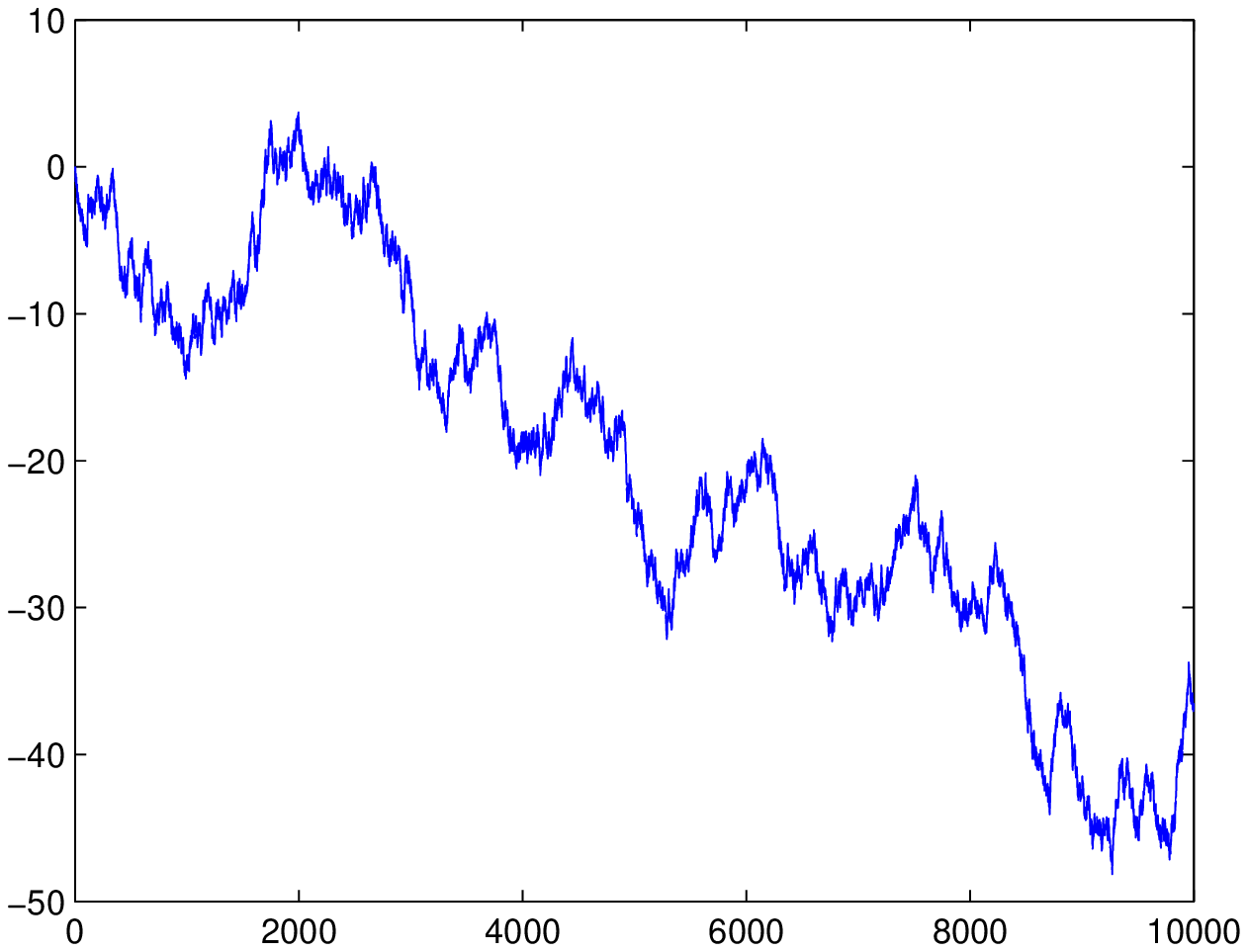}
\includegraphics[width=4.5cm, height=4.5cm]{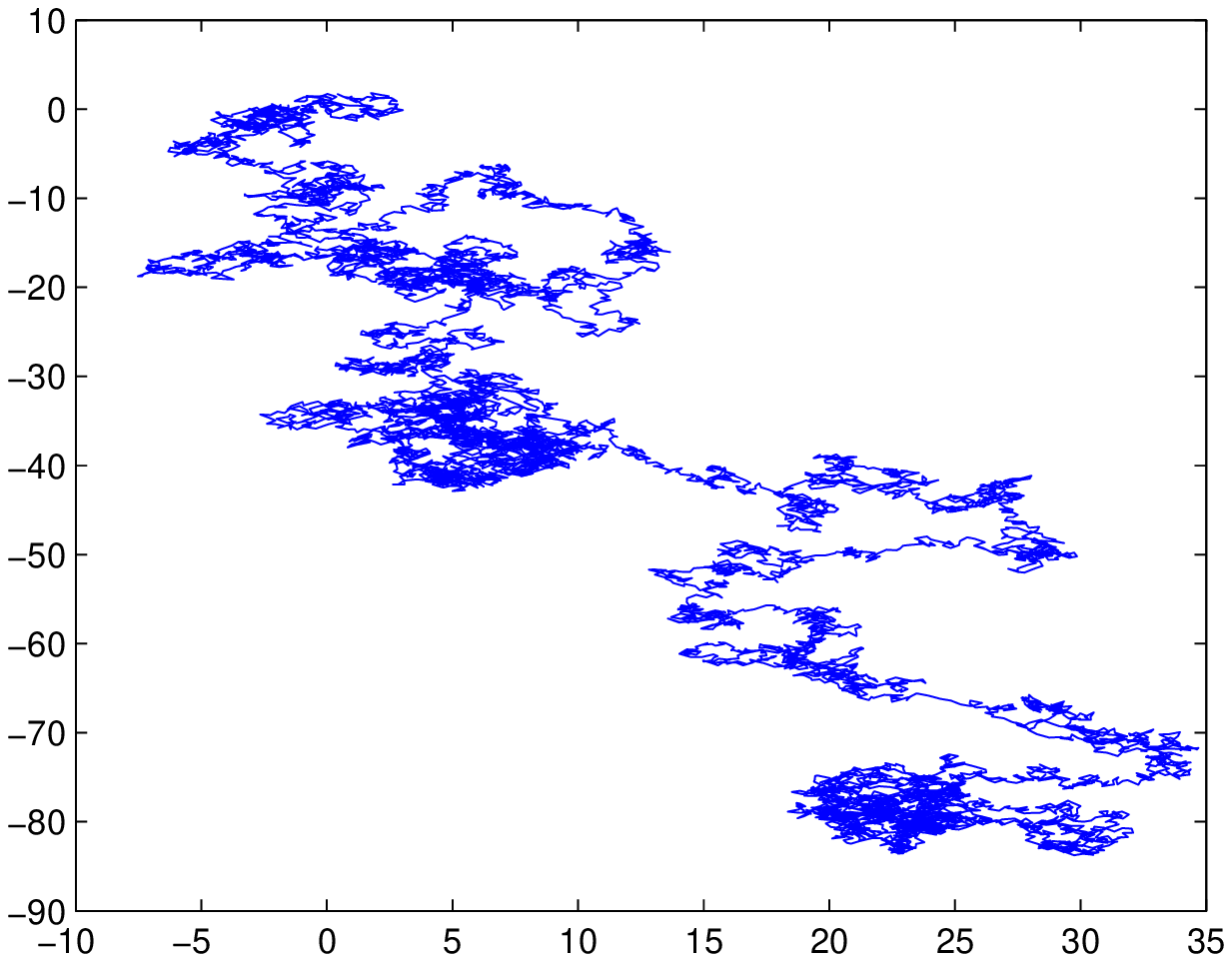}
\includegraphics[width=4.5cm, height=4.5cm]{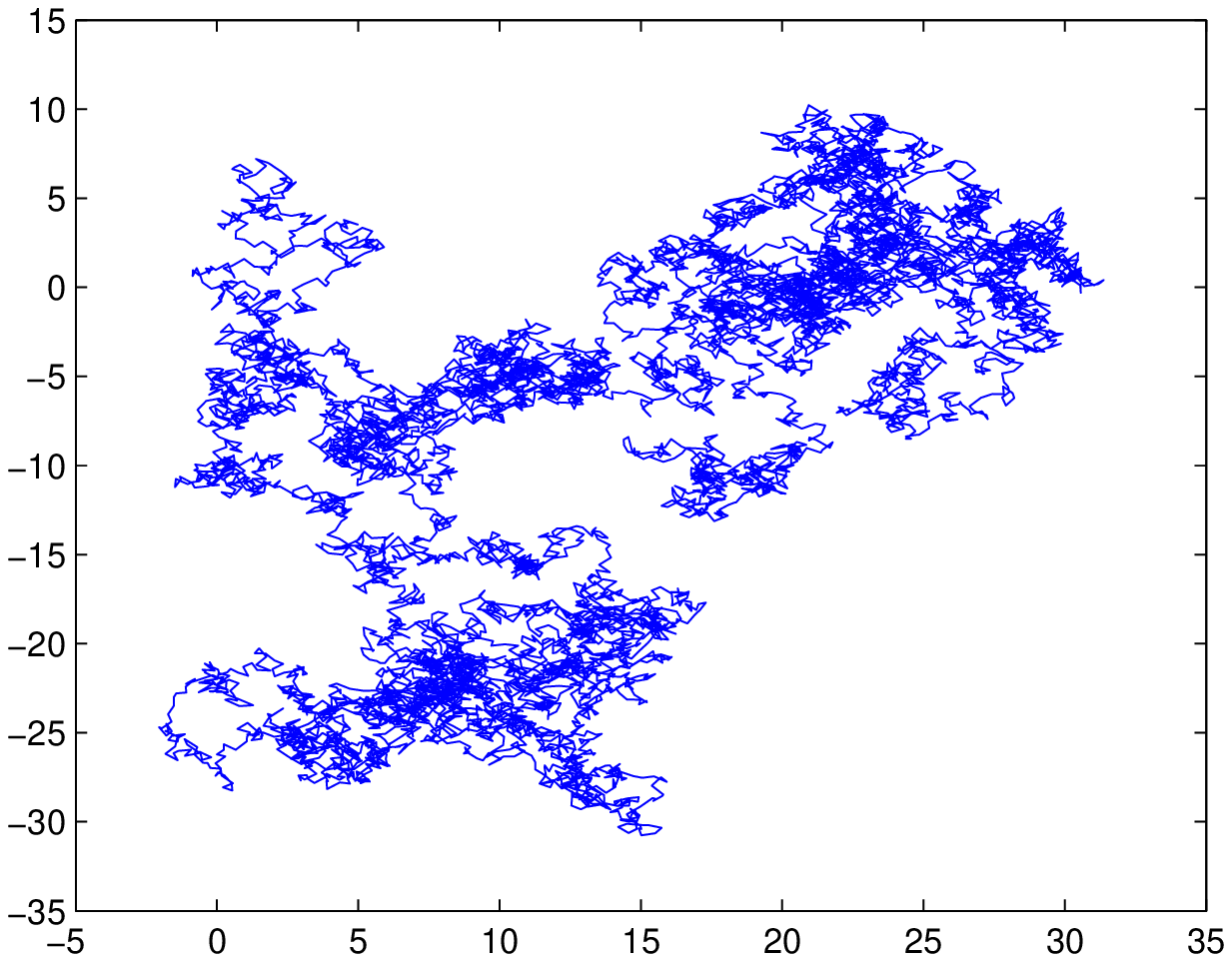}
\includegraphics[width=4.5cm, height=4.5cm]{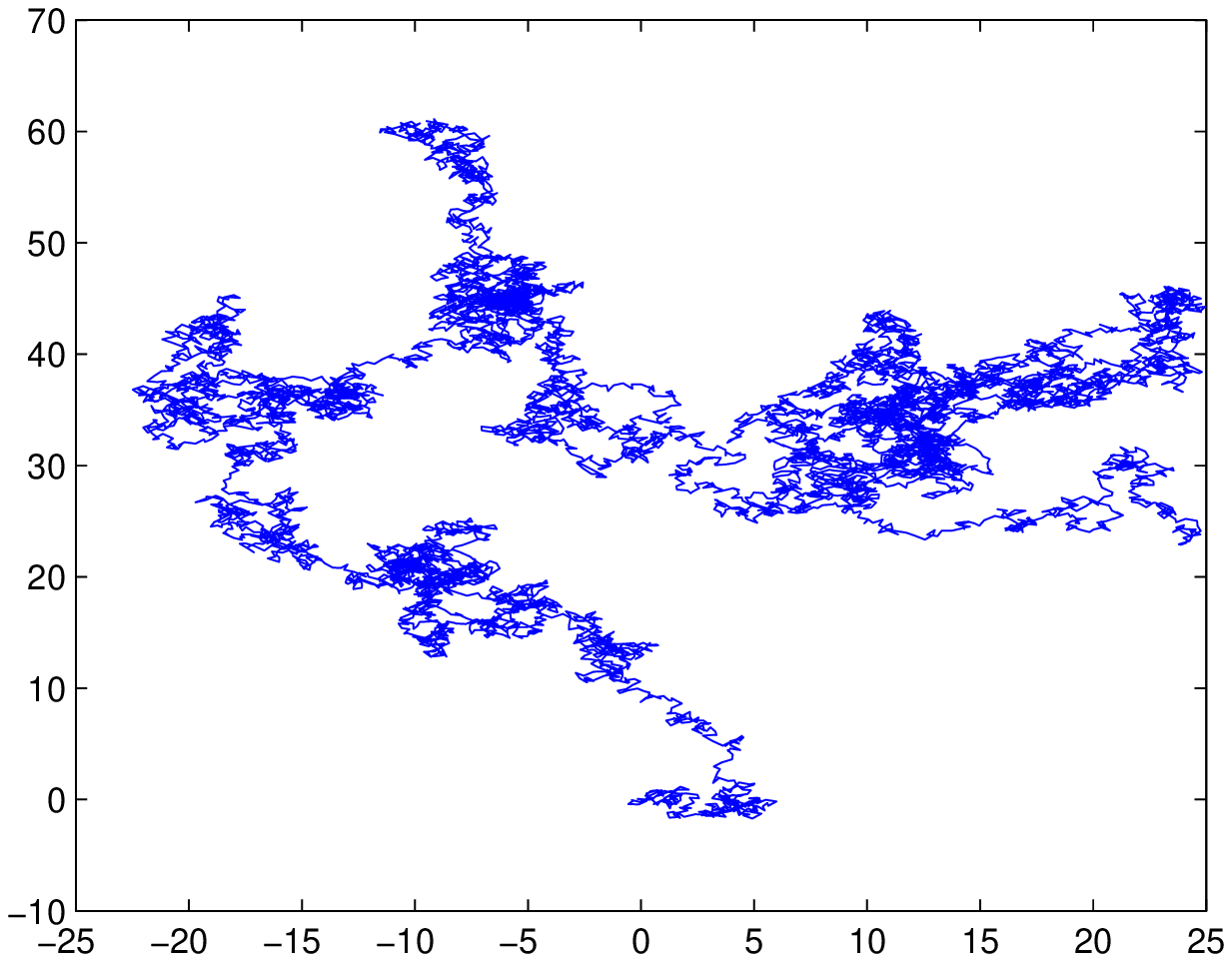}
\caption{\small Illustration of a Brownian motion. Upper figures:
one--dimensional Brownian motion -- three realizations of 
$x(t)$ as a function of $t$. Lower figures: two--dimensional Brownian motion
-- three realizations of $\vec{\br}(t)=[x(t),y(t)]$. All realizations start
at the origin.}
\label{brown}
\end{figure}

We may visualize each collision on the pollen grain as that of an impulse,
because the duration of each collision is extremely short. In other words, the
position of the particle $x(t)$ is responding to a sequence of (positive and
negative) impulses at
random times. Let
\begin{equation}
R_v(\tau)=\left<v(t)v(t+\tau)\right>=\left<\dot{x}(t)\dot{x}(t+\tau)\right>
\end{equation}
denote the autocorrelation of the random process $v(t)=\dot{x}(t)$
and let $S_v(\omega)=\calF\{R_v(\tau)\}$ be the power spectral
density.\footnote{To avoid confusion, it should be kept in mind that although $S_v(\omega)$ is
expressed as a function of the radial frequency $\omega$, which is measured in
radians per second, the physical units of the 
spectral density function itself here are
$\mbox{Volt}^2/\mbox{Hz}$ and not $\mbox{Volt}^2/[\mbox{radian per second}]$.
To pass to the latter, one should divide 
by $2\pi$. Thus, to calculate power, one must use
$\int_{-\infty}^{+\infty}S_v(2\pi f)\mbox{d}f$.}

Clearly, by the Langevin equation $\{v(t)\}$ is the response of a linear,
time--invariant linear system 
\begin{equation}
H(s)=\frac{1}{ms+\gamma}=\frac{1}{m(s+1/\theta)};
~~~~~~~h(t)=\frac{1}{m}e^{-t/\theta}u(t)
\end{equation}
to the random input process $\{F_r(t)\}$.
Assuming that the impulse process $\{F_r(t)\}$ is white noise, then
\begin{equation}
R_v(\tau)=\mbox{const}\cdot h(\tau)*h(-\tau)=\mbox{const}\cdot
e^{-|\tau|/\theta}=
R_v(0)e^{-|\tau|/\theta}=\frac{kT}{m}\cdot e^{-|\tau|/\theta}
\end{equation}
and
\begin{equation}
S_v(\omega)=\frac{2kT}{m}
\cdot\frac{\omega_0}{\omega^2+\omega_0^2},~~~~~\omega_0=\frac{1}{\theta}=\frac{\gamma}{m}
\end{equation}
that is, a Lorentzian spectrum.
We see that the relaxation time $\theta$ is indeed a measure of the
``memory'' of the particle and $\omega_0=1/\theta$ plays the role of 3dB
cutoff frequency of the spectrum of $\{v(t)\}$. What is the spectral density
of the driving input white noise process?
\begin{equation}
S_{F_r}(\omega)=\frac{S_v(\omega)}{|H(i\omega)|^2}=\frac{2kT\omega_0}{m(\omega^2+\omega_0^2)}\cdot
m^2(\omega^2+\omega_0^2)=2kTm\omega_0=2kT\gamma.
\end{equation}
This is a very important result. The spectral density of the white noise is
$2kT$ times the dissipative coefficient of the system, $\gamma$.
In other words, the dissipative element 
of the system is `responsible' for the noise.
At first glance, it may seem surprising: Why should the intensity of the
(external) driving force $F_r(t)$ be related to the dissipative coefficient
$\gamma$? The answer is that they are related via energy balance
considerations, since
we are assuming thermal equilibrium. Because the energy waste (dissipation) is proportional
to $\gamma$, the energy supply from $F_r(t)$ must also be proportional
to $\gamma$ in order to balance it.

\vspace{0.2cm}

\noindent
{\small {\it Example 8.2 -- Energy balance for the Brownian particle.}
The friction force $F_{friction}(t)=-\gamma v(t)$ causes the particle to loose kinetic energy
at the rate of
$$P_{loss}=\left<F_{friction}(t)v(t)\right>=-\gamma\left<v^2(t)\right>=-\gamma\cdot\frac{kT}{m}=
-\frac{kT}{\theta}.$$
On the other hand,
the driving force $F_r(t)$ injects kinetic energy at the rate of
\begin{eqnarray}
P_{injected}&=&\left<F_r(t)v(t)\right>\\
&=&\left<F_r(t)\int_{0}^{\infty}
\mbox{d}\tau h(\tau)F_r(t-\tau)\right>\\
&=&2kT\gamma\int_{0}^{\infty}
\mbox{d}\tau h(\tau)\delta(\tau)\\
&=&kT\gamma h(0)=\frac{kT\gamma}{m}=\frac{kT}{\theta},
\end{eqnarray}
which exactly balances the loss. Here, we used that fact that $\int_0^\infty
\mbox{d}\tau h(\tau)\delta(\tau)=h(0)/2$ since only ``half'' of the delta
function is ``alive'' where $h(\tau)>0$.\footnote{More rigorously, think of the
delta function here as the limit of narrow (symmetric) autocorrelation functions
which all integrate to unity.}
\underline{Exercise 8.2}: What happens if $\gamma=0$, yet
$F_r(t)$ has spectral density $N_0$? Calculate the rate of kinetic energy increase
in two different ways: (i) Show that $\frac{m}{2}\left<v^2(t)\right>$ is
linear in $t$ and find the constant of proportionality. (ii) calculate
$\left<F_r(t)v(t)\right>$ for this case.}

These principles apply not only to a Brownian particle in a liquid, but to any linear
system that obeys a first order stochastic differential equation with a white noise
input, provided that the energy equipartition theorem applies.
An obvious electrical analogue of this is a simple electric circuit
where a resistor $R$ and a capacitor $C$ are connected to one another (see
Fig.\ \ref{rc}).
\begin{figure}[ht]
\hspace*{5cm}\input{rc.pstex_t}
\caption{\small An $R$--$C$ circuit.}
\label{rc}
\end{figure}
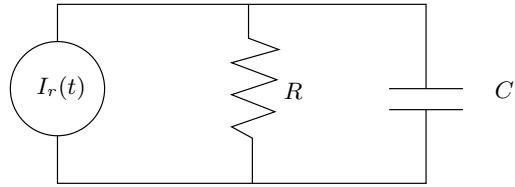
The thermal noise generated by the resistor (due to the thermal 
random motion of the colliding free
electrons in the conductor with extremely short mean time
between collisions), a.k.a.\ the {\it Johnson--Nyquist noise},
is modeled as a current source connected in
parallel to the resistor (or as an equivalent voltage source connected in series to the
resistor), which generates a white noise current process $I_r(t)$.
The differential equation pertaining to Kirchoff's current law is
\begin{equation}
C\dot{V}(t)+\frac{V(t)}{R}=I_r(t)
\end{equation}
where $V(t)$ is the voltage across the resistor as well as the parallel
capacitor. Now, this is exactly the same differential equation as before,
where  $I_r(t)$ plays the role of the driving force, $V(t)$ is replacing
$\dot{x}(t)$, 
$C$ substitutes $m$, and $1/R$ is the dissipative coefficient instead of
$\gamma$. Thus, the spectral density of the current is
\begin{equation}
S_{I_r}(\omega)=\frac{2kT}{R}.
\end{equation}
Alternatively, if one adopts the equivalent serial voltage source model then
$V_r(t)=RI_r(t)$ and so
\begin{equation}
S_{V_r}(\omega)=\frac{2kT}{R}\cdot R^2=2kTR.
\end{equation}
These results are well--knon from elementrary courses on random processes.

Finally, note that here we have something similar the
ultraviolet catastrophe: White noise has infinite power, which is unphysical. Once again,
this happens because we are working in the classical regime and we have not
addressed quantum effects pertaining to very high frequencies ($\hbar\omega \gg
kT$), which similarly as in black--body radiation, cause a sharp (exponential) decay in the spectrum
beyond a cutoff frequency of the order of magnitude of $kT/\hbar$
(which is a huge frequency at room temperature).
We will get back to this later on in Subsection 8.5.

\subsection{Diffusion and the Fokker--Planck Equation}

In this subsection, we consider the temporal 
evolution of the probability density
function of $x(t)$ (and not only its second order statistics,
as in the previous subsection), under quite general conditions.

The first successful treatment of Brownian motion was due to Einstein, who as
mentioned earlier, reduced the problem to one of diffusion. It might be
speculated that his motivation was that the diffusion equation is the simplest
differential equation in mathematical physics, which is asymmetric in
time:\footnote{This is different, for example, from the wave equation that
contains a {\it second} partial derivative w.r.t.\ time, and hence insensitive
to the direction of flow of time.}
Since it includes a first order derivative w.r.t.\ $t$, replacing 
$t$ by $-t$ changes the equation.
Because the Brownian motion is irreversible, it was important to include this
feature in the analysis. 
Einstein's argument can be summarized as follows:
Assume that each particle moves 
independently of all other particles. The relaxation time is small compared
to the observation time, but is long enough for the motions of a particle in
two consecutive intervals of $\theta$ to be independent.

Let the number of suspended grains be $N$ and let the $x$ coordinate change by
$\Delta$ in one relaxation time. $\Delta$ is a random variable,
symmetrically distributed around $\Delta=0$.
The number of particles $\mbox{d}N$ which are displaced by more than $\Delta$  
but less than $\Delta+\mbox{d}\Delta$ is 
\begin{equation}
\mbox{d}N=Np(\Delta)\mbox{d}\Delta
\end{equation}
where $p(\Delta)$ is the pdf of $\Delta$. Since only small displacements are
likely to occur, $p(\Delta)$ is sharply peaked at the origin.
Let $\rho(x,t)$ denote the density of particles at position $x$, at time $t$.
The number of particles in the interval $[x,
x+\mbox{d}x]$ at time $t+\delta$ ($\delta$ small) is
\begin{equation}
\label{kolmogorov}
\rho(x,t+\delta)\mbox{d}x=\mbox{d}x\int_{-\infty}^{+\infty}\rho(x-\Delta,t)p(\Delta)\mbox{d}\Delta
\end{equation}
This equation simply tells that the probability of finding the particle around
$x$ at time $t+\delta$ is made of contributions of finding it in $x-\Delta$ at
time $t$, and then moving by $\Delta$ within duration $\delta$ to arrive at $x$
at time $t+\delta$. Here we assumed independence between the location
$x-\Delta$ at time $t$ and the probability distribution of the displacement
$\Delta$, as $p(\Delta)$ is independent of $x-\Delta$.
Since $\theta$ is small, we take the liberty to use the Taylor series
expansion
\begin{equation}
\rho(x,t+\delta)\approx
\rho(x,t)+\delta\cdot\frac{\partial\rho(x,t)}{\partial t}
\end{equation}
Also, for small $\Delta$, we approximate $\rho(x-\Delta,t)$, this time to the
second order:
\begin{equation}
\rho(x-\Delta,t)\approx
\rho(x,t)-\Delta\cdot\frac{\partial\rho(x,t)}{\partial x}+
\frac{\Delta^2}{2}\cdot\frac{\partial^2\rho(x,t)}{\partial x^2}
\end{equation}
Putting these in eq.\ (\ref{kolmogorov}), we get
\begin{eqnarray}
\rho(x,t)+\delta\cdot\frac{\partial\rho(x,t)}{\partial
t}&=&\rho(x,t)\int_{-\infty}^{+\infty}p(\Delta)\mbox{d}\Delta-
\frac{\partial\rho(x,t)}{\partial
x}\int_{-\infty}^{+\infty}\Delta p(\Delta)\mbox{d}\Delta+\nonumber\\
& &\frac{1}{2}\cdot\frac{\partial^2\rho(x,t)}{\partial x^2}
\int_{-\infty}^{+\infty}\Delta^2 p(\Delta)\mbox{d}\Delta
\end{eqnarray}
or
\begin{equation}
\frac{\partial\rho(x,t)}{\partial t}
=\frac{1}{2\delta}\cdot\frac{\partial^2\rho(x,t)}{\partial x^2}
\int_{-\infty}^{+\infty}\Delta^2 p(\Delta)\mbox{d}\Delta
\end{equation}
which is the diffusion equation
\begin{equation}
\frac{\partial\rho(x,t)}{\partial t}=D\cdot\frac{\partial^2\rho(x,t)}{\partial x^2}
\end{equation}
with the diffusion coefficient being
\begin{equation}
D=\lim_{\delta\to 0}\frac{\left<\Delta^2\right>}{2\delta}=\lim_{\delta\to
0}\frac{\left<[x(t+\delta)-x(t)]^2\right>}{2\delta}.
\end{equation}
To solve the diffusion equation, define $\varrho(\kappa,t)$ as
the Fourier transform of $\rho(x,t)$ w.r.t.\ the variable $x$, i.e.,
\begin{equation}
\varrho(\kappa,t)=\int_{-\infty}^{+\infty}\mbox{d}x\cdot e^{-i\kappa x}
\rho(x,t).
\end{equation}
Then, the diffusion equation becomes an ordinary differential equation
w.r.t.\ $t$:
\begin{equation}
\frac{\varrho(\kappa,t)}{\partial t}=D(i\kappa)^2\varrho(\kappa,t)\equiv
-D\kappa^2\varrho(\kappa,t)
\end{equation}
whose solution is easily found to be
\begin{equation}
\varrho(\kappa,t)=C(\kappa)e^{-D\kappa^2t}.
\end{equation}
Assuming that the particle is initially located at the origin,
i.e., $\rho(x,0)=\delta(x)$, this means that the initial condition is
$C(\kappa)=\varrho(\kappa,0)=1$ for all $\kappa$, and so
\begin{equation}
\varrho(\kappa,t)=e^{-D\kappa^2t}.
\end{equation}
The density $\rho(x,t)$ is now obtained by the inverse Fourier transform,
which is
\begin{equation}
\rho(x,t)=\frac{e^{-x^2/(4Dt)}}{\sqrt{4\pi Dt}},
\end{equation}
and so $x(t)$ is zero--mean Gaussian with variance $\left<x^2(t)\right>=2Dt$.
Of course, any other initial location $x_0$ would yield a Gaussian with
the same variance $2Dt$, but the mean would be $x_0$.
Comparing the variance $2Dt$ with eq.\ (\ref{variance}), we have
\begin{equation}
D=\frac{kT}{\gamma}
\end{equation}
which is known as the {\it Einstein
relation},\footnote{Note that here $D$ is proportional to $T$, whereas in the
model of the
last section of Chapter 2 (which was an
oversimplified caricature of a diffusion model) it was
proportional to $\sqrt{T}$.}
widely used in semiconductor physics.

The analysis thus far assumed that $\left<\Delta\right>=0$, namely, there is no
drift to either the left or the right
direction. We next drop this assumption. In this case, the diffusion equation
generalizes to
\begin{equation}
\label{fokkerplanck}
\frac{\partial\rho(x,t)}{\partial
t}=-v\cdot\frac{\partial\rho(x,t)}{\partial x}+
D\cdot\frac{\partial^2\rho(x,t)}{\partial x^2}
\end{equation}
where
\begin{equation}
v=\lim_{\delta\to 0}\frac{\left<\Delta\right>}{\delta}
=\lim_{\delta\to
0}\frac{\left<x(t+\delta)-x(t)\right>}{\delta}=\frac{\mbox{d}}{\mbox{d}t}\left<\dot{x}(t)\right>
\end{equation}
has the obvious meaning of the average velocity.
Eq.\ (\ref{fokkerplanck}) is well known as the {\it Fokker--Planck equation}.
The diffusion equation and 
the Fokker--Planck equation are very central
in physics. As mentioned already in Chapter 2, 
they are fundamental in semiconductor physics, as
they describe processes of propagation of concentrations of electrons and
holes in semiconductor materials.

\vspace{0.2cm}

\noindent
{\small {\it Exercise 8.3}. Solve the Fokker--Planck
equation and show that the solution is $\rho(x,t)=\calN(vt,2Dt)$.
Explain the intuition.}

It is possible to further extend the Fokker--Planck equation so as to allow
the pdf of $\Delta$ to be location--dependent, that is, $p(\Delta)$ would be
replaced by $p_x(\Delta)$, but the important point to retain is that given the present
location $x(t)=x$, $\Delta$ would be independent of the earlier history of
$\{x(t'), t'< t\}$, which is to say that $\{x(t)\}$ should be a {\it Markov
process}. Consider then a general continuous--time Markov process defined by the
transition probability density function $W_\delta(x'|x)$, which denotes the pdf of
$x(t+\delta)$ at $x'$ given that $x(t)=x$.
A straightforward extension of the earlier derivation would lead to
the following more general form\footnote{Exercise 8.4: prove it.}
\begin{equation}
\frac{\partial\rho(x,t)}{\partial
t}=-\frac{\partial}{\partial x}[v(x)\rho(x,t)]+\frac{\partial^2}{\partial
x^2}[D(x)\rho(x,t)],
\end{equation}
where
\begin{equation}
v(x)=\lim_{\delta\to
0}\frac{1}{\delta}\int_{-\infty}^{+\infty}(x'-x)W_\delta(x'|x)\mbox{d}x'
=\bE[\dot{x}(t)|x(t)=x].
\end{equation}
is the local average velocity and
\begin{equation}
D(x)=\lim_{\delta\to
0}\frac{1}{2\delta}\int_{-\infty}^{+\infty}(x'-x)^2W_\delta(x'|x)\mbox{d}x'
=\lim_{\delta\to 0}\frac{1}{2\delta}\bE\{[x(t+\delta)-x(t)]^2|x(t)=x\}
\end{equation}
is the local diffusion coefficient. 

\vspace{0.2cm}

\noindent
{\small {\it Example 8.3.} Consider the stochastic differential equation
$$\dot{x}(t)=-ax(t)+n(t),$$
where $n(t)$ is Gaussian white noise with spectral density $N_0/2$. From the 
solution of this differential equation, it is easy to see that 
$$x(t+\delta)=x(t)e^{-a\delta}+e^{-a(t+\delta)}\int_t^{t+\delta}\mbox{d}\tau
n(\tau)e^{a\tau}.$$
This relation, between $x(t)$ and $x(t+\delta)$, 
can be used to derive the first and the second moments of $[x(t+\delta)-x(t)]$ for small
$\delta$, and to find that $v(x)=-ax$ and $D(x)=N_0/4$ (Exercise 8.5: Show this). Thus,
the Fokker--Planck equation, in this example, reads
$$\frac{\partial\rho(x,t)}{\partial t}=a\cdot\frac{\partial}{\partial x}
[x\cdot\rho(x,t)]+\frac{N_0}{4}\cdot\frac{\partial^2\rho(x,t)}{\partial
x^2}.$$
It is easy to check that the r.h.s.\ vanishes for $\rho(x,t)\propto
e^{-2ax^2/N_0}$ (independent of $t$),
which means that in equilibrium, $x(t)$ is Gaussian with
zero mean and variance $N_0/4a$. This is in agreement with the fact that,
as $x(t)$ is the response of the linear system $H(s)=1/(s+a)$
(or in the time domain, $h(t)=e^{-at}u(t)$) to $n(t)$, its
variance is indeed (as we know from ``Random Signals''):
$$\frac{N_0}{2}\int_0^\infty h^2(t)\mbox{d}t=
\frac{N_0}{2}\int_0^\infty e^{-2at}\mbox{d}t=\frac{N_0}{4a}.$$
Note that if $x(t)$ is the voltage across the capacitor in a simple
$R$--$C$ network, then $a=1/RC$, and since
$\calE(x)=Cx^2/2$, then in equilibrium we have the Boltzmann weight
$\rho(x)\propto \exp(-Cx^2/2kT)$, which is again, a zero--mean Gaussian.
Comparing the exponents, we immediately obtain $N_0/2=2kT/RC^2$.
{\it Exercise 8.6:} Find the solution $\rho(x,t)$ for all $x$ and $t$ subject
to the initial condition $\rho(x,0)=\delta(x)$.} $\Box$

A slightly different
representation of the Fokker--Planck equation is the following:
\begin{equation}
\frac{\partial\rho(x,t)}{\partial t}=-\frac{\partial}{\partial x}\left\{
v(x)\rho(x,t)-\frac{\partial}{\partial x}[D(x)\rho(x,t)]\right\},
\end{equation}
Now, $v(x)\rho(x,t)$ has the obvious interpretation of the {\it drift current
density} $J_{\mbox{\tiny drift}}(x,t)$ of a `mass'
whose density is $\rho(x,t)$ (in this case, it is a probability mass), 
whereas 
$$J_{\mbox{\tiny diffusion}}(x,t)=
-\frac{\partial}{\partial x}[D(x)\rho(x,t)]$$ 
is the {\it diffusion
current density}.\footnote{This generalizes Fick's law that we have seen in 
Chapter 2. There, $D$ was fixed (independent of $x$), and so the diffusion
current was proportional to the negative gradient of the density.}
Thus, 
\begin{equation}
\label{cont}
\frac{\partial\rho(x,t)}{\partial t}=
-\frac{\partial J_{\mbox{\tiny total}}(x,t)}{\partial x}
\end{equation}
where 
$$J_{\mbox{\tiny total}}(x,t)=J_{\mbox{\tiny
drift}}(x,t)+J_{\mbox{\tiny diffusion}}(x,t).$$ 
Eq.\ (\ref{cont}) is the {\it equation of continuity}, which we saw in Chapter 2. 
In steady state, when
$\rho(x,t)$ is time--invariant, the total current may vanish. The
drift current and the diffusion current balance one another, or at least
the net current is homogenous (independent of $x$), so no mass accumulates anywhere.

\vspace{0.2cm}

\noindent
{\it Comment.}
It is interesting to relate
the diffusion constant $D$ to the mobility of the electrons in the
context of electric conductivity. The mobility
$\mu$ is defined according to
$v=\mu E$,
where $E$ is the electric field. According to Fick's law, the
current density
is proportional to the negative gradient of the concentration, and $D$ is
defined as the constant of proportionality, i.e.,
$J_{\mbox{\tiny electrons}}=-D\partial\rho/\partial x$ or $J=-Dq_e\partial\rho/\partial x$.
If one sets up a field $E$ in an open circuit, the diffusion current cancels
the drift current, that is
\begin{equation}
J=\rho q_e\mu E-Dq_e\frac{\partial\rho}{\partial x} =0
\end{equation}
which would give
$\rho(x)\propto e^{\mu Ex/D}$. On the other hand, under thermal equilibrium,
with potential energy $V(x)=-q_e Ex$, we also have $\rho(x)\propto
e^{-V/kT}=e^{q_e Ex/kT}$. Upon comparing the exponents, we readily obtain
the Einstein relation\footnote{
This is a relation that is studied in courses on physical principles of
seminconductor devices.}
$D=kT\mu/q_e$. Note that $\mu/q_e=v/q_eE$ is related to the admittance
(the dissipative coefficient)
since $v$ is proportional to the current and $E$ is proportional to
the voltage.

\subsection{The Fluctuation--Dissipation Theorem}

We next take another point of view on stochastic dynamics of a physical system:
Suppose that a system (not necessarily
a single particle as in the previous subsections)
is initially in equilibrium of the canonical ensemble, but at a certain time
instant, it is subjected to an abrupt, yet small change in a certain parameter that controls it
(say, a certain force, like pressure, magnetic field, etc.). 
Right after this abrupt change in the parameter, the system is, of course, no longer in equilibrium,
but it is not far, since the change is assumed small. How fast does the system
re--equilibrate and what is its dynamical behavior in the course of the
passage to the new equilibrium? Also, since the change was abrupt and not
quasi--static, how is energy dissipated? Quite remarkably, it turns out that
the answers to both questions are
related to the {\it equilibrium fluctuations} of the system. Accordingly, the
principle that quantifies and characterizes this relationship is called the
{\it fluctuation--dissipation theorem} and this is the subject of this
subsection. We shall also relate it to the derivations of the previous
subsection.

Consider a physical system, which in the absence of any applied external field, has an 
Hamiltonian $\calE(x)$, where here $x$ denotes the microstate, and so, its
equilibrium distribution is the Boltzmann distribution with a partition function given by:
\begin{equation}
Z(\beta)=\sum_x e^{-\beta\calE(x)}.
\end{equation}
Now, let $w_x$ be an observable (a measurable physical quantity that depends
on the microstate), which has
a conjugate force $F$, so that when $F$ is applied,
the change in the Hamiltonian is $\Delta\calE(x)=-Fw_x$. Next suppose that
the external force
is time--varying according to 
a certain waveform $\{F(t),~-\infty < t < \infty\}$.
As in the previous subsection, it should be kept in mind that the
overall effective force can be 
thought of as a superposition of two contributions, a deterministic
contribution, which is the above 
mentioned $F(t)$ -- the external field that the experimentalist 
applies on purpose and fully controls, and a random part $F_r(t)$, which
pertains to interaction with the environment (or the heat bath at temperature
$T$).\footnote{For example, 
think again of the example a Brownian particle colliding with
other particles. The other particles can be thought of as the environment in
this case.}
The random 
component $F_r(t)$ is
responsible for the randomness of the 
microstate $x$ and hence also the randomness of the observable. We shall
denote the random variable corresponding to the observable at time $t$ by
$W(t)$. Thus, $W(t)$ is random variable, which takes values in the set
$\{w_x,~x\in\calX\}$, where $\calX$ is the space of microstates.
When the external deterministic field is kept fixed
($F(t)\equiv \mbox{const.}$), the system is expected to converge to equilibrium
and eventually obey the Boltzmann law. 
While in the section on Brownian motion, we
focused only on the contribution of the random part, $F_r(t)$, now
let us refer only to the deterministic part, $F(t)$. 
We will get back to
the random part later on.

Let us assume first that $F(t)$ was switched on 
to a small level $\epsilon$ at time $-\infty$, and then
switched off at time $t=0$, in other words, 
$F(t)=\epsilon U(-t)$, where $U(\cdot)$ is the unit step function
(a.k.a.\ the Heaviside function). 
We are interested in the behavior of the mean of the
observable $W(t)$ at time $t$, which we 
shall denote by $\left<W(t)\right>$, for $t > 0$. Also,
$\left<W(\infty)\right>$ will denote the 
limit of $\left<W(t)\right>$ as $t\to\infty$, namely, the
equilibrium mean of the observable in the absence of an external field.
Define now the (negative) step response function as
\begin{equation}
\zeta(t)=\lim_{\epsilon\to 0}\frac{\left<W(t)\right>-\left<W(\infty)\right>}{\epsilon}
\end{equation}
and the 
auto--covariance function pertaining to the final equilibrium as
\begin{equation}
R_{W}(\tau) \dfn \lim_{t\to\infty}\left<W(t)W(t+\tau)\right>-\left<W(\infty)\right>^2,
\end{equation}
Then, the {\it fluctuation--dissipation theorem} (FDT) asserts that
\begin{equation}
R_{W}(\tau)=kT\cdot \zeta(\tau)
\end{equation}
The FDT then relates between the linear transient response of the system
to a small excitation (after it has been removed) and
the autocovariance of the observable in equilibrium. 
The transient response, that fades away is the
dissipation, whereas the autocovariance is the fluctuation.
Normally, $R_{W}(\tau)$ decays for large $\tau$ and so $\left<W(t)\right>$
converges to $\left<W(\infty)\right>$
at the same rate (see Fig.\ \ref{linearresponse}).
\begin{figure}[ht]
\hspace*{4cm}\input{linearresponse.pstex_t}
\caption{\small Illustration of the response 
of $\left<W(t)\right>$ to a step function at the
input force $F(t)=\epsilon U(-t)$. According 
to the FDT, the response (on top of
the asymptotic level $\left<W(\infty)\right>$) is
proportional to the equilibrium autocorrelation function $R_W(t)$, which
in turn may
decay either monotonically (solid curve) or 
in an oscillatory manner (dashed curve).}
\label{linearresponse}
\end{figure}
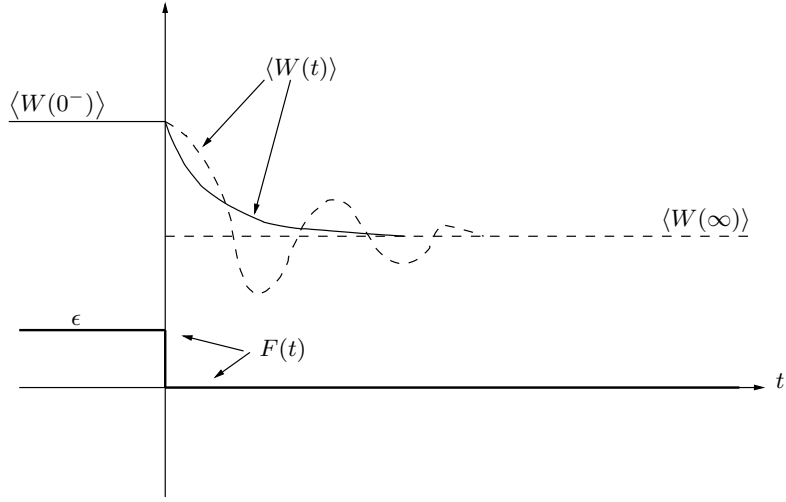

To prove this result, we proceed as follows: First, we have by definition:
\begin{equation}
\left<W(\infty)\right> = \frac{\sum_x w_x e^{-\beta\calE(x)}}{\sum_{x} e^{-\beta\calE(x)}}
\end{equation}
Now, for $t < 0$, we have
\begin{equation}
P(x)=\frac{e^{-\beta\calE(x)-\beta\Delta\calE(x)}}{\sum_{x}
e^{-\beta\calE(x)-\beta\Delta\calE(x)}}.
\end{equation}
Thus, for all negative times, and for $t =0^-$ in particular, we have
\begin{equation}
\left<W(0^-)\right> = \frac{\sum_xw_xe^{-\beta\calE(x)-\beta\Delta\calE(x)}}
{\sum_{x} e^{-\beta\calE(x)-\beta\Delta\calE(x)}}.
\end{equation}
Let $P_t(x'|x)$ denote the probability that the system would be at state $x'$ at time $t$ ($t > 0$)
given that it was in state $x$ at time $t=0^-$. 
This probability depends on the dynamical properties of
the system (in the absence of the perturbing force). 
Let us define $\left<W(t)\right>_x=\sum_{x'} w_{x'}P_t(x'|x)$,
which is the expectation
of $W(t)$ ($t > 0$)
given that the system was in state $x$ at $t=0^-$.
Now, 
\begin{eqnarray}
\left<W(t)\right>
&=&\frac{\sum_x\left<W(t)\right>_xe^{-\beta\calE(x)-\beta\Delta\calE(x)}}
{\sum_{x} e^{-\beta\calE(x)-\beta\Delta\calE(x)}}\nonumber\\
&=&\frac{\sum_x\left<W(t)\right>_xe^{-\beta\calE(x)+\beta\epsilon w_x}}
{\sum_{x} e^{-\beta\calE(x)+\beta\epsilon w_{x}}}
\end{eqnarray}
and $\left<W(\infty)\right>$ can be seen as a special case of this quantity
for $\epsilon=0$ (no perturbation at all). Thus,
$\zeta(t)$ is, by definition, nothing but the derivative of
$\left<W(t)\right>$ w.r.t.\ $\epsilon$, computed at $\epsilon=0$.
I.e.,
\begin{eqnarray}
\zeta(\tau)&=&\frac{\partial}{\partial\epsilon}\left[
\frac{\sum_x\left<W(\tau)\right>_xe^{-\beta\calE(x)+\beta\epsilon w_x}}
{\sum_{x} e^{-\beta\calE(x)+\beta\epsilon
w_{x}}}\right]_{\epsilon=0}\nonumber\\
&=&\beta\cdot\frac{\sum_x\left<W(\tau)\right>_xw_xe^{-\beta\calE(x)}
}{\sum_{x} e^{-\beta\calE(x)}}
-\beta\cdot\frac{\sum_x\left<W(\tau)\right>_xe^{-\beta\calE(x)}}
{\sum_{x} e^{-\beta\calE(x)}}\cdot
\frac{\sum_xw_xe^{-\beta\calE(x)}}
{\sum_{x} e^{-\beta\calE(x)}}\nonumber\\
&=&\beta\left[\lim_{t\to\infty}\left<W(t)W(t+\tau)\right>-\left<W(\infty)\right>^2\right]\nonumber\\
&=&\beta R_W(\tau),
\end{eqnarray}
where we have used 
the fact that the dynamics of $\{P_t(x'|x)\}$ preserve the equilibrium
distribution. 

\vspace{0.2cm}

\noindent
{\small {\it Exercise 8.7.} Extend the FDT to account for a situation
where the force $F(t)$ is not conjugate to $W(t)$, but to another
physical quantity $V(t)$.}

\vspace{0.2cm}

While $\zeta(t)$ is essentially the 
response of the system to a (negative) step function
in $F(t)$, then obviously, 
\begin{equation}
h(t)=\left\{\begin{array}{cc}
0 & t< 0\\
-\dot{\zeta}(t) & t\ge 0\end{array}\right.
=\left\{\begin{array}{cc}
0 & t < 0\\
-\beta\dot{R}_W(t) & t\ge 0\end{array}\right.
\end{equation}
would be the impulse response.
Thus, we can now express the response of $\left<W(t)\right>$ to a general
signal $F(t)$ that vanishes for $t>0$ to be
\begin{eqnarray}
\left<W(t)\right>-\left<W(\infty)\right> &\approx&
-\beta\int_{-\infty}^0\dot{R}_{W}(t-\tau)F(\tau)\mbox{d}\tau\nonumber\\
&=&-\beta\int_{-\infty}^0R_{W}(t-\tau)\dot{F}(\tau)\mbox{d}\tau\nonumber\\
&=&-\beta R_{W}\otimes\dot{F},
\end{eqnarray}
where the second passage is from integration by parts and where
$\otimes$ denotes convolution. Indeed, in our first example, 
$\dot{F}(t)=-\epsilon\delta(t)$
and we are back to the result $\left<W(t)\right>-\left<W(\infty)\right>=\beta\epsilon R_{W}(t)$.

It is instructive to look at these relations 
also in the frequency domain. Applying the
one sided Fourier transform on both 
sides of the relation $h(t)=-\beta\dot{R}_{W}(t)$
and taking the complex 
conjugate (i.e., multiplying by $e^{i\omega t}$ and integrating
over $t > 0$), we get
\begin{equation}
H(-i\omega)\equiv \int_0^\infty h(t)e^{i\omega t}\mbox{d}t=
-\beta\int_0^\infty\dot{R}_{W}(t)e^{i\omega t}\mbox{d}t
=\beta i\omega \int_0^\infty R_{W}(t)e^{i\omega t}\mbox{d}t+\beta R_{W}(0),
\end{equation}
where the last step is due to integration by parts.
Upon taking the imaginary parts of both sides, we get:
\begin{equation}
\mbox{Im}\{H(-i\omega)\}=\beta\omega\int_0^\infty R_{W}(t)\cos(\omega
t)\mbox{d}t=\frac{1}{2}\beta\omega
S_{W}(\omega),
\end{equation}
where $S_{W}(\omega)$ is the power 
spectrum of $\{W(t)\}$ in equilibrium, that is,
the Fourier transform of $R_{W}(\tau)$. Equivalently, we have:
\begin{equation}
\label{SW}
S_{W}(\omega)=2kT\cdot\frac{\mbox{Im}\{H(-i\omega)\}}{\omega}
=-2kT\cdot\frac{\mbox{Im}\{H(i\omega)\}}{\omega}
\end{equation}

\vspace{0.2cm}

\noindent
{\small {\it Example 8.4 An electric circuit.}
Consider the circuit in Fig.\ \ref{rc2}.
The driving force is the voltage source $V(t)$ and the conjugate variable is
$Q(t)$ the of the capacitor. The resistors are considered part of thermal
environment. The voltage waveform is $V(t)=\epsilon U(-t)$. 
At time $t=0^-$, the voltage across the capacitor is
$\epsilon/2$ and the energy is $\frac{1}{2}C(V_r+\frac{\epsilon}{2})^2$, whereas
for $t\to \infty$, it is $\frac{1}{2}CV_r^2$, so the difference is
$\Delta\calE=\frac{1}{2}CV_r\epsilon=\frac{1}{2}Q_r\epsilon$, neglecting the $O(\epsilon^2)$
term. According to the FDT then, $\zeta(t)=\frac{1}{2}\beta R_Q(t)$, where
the factor of $1/2$ follows the one in $\Delta\calE$. This then gives
\begin{equation}
S_Q(\omega)=4kT\cdot\frac{\mbox{Im}\{H(-i\omega)\}}{\omega}.
\end{equation}
In this case,
\begin{equation}
H(i\omega)=\frac{(R\|1/[i\omega C])\cdot C}{R+(R\|1/[i\omega
C])}=\frac{C}{2+i\omega RC}
\end{equation}
for which
\begin{equation}
\mbox{Im}\{H(-i\omega)\}=\frac{\omega RC^2}{4+(\omega RC)^2}
\end{equation}
and finally,
\begin{equation}
S_Q(\omega)=\frac{4kTRC^2}{4+(\omega RC)^2}
\end{equation}
Thus, the thermal noise voltage across the capacitor is $4kTR/[4+(\omega
RC)^2]$. The same result can be obtained, of course, using the method studied
in ``Random Signals'':
\begin{equation}
2kT\cdot\mbox{Re}\left\{R\|R\|\frac{1}{i\omega
C}\right\}=\frac{4kTR}{4+(\omega RC)^2}.
\end{equation} 
This concludes Example 8.4. $\Box$}

\begin{figure}[ht]
\hspace*{5cm}\input{rc2.pstex_t}
\caption{\small Electric circuit for Example 8.4.}
\label{rc2}
\end{figure}
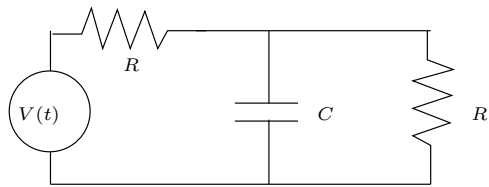

Earlier, we have seen that $\left<W(t)\right>-\left<W(\infty)\right>$ responds to a
determinstic (small) waveform $F(t)$ via a linear(ized) system with an impulse response
$h(t)$. By the same token, we can think of the random part around the mean,
$W(t)-\left<W(\infty)\right>$, as the response of the same system to a random
input $F_r(t)$ (thus, the total response is the superposition).
If we wish to 
think of our physical system in equilibrium as a
linear(ized) system with input $F_r(t)$ and output $W(t)$, then what would
should the spectrum of the input process $\{F_r(t)\}$ be in order to comply
with the last result? Denoting by $S_{F_r}(\omega)$
the spectrum of the input process, we know from ``Random Signals'' that
\begin{equation}
S_W(\omega)=S_{F_r}(\omega)\cdot|H(i\omega)|^2
\end{equation}
and so comparing with (\ref{SW}), we have
\begin{equation}
\label{noisespectrum}
S_{F_r}(\omega)=2kT\cdot\frac{\mbox{Im}\{H(-i\omega)\}}
{\omega\cdot|H(i\omega)|^2}.
\end{equation}
This extends our earlier result concerning the spectrum of the driving white
noise in case the Brownian particle, where we obtained a spectral density of
$2kT\gamma$. 

\vspace{0.2cm}

\noindent
{\small {\it Example 8.5 -- Second order linear system.}
For a second order linear system (e.g., a damped harmonic oscillator),
\begin{equation}
m\ddot{W}(t)+\gamma\dot{W}(t)+KW(t)=F_r(t)
\end{equation}
the force $F_r(t)$ is indeed conjugate to the variable $W(t)$, which
is the location, as required by the FDT.
Here, we have
\begin{equation}
H(i\omega)=\frac{1}{m(i\omega)^2+\gamma i\omega+K}=
\frac{1}{K-m\omega^2+\gamma i\omega}.
\end{equation}
In this case,
\begin{equation}
\mbox{Im}\{H(-i\omega)\}=\frac{\gamma\omega}{(K-m\omega^2)^2+\gamma^2\omega^2}=
\gamma\omega|H(i\omega)|^2
\end{equation}
and so, we readily obtain
\begin{equation}
S_{F_r}(\omega)=2kT\gamma,
\end{equation}
recovering the principle that the spectral density of the
noise process is $2kT$ times the dissipative coefficient $\gamma$ of the
system, which
is responsible to the irreversible component. The difference between this and
the earlier derivation is that earlier, we {\it assumed in advance} that the input noise
process is white and we only computed its spectral level, whereas now, we have
actually shown that at least for a second order linear system like this, it must be
white noise (as far as the classical approximation holds).}$\Box$

From eq.\ (\ref{noisespectrum}), we see that the thermal interaction with the environment, when
{\it referred to the input} of the system, 
has a spectral density of the form that
we can calculate.
In general, it does {\it not} necessarily have to be a flat spectrum.
Consider for example, an arbitrary electric network consisting of one voltage
source (in the role of $F(t)$) and a bunch of resistors and capacitors.
Suppose that our observable $W(t)$ is the voltage across one of the
capacitors. Then, there is a certain 
transfer function $H(i\omega)$ from the voltage source
to $W(t)$. The thermal noise process is 
contributed by all resistors by considering
equivalent noise sources (parallel current sources or serial voltage
sources) attached to each resistor. However, in order to refer the
contribution of these noise sources to the 
input $F(t)$, we must calculate equivalent noise
sources which are in series with the given 
voltage source $F(t)$. These equivalent
noise sources will no longer generate white noise processes, in general.
For example, in the circuit of Fig.\ \ref{rc2}, if an extra capacitor $C$
would be connected in series to one of the resistors,
then, the contribution of the right resistor referred to the
left one is not white noise.\footnote{Exercise 8.8: Calculate its spectrum.}

\subsection{Johnson--Nyquist Noise in the Quantum--Mechanical Regime}

As promised at the end of Subsection 8.3, we now return to the problematics
of the formula $S_{V_r}(\omega)=2kTR$ when it comes to very high frequencies,
namely, the electrical analogue to the ultraviolet catastrophe. At very
high frequencies, we are talking about very short waves, much shorter than the
physical sizes of the electric circuit.

The remedy to the unreasonable classical results in the high frequency range, 
is to view the motion of electrons in a resistor as 
an instance of black--body radiation, but instead of the three--dimensional
case that we studied earlier, this time, we are talking about the
one--dimensional case. The difference is mainly the calculation of the density of states.
Consider a long transmission line with characteristic impedance $R$
of length $L$, terminating at both ends by
resistances $R$ (see Fig.\ \ref{transline}), so that the impedances are
matched at both ends. Then any voltage wave propagating along the transmission
line is fully absorbed by the terminating resistor without reflection.
\begin{figure}[ht]
\hspace*{4cm}\input{transline.pstex_t}
\caption{\small Transmission line of length $L$, terminated by resistances $R$
at both ends.}
\label{transline}
\end{figure}
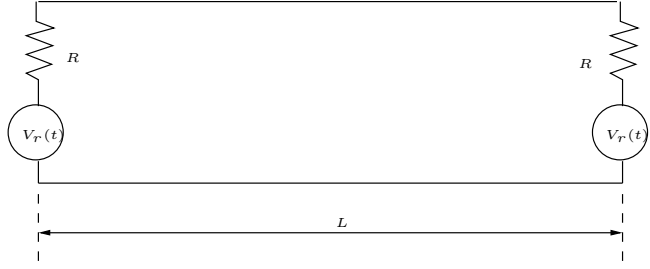
The system resides in thermal equilibrium at temperature $T$.
The resistor then can be thought of as a black--body radiator in one
dimension. A voltage wave of the form $V(x,t)=V_0\exp[i(\kappa x-\omega t)]$
propagates along the transmission line with velocity $v=\omega/\kappa$, which
depends on the capacitance and the inductance of the transmission line per
unit length.
To count the possible modes, let us impose the periodic boundary condition
$V(0,t)=V(L,t)$. Then $\kappa L=2\pi n$ for any positive integer $n$.
Thus, there are $\Delta n=\Delta\kappa/2\pi$ such modes per unit length
in the frequency range between $\omega=v\kappa$ and $\omega+\Delta\omega=
v(\kappa+\Delta\kappa)$. The mean energy of such a mode is given by
\begin{equation}
\epsilon(\omega)=\frac{\hbar\omega}{e^{\hbar\omega/kT}-1}.
\end{equation}
Since there are $\Delta n=\Delta\omega/(2\pi v)$ propagating modes per unit
length in this frequency range, the mean energy per unit time (i.e., the
power) incident
upon a resistor in this frequency range is
\begin{equation}
P=\frac{L}{L/v}\cdot\frac{\Delta\omega}{2\pi v}\cdot\epsilon(\omega)=
\frac{1}{2\pi}\cdot\frac{\hbar\omega\Delta\omega}{e^{\hbar\omega/kT}-1},
\end{equation}
where $L/v$ at the denominator is the travel time of the wave along the
transmission line.
This is the radiation power absorbed by the resistor, which must be equal to
the power emitted by the resistor in this frequency range. Let the thermal
voltage generated by the resistor in the frequency
range $[\omega,\omega+\Delta\omega]$ be denoted by $V_r(t)[\omega,\omega+\Delta\omega]$. 
This voltage sets up a current
of $V_r(t)[\omega,\omega+\Delta\omega]/2R$ and hence an average power of 
$\left<V_r^2(t)[\omega,\omega+\Delta\omega]\right>/4R$.
Thus, the balance between the absorbed and the emitted power gives
\begin{equation}
\frac{\left<V_r^2(t)[\omega,\omega+\Delta\omega]\right>}{4R}=
\frac{1}{2\pi}\cdot\frac{\hbar\omega\cdot\Delta\omega}
{e^{\hbar\omega/kT}-1},
\end{equation}
which is
\begin{equation}
\frac{\left<V_r^2(t)[\omega,\omega+\Delta\omega]\right>}{\Delta\omega}=
\frac{4R}{2\pi}\cdot\frac{\hbar\omega}{e^{\hbar\omega/kT}-1}
\end{equation}
or
\begin{equation}
\frac{\left<V_r^2(t)[f,f+\Delta f]\right>}{\Delta f}=4R\cdot\frac{\hbar\omega}{e^{\hbar\omega/kT}-1}
\end{equation}
Taking the limit $\Delta f\to 0$,
the left--hand side becomes the one--sided spectral density of the
thermal noise, and so the two--sided spectral density is
\begin{equation}
S_{V_r}(\omega)=2R\cdot\frac{\hbar\omega}{e^{\hbar\omega/kT}-1}.
\end{equation}
We see then that when quantum--mechanical considerations are incorporated,
the noise spectrum is no longer strictly flat. As long as $\hbar\omega \ll kT$, the
denominator is very well approximated by $\hbar\omega/kT$, and so we recover the formula
$2kTR$, but for frequencies of the order of magnitude of $\omega_c\dfn kT/\hbar$, 
the spectrum falls off exponentially
rapidly. So the quantum--mechanical correction is to replace
\begin{equation}
kT~\Longrightarrow~\frac{\hbar\omega}{e^{\hbar\omega/kT}-1}.
\end{equation}
At room temperature ($T=300^o K$), the cutoff frequency is
$f_c=\omega_c/2\pi\approx 6.2\mbox{THz}$,\footnote{Recall that $1\mbox{THz}=10^{12}\mbox{Hz}$} 
so the spectrum
can be safely considered $2kTR$ flat over any frequency range of practical
interest from the engineering perspective. 

What is the total RMS noise voltage generated by a resistor $R$ at temperature
$T$? The total mean square noise voltage is
\begin{eqnarray}
\left<V_r^2(t)\right>&=&4R\int_0^\infty
\frac{hf\mbox{d}f}{e^{hf/kT}-1}\nonumber\\
&=&\frac{4R(kT)^2}{h}\int_0^\infty \frac{x\mbox{d}x}{e^{x}-1}\nonumber\\
&=&\frac{4R(kT)^2}{h}\int_0^\infty \frac{xe^{-x}\mbox{d}x}{1-e^{-x}}\nonumber\\
&=&\frac{4R(kT)^2}{h}\sum_{n=1}^\infty\int_0^\infty xe^{-nx}\mbox{d}x\nonumber\\
&=&\frac{4R(kT)^2}{h}\sum_{n=1}^\infty\frac{1}{n^2}\nonumber\\
&=&\frac{2R(\pi kT)^2}{3h},
\end{eqnarray}
which is quadratic in $T$ since both the (low frequency) spectral
density and the effective bandwidth are linear in $T$.
The RMS is then
\begin{equation}
V_{RMS}=\sqrt{\left<V_r^2(t)\right>}=\sqrt{\frac{2R}{3h}}\cdot\pi kT,
\end{equation}
namely, proportional to temperature and to the square root of the resistance.
To get an idea of the order of magnitude,
a resistor of $100\Omega$ at temperature $T=300^{o}$K generates an RMS
thermal noise of about $4$mV. The equivalent noise bandwidth is
\begin{equation}
B_{eq}=\frac{2R(\pi kT)^2/3h}{2kTR}=\frac{\pi^2kT}{3h}=\frac{\pi^2}{3}
\cdot f_c.
\end{equation}

\vspace{0.2cm}

\noindent
{\small {\it Exercise 8.9:} Derive an expression for the autocorrelation function
of the Johnson--Nyquist noise in the quantum mechnical regime.}

\subsection{Other Noise Sources}

In addition to thermal noise, that we have discussed thus far, there
are other physical mechanisms that generate noise in Nature in general, and
in electronic circuits, in particular. We will not cover them in great depth here,
but only provide short descriptions.

\vspace{0.2cm}

\noindent
{\bf Flicker noise},
also known as {\bf $1/f$ noise}, is a random process with a
spectrum that falls off steadily into the higher frequencies, with a
pink spectrum. It occurs in almost all electronic devices, and results from a
variety of effects, though always related to a direct current. According to
the underlying theory, there are fluctuations
in the conductivity due 
to the superposition of many independent thermal processes of alternate
excitation and relaxation of certain defects (e.g., dopant atoms or vacant
lattice sites). This means that 
every once in a while, a certain lattice site or a dopant
atom gets excited and it 
moves into a state of higher energy for some time, and then 
it relaxes back to the lower energy state until the next excitation.
Each one of these excitation/relaxation processes
can be modeled as a random telegraph signal 
with a different time constant 
$\theta$ (due to different physical/geometric characteristics)
and hence contributes a Lorentzian spectrum parametrized by $\theta$.
The superposition of these processes, whose
spectrum is given by the integral of
the Lorentzian function over a range of
values of $\theta$ (with a certain weight),
gives rise to the $1/f$
behavior over a wide range of frequencies.
To see this more concretely in the mathematical language,
a random telegraph
signal $X(t)$ is given by $X(t)=(-1)^{N(t)}$, where $N(t)$ is a Poisson
process of rate $\lambda$. It is a binary signal where the level $+1$
can symbolize excitation and the level $-1$ designates relaxation. Here
the dwell times between jumps are exponentially distributed. The
autocorrelation function is given by
\begin{eqnarray}
\left<X(t)X(t+\tau)\right>&=&\left<(-1)^{N(t)+N(t+\tau)}\right>\nonumber\\
&=&\left<(-1)^{N(t+\tau)-N(t)}\right>\nonumber\\
&=&\left<(-1)^{N(\tau)}\right>\nonumber\\
&=&e^{-\lambda\tau}\sum_{k=0}^\infty\frac{(\lambda\tau)^k}{k!}\cdot(-1)^k\nonumber\\
&=&e^{-\lambda\tau}\sum_{k=0}^\infty\frac{(-\lambda\tau)^k}{k!}\nonumber\\
&=&e^{-2\lambda\tau}
\end{eqnarray}
and so the spectrum is Lorentzian:
\begin{equation}
S_X(\omega)=\calF\{e^{-2\lambda|\tau|}\}=\frac{4\lambda}{\omega^2+4\lambda^2}
=\frac{2\theta}{1+(\omega\theta)^2},
\end{equation}
where the time constant is $\theta=1/2\lambda$ and the cutoff frequency is
$\omega_c=2\lambda$. Now, calculating the integral
$$\int_{\theta_{\min}}^{\theta_{\max}}\mbox{d}\theta
\cdot g(\theta)\cdot\frac{2\theta}{1+(\omega\theta)^2}$$
with $g(\theta)=1/\theta$, yields a composite spectrum
that is proportional to 
$$\frac{1}{\omega}\tan^{-1}(\omega\theta_{\max})-
\frac{1}{\omega}\tan^{-1}(\omega\theta_{\min}).$$
For $\omega \ll 1/\theta_{\max}$, using
the approximation $\tan^{-1}(x)\approx x$ ($|x|\ll 1$), this is approximately 
a constant. 
For $\omega \gg 1/\theta_{\min}$, using the approximation 
$\tan^{-1}(x)\approx \frac{\pi}{2}-\frac{1}{x}$ ($|x|\gg 1$), this is 
approximately proportional to $1/\omega^2$. In between, in the range
$1/\theta_{\max}\ll \omega \ll 1/\theta_{\min}$ (assuming that
$1/\theta_{\max}\ll 1/\theta_{\min}$), the behavior is accordiong to
$$\frac{1}{\omega}\left(\frac{\pi}{2}-\frac{1}{\omega\theta_{\max}}\right)-\theta_{\min}=
\frac{1}{\omega}\left(\frac{\pi}{2}-\frac{1}{\omega\theta_{\max}}-\omega\theta_{\min}\right)\approx
\frac{\pi}{2\omega},$$
which is the $1/f$ behavior in this wide
range of frequencies.
There are several theories why $g(\theta)$
should be inversely proportional to $\theta$, 
but the truth is that they are not perfect, and
the issue of $1/f$ noise is not yet perfectly (and universally) understood.

\vspace{0.2cm}

\noindent
{\bf Shot noise}
in electronic devices consists of unavoidable random statistical
fluctuations of the electric current in an electrical conductor. Random
fluctuations are inherent when current flows, as the current is a flow of
discrete charges (electrons). The the derivation of the spectrum of shot
noise is normally studied in courses on random procsses. 

\vspace{0.2cm}

\noindent
{\bf Burst noise}
consists of sudden step--like transitions between two or more
levels (non-Gaussian), as high as several hundred microvolts, at random and
unpredictable times. Each shift in offset voltage or current lasts for several
milliseconds, and the intervals between pulses tend to be in the audio range
(less than 100 Hz), leading to the term popcorn noise for the popping or
crackling sounds it produces in audio circuits.

\vspace{0.2cm}

\noindent
{\bf Avalanche noise}
is the noise produced when a junction diode is operated at the
onset of avalanche breakdown, a semiconductor junction phenomenon in which
carriers in a high voltage gradient develop sufficient energy to dislodge
additional carriers through physical impact, creating ragged current flows.

\newpage
\section{A Brief Touch on Information Theory$^*$}

\subsection{Introduction -- What is Information Theory About?}

Our last topic in this course is about a very brief and concise description on
the relation between statistical physics and information theory, a research
field pioneered by Claude Elwood Shannon (1917--2001), whose seminal paper
``A Mathematical Theory of Communications'' (1948) has established the
landmark of this field. 

In a nutshell, information theory is a science that focuses on the
{\it fundamental limits}, on the one hand, the 
{\it achievable performance}, on the other hand, concerning various information
processing tasks, including most notably:
\begin{enumerate}
\item Data compression (lossless/lossy).
\item Error correction coding (coding for protection against errors due to
channel noise.
\item Encryption.
\end{enumerate}
There are also additional tasks of information processing
that are considered to belong under the umbrella of information theory,
like: signal detection, estimation (parameter
estimation, filtering/smoothing, prediction), information embedding, process simulation,
extraction of random bits, information relaying, and more.

Core information theory, which is called {\it Shannon
theory} in the jargon of the professionals, is about {\it coding theorems}.
It is associated with the development of computable formulas that
characterize the best performance that can possibly be achieved in these
information processing tasks under some (usually simple) assumptions on 
the probabilistic models that govern the data, the channel noise, the side
information, the jammers if
applicable, etc.
While in most cases, this theory does not suggest constructive communication
systems, it certainly provides insights concerning the features that an
optimal (or nearly optimal) communication system must have.
Shannon theory serves, first and foremost, as the theoretical basis for
modern digital communication engineering. That being said,
much of the modern research activity in 
information theory evolves, not only around
Shannon theory, but also on the never-ending efforts to develop methodologies (mostly, specific code
structures and algorithms) for designing very efficient communication systems,
which hopefully come close to the bounds and the fundamental performance
limits.

But the scope of information theory
it is not limited merely to communication engineering: it plays a role
also in computer science, and many other disciplines, one of them is
thermodynamics and statistical mechanics, which is the focus of these
remaining lectures. Often, information--theoretic problems are well approached
from a statistical--mechanical point of view. We will taste this very briefly
in two examples of problems.

In the framework of this course, we will not delve into information theory too
deeply, but our purpose would be just 
to touch upon the
interface of these two fields. 

\subsection{Entropy in Info Theory and in Thermo \& Statistical
Physics}

Perhaps the first relation that crosses one's mind is that in
both fields there is a fundamental notion of {\it entropy}.
Actually, in information
theory, the term entropy was coined in the footsteps of the
thermodynamic/statistical--mechanical entropy. Along this course, we have seen
already three (seemingly) different forms of the entropy:
The first is the thermodynamic entropy defined, in its differential form as 
\begin{equation}
\delta S=\delta Q/T, 
\end{equation}
which was first introduced by Clausius in 1850.
The second is the statistical entropy 
\begin{equation}
S=k\ln\Omega, 
\end{equation}
which was defined by Boltzmann
in 1872. The third is yet another formula for the entropy -- the
Gibbs formula for the entropy of the canonical ensemble:
\begin{equation}
S=-k\sum_x P(x)\ln P(x)=-k\left<\ln P(x)\right>,
\end{equation}
which we encountered in Chapter 3.

It is virtually impossible to miss the functional resemblance between the last
form above and the information--theoretic entropy, a.k.a.\ the {\it Shannon
entropy}, which is simply 
\begin{equation}
H= -\sum_xP(x)\log_2 P(x)
\end{equation}
namely, the same expression as above exactly, just without the factor $k$ 
and with the basis of the logarithm being
2 rather than $e$. So they simply differ by an immaterial constant factor.
Indeed, this clear analogy was recognized already by Shannon and von
Neumann. The well--known anecdote on this tells that von Neumann advised
Shannon to adopt this term because it would provide him with {\it ``... a
great edge in debates because nobody really knows what entropy is anyway.''}

What is the information--theoretic meaning of entropy? It turns out that it
has many information--theoretic meanings, but the most fundamental one
concerns {\it optimum
compressibility} of data. Suppose that we have a bunch of i.i.d.\ random
variables, $x_1,x_2,\ldots,x_N$, taking values in a discrete set, say, the
components of the microstate in a quantum system of non--interacting
particles, and we want to represent the microstate information digitally (in
bits) as compactly as possible, without losing any information -- in other
words, we require the ability to fully
reconstruct the data from the compressed binary representation. How short can
this binary representation be?

Let us look at the following example. Suppose that each $x_i$ takes values in the set $\{A,B,C,D\}$,
independently with probabilities
$$P(A)=1/2;~~~~P(B)=1/4;~~~P(C)=1/8;~~~P(D)=1/8.$$
Clearly, when translating the letters into bits, the naive approach would be
to say the following: we have 4 letters, so it takes 2 bits to distinguish between them,
by mapping, say lexicographically, as follows:
$$A\to 00;~~~~B\to 01;~~~C\to 10;~~~D\to 11.$$
This would mean representing the list of $x$'s using 2 bits per--symbol.
Very simple. But is this the best thing one can do?

It turns out that the answer is negative. Intuitively, if we can assign
variable--length codewords to the various letters, using shorter codewords
for more probably symbols and longer codewords for the less frequent ones,
we might be able to gain something. In our example, $A$ is most probable,
while $C$ and $D$ are the least probable, so how about the following solution:
$$A\to 0;~~~~B\to 10;~~~C\to 110;~~~D\to 111.$$
Now the average number of bits per symbol is:
$$\frac{1}{2}\cdot 1+\frac{1}{4}\cdot 2+\frac{1}{8}\cdot 3+\frac{1}{8}\cdot
3=1.75.$$
We have improved the average bit rate by 12.5\%. Splendid. Is this now the best one can
do or can we improve even further?

It turns out that this time, the answer is affirmative.
Note that in this solution, each letter has a probability of the form
$2^{-\ell}$ ($\ell$ -- positive integer) and the length of the assigned
codeword is exactly $\ell$ (for $A$, $\ell=1$, for $B$ -- $\ell=2$, and for
$C$ and $D$, $\ell=3$). In other words, the length of the codeword for each
letter is the negative logarithm of its probability, so the average number of
bits per symbol is $\sum_{x\in\{A,B,C,D\}} P(x)[-\log_2 P(x)]$, which is
exactly  the entropy $H$ of the information source. One of the
basic coding theorems of information theory tells
us that we cannot compress to any coding rate below
the entropy and still expect to be able to reconstruct the $x$'s perfectly. 
But why is this true?

We will not get into a rigorous proof of this statement, but we will make an
attempt to give a statistical--mechanical insight into it. Consider the microstate
$\bx=(x_1,\ldots,x_N)$ and let us think of the probability function
\begin{equation}
P(x_1,\ldots,x_N)=\prod_{i=1}^N P(x_i)=\exp\left\{-(\ln
2)\sum_{i=1}^N\log_2[1/P(x_i)]\right\}
\end{equation}
as an instance of the canonical ensemble at inverse temperature
$\beta=\ln 2$,
where Hamiltonian is additive, namely,
$\calE(x_1,\ldots,x_N)=\sum_{i=1}^N\epsilon(x_i)$, with
$\epsilon(x_i)=-\log_2 P(x_i)$.
Obviously, $Z(\beta)=Z(\ln 2)=1$, so the free energy is exactly zero here.
Now, by the weak law of large numbers, for most realizations of the microstate $\bx$, 
\begin{equation}
\frac{1}{N}\sum_{i=1}^N\epsilon(x_i) \approx \left<\epsilon(x_i)\right>=
\left<-\log_2 P(x_i)\right>=H,
\end{equation}
so the average `internal energy' is $NH$.
It is safe to consider instead the corresponding {\it microcanonical ensemble}, which
is equivalent as far as macroscopic averages go. 
In the microcanonical
ensemble, we would then have:
\begin{equation}
\frac{1}{N}\sum_{i=1}^N\epsilon(x_i) = H
\end{equation}
for {\it every} realization of $\bx$. How many bits would it take us to
represent $\bx$ in this microcanonical ensemble? Since all $\bx$'s are
equiprobable in the microcanonical ensemble, we assign to all $\bx$'s binary codewords of the
same length, call it $L$. In order to have a one--to--one mapping between
the set of accessible $\bx$'s and binary strings of representation, $2^L$, which is the number of
binary strings of length $L$, should be no less than the number of microstates
$\{\bx\}$ of the microcanonical ensemble. Thus,
\begin{equation}
\label{converse}
L\ge\log_2\bigg|\left\{\bx:~\sum_{i=1}^N\epsilon(x_i) =
NH\right\}\bigg|\dfn \log_2\Omega(H),
\end{equation}
but the r.h.s.\ is exactly related (up to a constant factor) 
to Boltzmann's entropy associated with
`internal energy' at the level of $NH$. Now, observe that the free energy of
the original,
canonical ensemble, which is zero, is related to the entropy $\ln\Omega(H)$
via the Legendre relation $\ln Z\approx\ln\Omega-\beta E$, which is
\begin{equation}
0=\ln Z(\ln 2)\approx\ln\Omega(H)-NH\ln 2
\end{equation}
and so,
\begin{equation}
\ln\Omega(H)\approx NH\ln 2
\end{equation}
or
\begin{equation}
\log_2\Omega(H)\approx NH,
\end{equation}
and therefore, by (\ref{converse}):
\begin{equation}
L\ge \log_2\Omega(H)\approx NH,
\end{equation}
which means that the length of the binary representation essentially cannot be less than
$NH$, namely, a compression rate of $H$ bits per component of $\bx$. 
So we have seen that the entropy has a very concrete information--theoretic
meaning, and in fact, it is not the only one, but we will not delve into this
any further here. 

\subsection{Statistical Physics of Optimum Message Distributions}

We next study another, 
very simple paradigm of a communication system, studied by Reiss
\cite{Reiss69} and Reiss and
Huang \cite{RH71}. The analogy and the parallelism to the basic concepts of
statistical mechanics, that were introduced earlier, will be
quite evident from the choice of the notation, which is deliberately chosen
to correspond to that of analogous physical quantities.

Consider a continuous--time communication system that
includes a noiseless channel, with capacity
\begin{equation}
C=\lim_{E\to\infty}\frac{\log_2 M(E)}{E},
\end{equation}
where $M(E)$ is the number of distinct messages 
(and $\log_2$ of this is the number of bits)
that can be transmitted
over a time interval of $E$ seconds.
Over a duration of $E$
seconds, $L$ information symbols are conveyed, so that the average transmission
time per symbol is $\sigma=E/L$ seconds per symbol.
In the absence of any constraints on the structure of the encoded messages,
$M(E)=r^L=r^{E/\sigma}$, where $r$ is the channel input--output alphabet size.
Thus, $C=(\log r)/\sigma$ bits per second.

Consider now the thermodynamic limit of $L\to\infty$. Suppose that the $L$
symbols of duration $E$ form $N$ {\it words}, where by `word', we
mean a certain variable--length string of channel symbols. The average
transmission time per
word is then $\epsilon=E/N$. Suppose further that the code defines a
certain set of word transmission times: Word number $i$ takes $\epsilon_i$
seconds to transmit. What is the optimum allocation of word probabilities
$\{P_i\}$
that would support full utilization of the channel capacity? Equivalently,
given the probabilities $\{P_i\}$, what are the optimum transmission times
$\{\epsilon_i\}$? For simplicity, we will assume that $\{\epsilon_i\}$ are all
distinct. Suppose that each word appears $N_i$ times in the entire message.
Denoting $\bN=(N_1,N_2,\ldots)$, $P_i=N_i/N$, and $\bP=(P_1,P_2,\ldots)$,
the total number of messages pertaining to a given $\bN$ is
\begin{equation}
\Omega(\bN)=\frac{N!}{\prod_i N_i!}\exe \exp\{N\cdot
H(\bP)\}
\end{equation}
where $H(\bP)$ is the Shannon entropy pertaining to the probability
distribution $\bP$. Now,
\begin{equation}
M(E)=\sum_{\bN:~\sum_iN_i\epsilon_i=E}\Omega(\bN).
\end{equation}
This sum is dominated by the maximum term, namely, the maximum--entropy
assignment of relative frequencies
\begin{equation}
P_i=\frac{e^{-\beta\epsilon_i}}{Z(\beta)}
\end{equation}
where $\beta > 0$ is a Lagrange multiplier chosen such that
$\sum_iP_i\epsilon_i=\epsilon$, which gives
\begin{equation}
\epsilon_i=-\frac{\ln[P_iZ(\beta)]}{\beta}.
\end{equation}
For $\beta=1$, this is in agreement with our earlier observation that the optimum
message length assignment in variable--length lossless data compression
is according to the negative logarithm of the probability. 

Suppose now that $\{\epsilon_i\}$ are kept fixed and consider
a small perturbation in $P_i$, denoted $\dd P_i$. Then
\begin{align}
\dd \epsilon&=\sum_i\epsilon_i\dd P_i\nonumber\\
&=-\frac{1}{\beta}\sum_i(\dd P_i)\ln [P_iZ(\beta)]\nonumber\\
&=-\frac{1}{\beta}\sum_i(\dd P_i)\ln P_i
-\frac{1}{\beta}\sum_i(\dd P_i)\ln Z(\beta)\nonumber\\
&=-\frac{1}{\beta}\sum_i(\dd P_i)\ln P_i\nonumber\\
&=\frac{1}{k\beta}\dd\left(-k\sum_iP_i\ln P_i\right)\nonumber\\
&\dfn T\dd s,
\end{align}
where we have defined $T=1/(k\beta)$ and $s=-k\sum_iP_i\ln
P_i$. The free energy per particle is given by
\begin{equation}
f=\epsilon-Ts=-kT\ln Z,
\end{equation}
which is related to the redundancy of the code. 

In \cite{Reiss69}, there is also an extension
of this setting to the case where $N$ is not fixed, with correspondence to the
grand---canonical ensemble. However, we will not include it here.

\end{document}

%% file: collision.pstex_t
\begin{picture}(0,0)%
\includegraphics{collision.pstex}%
\end{picture}%
\setlength{\unitlength}{3947sp}%
\begingroup\makeatletter\ifx\SetFigFont\undefined%
\gdef\SetFigFont#1#2#3#4#5{%
  \reset@font\fontsize{#1}{#2pt}%
  \fontfamily{#3}\fontseries{#4}\fontshape{#5}%
  \selectfont}%
\fi\endgroup%
\begin{picture}(5483,1626)(214,-73)
\put(1504,1125){\makebox(0,0)[lb]{\smash{{\SetFigFont{10}{12.0}{\rmdefault}{\mddefault}{\itdefault}{$A$}%
}}}}
\put(2462,304){\makebox(0,0)[lb]{\smash{{\SetFigFont{10}{12.0}{\rmdefault}{\mddefault}{\itdefault}{$\vec{v}_2$}%
}}}}
\put(2781,1307){\makebox(0,0)[lb]{\smash{{\SetFigFont{10}{12.0}{\rmdefault}{\mddefault}{\itdefault}{$\vec{v}_1$}%
}}}}
\put(545,1125){\makebox(0,0)[lb]{\smash{{\SetFigFont{10}{12.0}{\rmdefault}{\mddefault}{\itdefault}{$2b$}%
}}}}
\put(2918,851){\makebox(0,0)[lb]{\smash{{\SetFigFont{10}{12.0}{\rmdefault}{\mddefault}{\itdefault}{$c$}%
}}}}
\put(4104,532){\makebox(0,0)[lb]{\smash{{\SetFigFont{10}{12.0}{\rmdefault}{\mddefault}{\itdefault}{$B$}%
}}}}
\put(5473,532){\makebox(0,0)[lb]{\smash{{\SetFigFont{10}{12.0}{\rmdefault}{\mddefault}{\itdefault}{$2a$}%
}}}}
\end{picture}%

%% file: leak.pstex_t
\begin{picture}(0,0)%
\includegraphics{leak.pstex}%
\end{picture}%
\setlength{\unitlength}{3947sp}%
\begingroup\makeatletter\ifx\SetFigFont\undefined%
\gdef\SetFigFont#1#2#3#4#5{%
  \reset@font\fontsize{#1}{#2pt}%
  \fontfamily{#3}\fontseries{#4}\fontshape{#5}%
  \selectfont}%
\fi\endgroup%
\begin{picture}(3590,2024)(323,-1473)
\put(2534,-1328){\makebox(0,0)[lb]{\smash{{\SetFigFont{9}{10.8}{\rmdefault}{\mddefault}{\itdefault}{$P_1 > P_2$}%
}}}}
\put(701,-961){\makebox(0,0)[lb]{\smash{{\SetFigFont{9}{10.8}{\rmdefault}{\mddefault}{\itdefault}{$\rho_1$}%
}}}}
\put(1501,-961){\makebox(0,0)[lb]{\smash{{\SetFigFont{9}{10.8}{\rmdefault}{\mddefault}{\itdefault}{$P_1$}%
}}}}
\put(2368,-961){\makebox(0,0)[lb]{\smash{{\SetFigFont{9}{10.8}{\rmdefault}{\mddefault}{\itdefault}{$\rho_2$}%
}}}}
\put(3268,-961){\makebox(0,0)[lb]{\smash{{\SetFigFont{9}{10.8}{\rmdefault}{\mddefault}{\itdefault}{$P_2$}%
}}}}
\put(2335,-94){\makebox(0,0)[lb]{\smash{{\SetFigFont{9}{10.8}{\rmdefault}{\mddefault}{\itdefault}{area $A$}%
}}}}
\put(1158,-318){\makebox(0,0)[lb]{\smash{{\SetFigFont{9}{10.8}{\rmdefault}{\mddefault}{\itdefault}{net flow}%
}}}}
\end{picture}%

%% file: schottky.pstex_t
\begin{picture}(0,0)%
\includegraphics{schottky.pstex}%
\end{picture}%
\setlength{\unitlength}{3947sp}%
\begingroup\makeatletter\ifx\SetFigFont\undefined%
\gdef\SetFigFont#1#2#3#4#5{%
  \reset@font\fontsize{#1}{#2pt}%
  \fontfamily{#3}\fontseries{#4}\fontshape{#5}%
  \selectfont}%
\fi\endgroup%
\begin{picture}(2972,1922)(1289,-1472)
\end{picture}%

%% file: legendre.pstex_t
\begin{picture}(0,0)%
\includegraphics{legendre.pstex}%
\end{picture}%
\setlength{\unitlength}{3947sp}%
\begingroup\makeatletter\ifx\SetFigFont\undefined%
\gdef\SetFigFont#1#2#3#4#5{%
  \reset@font\fontsize{#1}{#2pt}%
  \fontfamily{#3}\fontseries{#4}\fontshape{#5}%
  \selectfont}%
\fi\endgroup%
\begin{picture}(6474,4515)(289,-3749)
\put(3689,-552){\makebox(0,0)[lb]{\smash{{\SetFigFont{9}{10.8}{\rmdefault}{\mddefault}{\itdefault}{$\calL_E(\beta;V,N)[S]$}%
}}}}
\put(1852,-2022){\makebox(0,0)[lb]{\smash{{\SetFigFont{9}{10.8}{\rmdefault}{\mddefault}{\itdefault}{$V\to P$}%
}}}}
\put(2260,-3695){\makebox(0,0)[lb]{\smash{{\SetFigFont{9}{10.8}{\rmdefault}{\mddefault}{\itdefault}{$\calL_A(a;B,C)[f]\dfn\inf_{a\ge 0}[a\cdot A-f(a,B,C)]$}%
}}}}
\put(2465,-552){\makebox(0,0)[lb]{\smash{{\SetFigFont{9}{10.8}{\rmdefault}{\mddefault}{\itdefault}{$E\to \beta=\frac{1}{kT}$}%
}}}}
\put(2551,464){\makebox(0,0)[lb]{\smash{{\SetFigFont{9}{10.8}{\familydefault}{\mddefault}{\updefault}{Micro-canonical ensemble}%
}}}}
\put(2401,164){\makebox(0,0)[lb]{\smash{{\SetFigFont{9}{10.8}{\rmdefault}{\mddefault}{\itdefault}{$S(E,V,N)=k\ln\Omega(E,V,N)$}%
}}}}
\put(3893,-2022){\makebox(0,0)[lb]{\smash{{\SetFigFont{9}{10.8}{\rmdefault}{\mddefault}{\itdefault}{$N\to\mu$}%
}}}}
\put(2220,-2348){\makebox(0,0)[lb]{\smash{{\SetFigFont{9}{10.8}{\rmdefault}{\mddefault}{\itdefault}{$\calL_V(P;\beta,N)[F]$}%
}}}}
\put(5159,-2307){\makebox(0,0)[lb]{\smash{{\SetFigFont{9}{10.8}{\rmdefault}{\mddefault}{\itdefault}{$\calL_N(-\mu;T,V)[F]$}%
}}}}
\put(4126,-3061){\makebox(0,0)[lb]{\smash{{\SetFigFont{9}{10.8}{\rmdefault}{\mddefault}{\itdefault}{$J(T,V,\mu)=kT\ln\Xi(T,V,\mu)=PV$}%
}}}}
\put(4426,-2761){\makebox(0,0)[lb]{\smash{{\SetFigFont{9}{10.8}{\familydefault}{\mddefault}{\updefault}{Grand-canonical ensemble}%
}}}}
\put(376,-3061){\makebox(0,0)[lb]{\smash{{\SetFigFont{9}{10.8}{\rmdefault}{\mddefault}{\itdefault}{$G(T,P,N)=-kT\ln Y_N(T,P)=\mu N$}%
}}}}
\put(1051,-2761){\makebox(0,0)[lb]{\smash{{\SetFigFont{9}{10.8}{\familydefault}{\mddefault}{\updefault}{Gibbs ensemble}%
}}}}
\put(2401,-1486){\makebox(0,0)[lb]{\smash{{\SetFigFont{9}{10.8}{\rmdefault}{\mddefault}{\itdefault}{$F(T,V,N)=-kT\ln Z_N(T,V)$}%
}}}}
\put(2701,-1186){\makebox(0,0)[lb]{\smash{{\SetFigFont{9}{10.8}{\familydefault}{\mddefault}{\updefault}{Canonical ensemble}%
}}}}
\end{picture}%

%% file: combinatorics.pstex_t
\begin{picture}(0,0)%
\includegraphics{combinatorics.pstex}%
\end{picture}%
\setlength{\unitlength}{3947sp}%
\begingroup\makeatletter\ifx\SetFigFont\undefined%
\gdef\SetFigFont#1#2#3#4#5{%
  \reset@font\fontsize{#1}{#2pt}%
  \fontfamily{#3}\fontseries{#4}\fontshape{#5}%
  \selectfont}%
\fi\endgroup%
\begin{picture}(4869,570)(379,-123)
\end{picture}%

%% file: springs2d.pstex_t
\begin{picture}(0,0)%
\includegraphics{springs2d.pstex}%
\end{picture}%
\setlength{\unitlength}{3947sp}%
\begingroup\makeatletter\ifx\SetFigFont\undefined%
\gdef\SetFigFont#1#2#3#4#5{%
  \reset@font\fontsize{#1}{#2pt}%
  \fontfamily{#3}\fontseries{#4}\fontshape{#5}%
  \selectfont}%
\fi\endgroup%
\begin{picture}(4497,3054)(236,-2468)
\end{picture}%

%% file: spheres.pstex_t
\begin{picture}(0,0)%
\includegraphics{spheres.pstex}%
\end{picture}%
\setlength{\unitlength}{3947sp}%
\begingroup\makeatletter\ifx\SetFigFont\undefined%
\gdef\SetFigFont#1#2#3#4#5{%
  \reset@font\fontsize{#1}{#2pt}%
  \fontfamily{#3}\fontseries{#4}\fontshape{#5}%
  \selectfont}%
\fi\endgroup%
\begin{picture}(2689,2574)(364,-2023)
\end{picture}%

%% file: spinglass1.pstex_t
\begin{picture}(0,0)%
\includegraphics{spinglass1.pstex}%
\end{picture}%
\setlength{\unitlength}{3947sp}%
\begingroup\makeatletter\ifx\SetFigFont\undefined%
\gdef\SetFigFont#1#2#3#4#5{%
  \reset@font\fontsize{#1}{#2pt}%
  \fontfamily{#3}\fontseries{#4}\fontshape{#5}%
  \selectfont}%
\fi\endgroup%
\begin{picture}(3437,3662)(195,-2960)
\end{picture}%

%% file: fgraph.pstex_t
\begin{picture}(0,0)%
\includegraphics{fgraph.pstex}%
\end{picture}%
\setlength{\unitlength}{3947sp}%
\begingroup\makeatletter\ifx\SetFigFont\undefined%
\gdef\SetFigFont#1#2#3#4#5{%
  \reset@font\fontsize{#1}{#2pt}%
  \fontfamily{#3}\fontseries{#4}\fontshape{#5}%
  \selectfont}%
\fi\endgroup%
\begin{picture}(2738,1963)(514,-1568)
\put(882,303){\makebox(0,0)[lb]{\smash{{\SetFigFont{6}{7.2}{\rmdefault}{\mddefault}{\itdefault}{$f$}%
}}}}
\put(1159,-666){\makebox(0,0)[lb]{\smash{{\SetFigFont{6}{7.2}{\rmdefault}{\mddefault}{\itdefault}{$\epsilon_1^*$}%
}}}}
\put(1754,-666){\makebox(0,0)[lb]{\smash{{\SetFigFont{6}{7.2}{\rmdefault}{\mddefault}{\itdefault}{$\epsilon_2^*$}%
}}}}
\put(1754,283){\makebox(0,0)[lb]{\smash{{\SetFigFont{6}{7.2}{\rmdefault}{\mddefault}{\itdefault}{$T=0$}%
}}}}
\put(2169,-152){\makebox(0,0)[lb]{\smash{{\SetFigFont{6}{7.2}{\rmdefault}{\mddefault}{\itdefault}{$T_1$}%
}}}}
\put(2684,-211){\makebox(0,0)[lb]{\smash{{\SetFigFont{6}{7.2}{\rmdefault}{\mddefault}{\itdefault}{$T_2 > T_1$}%
}}}}
\put(3000,-566){\makebox(0,0)[lb]{\smash{{\SetFigFont{6}{7.2}{\rmdefault}{\mddefault}{\itdefault}{$T_2$}%
}}}}
\put(3159,-745){\makebox(0,0)[lb]{\smash{{\SetFigFont{6}{7.2}{\rmdefault}{\mddefault}{\itdefault}{$\epsilon$}%
}}}}
\end{picture}%

%% file: cw1.pstex_t
\begin{picture}(0,0)%
\includegraphics{cw1.pstex}%
\end{picture}%
\setlength{\unitlength}{3947sp}%
\begingroup\makeatletter\ifx\SetFigFont\undefined%
\gdef\SetFigFont#1#2#3#4#5{%
  \reset@font\fontsize{#1}{#2pt}%
  \fontfamily{#3}\fontseries{#4}\fontshape{#5}%
  \selectfont}%
\fi\endgroup%
\begin{picture}(5692,1506)(589,-1014)
\put(2870,-300){\makebox(0,0)[lb]{\smash{{\SetFigFont{7}{8.4}{\rmdefault}{\mddefault}{\itdefault}{$m$}%
}}}}
\put(1055,-949){\makebox(0,0)[lb]{\smash{{\SetFigFont{7}{8.4}{\rmdefault}{\mddefault}{\itdefault}{$y=m$}%
}}}}
\put(2092,-189){\makebox(0,0)[lb]{\smash{{\SetFigFont{7}{8.4}{\rmdefault}{\mddefault}{\itdefault}{$y=\tanh(\beta Jm)$}%
}}}}
\put(3976, 75){\makebox(0,0)[lb]{\smash{{\SetFigFont{7}{8.4}{\rmdefault}{\mddefault}{\itdefault}{$y=\tanh(\beta Jm)$}%
}}}}
\put(5219,-372){\makebox(0,0)[lb]{\smash{{\SetFigFont{7}{8.4}{\rmdefault}{\mddefault}{\itdefault}{$+m_0$}%
}}}}
\put(4374,-248){\makebox(0,0)[lb]{\smash{{\SetFigFont{7}{8.4}{\rmdefault}{\mddefault}{\itdefault}{$-m_0$}%
}}}}
\put(4174,-973){\makebox(0,0)[lb]{\smash{{\SetFigFont{7}{8.4}{\rmdefault}{\mddefault}{\itdefault}{$y=m$}%
}}}}
\put(6050,-304){\makebox(0,0)[lb]{\smash{{\SetFigFont{7}{8.4}{\rmdefault}{\mddefault}{\itdefault}{$m$}%
}}}}
\end{picture}%

%% file: cw2.pstex_t
\begin{picture}(0,0)%
\includegraphics{cw2.pstex}%
\end{picture}%
\setlength{\unitlength}{3947sp}%
\begingroup\makeatletter\ifx\SetFigFont\undefined%
\gdef\SetFigFont#1#2#3#4#5{%
  \reset@font\fontsize{#1}{#2pt}%
  \fontfamily{#3}\fontseries{#4}\fontshape{#5}%
  \selectfont}%
\fi\endgroup%
\begin{picture}(5727,2058)(214,-1434)
\put(2684,-1026){\makebox(0,0)[lb]{\smash{{\SetFigFont{10}{12.0}{\rmdefault}{\mddefault}{\itdefault}{$m$}%
}}}}
\put(1414,509){\makebox(0,0)[lb]{\smash{{\SetFigFont{10}{12.0}{\rmdefault}{\mddefault}{\itdefault}{$\psi(m)$}%
}}}}
\put(4986,-1065){\makebox(0,0)[lb]{\smash{{\SetFigFont{10}{12.0}{\rmdefault}{\mddefault}{\itdefault}{$+m_0$}%
}}}}
\put(3625,-1065){\makebox(0,0)[lb]{\smash{{\SetFigFont{10}{12.0}{\rmdefault}{\mddefault}{\itdefault}{$-m_0$}%
}}}}
\put(5680,-967){\makebox(0,0)[lb]{\smash{{\SetFigFont{10}{12.0}{\rmdefault}{\mddefault}{\itdefault}{$m$}%
}}}}
\put(4501,464){\makebox(0,0)[lb]{\smash{{\SetFigFont{10}{12.0}{\rmdefault}{\mddefault}{\itdefault}{$\psi(m)$}%
}}}}
\end{picture}%

%% file: cw3.pstex_t
\begin{picture}(0,0)%
\includegraphics{cw3.pstex}%
\end{picture}%
\setlength{\unitlength}{3947sp}%
\begingroup\makeatletter\ifx\SetFigFont\undefined%
\gdef\SetFigFont#1#2#3#4#5{%
  \reset@font\fontsize{#1}{#2pt}%
  \fontfamily{#3}\fontseries{#4}\fontshape{#5}%
  \selectfont}%
\fi\endgroup%
\begin{picture}(5295,1867)(514,-1502)
\put(1683,199){\makebox(0,0)[lb]{\smash{{\SetFigFont{9}{10.8}{\rmdefault}{\mddefault}{\itdefault}{$\psi(m)$}%
}}}}
\put(2674,-1075){\makebox(0,0)[lb]{\smash{{\SetFigFont{9}{10.8}{\rmdefault}{\mddefault}{\itdefault}{$m$}%
}}}}
\put(627,-1176){\makebox(0,0)[lb]{\smash{{\SetFigFont{9}{10.8}{\rmdefault}{\mddefault}{\itdefault}{$-m_0(B)$}%
}}}}
\put(5568,-1077){\makebox(0,0)[lb]{\smash{{\SetFigFont{9}{10.8}{\rmdefault}{\mddefault}{\itdefault}{$m$}%
}}}}
\put(4482,242){\makebox(0,0)[lb]{\smash{{\SetFigFont{9}{10.8}{\rmdefault}{\mddefault}{\itdefault}{$\psi(m)$}%
}}}}
\put(4929,-1167){\makebox(0,0)[lb]{\smash{{\SetFigFont{9}{10.8}{\rmdefault}{\mddefault}{\itdefault}{$+m_0(B)$}%
}}}}
\end{picture}%

%% file: cw4.pstex_t
\begin{picture}(0,0)%
\includegraphics{cw4.pstex}%
\end{picture}%
\setlength{\unitlength}{3947sp}%
\begingroup\makeatletter\ifx\SetFigFont\undefined%
\gdef\SetFigFont#1#2#3#4#5{%
  \reset@font\fontsize{#1}{#2pt}%
  \fontfamily{#3}\fontseries{#4}\fontshape{#5}%
  \selectfont}%
\fi\endgroup%
\begin{picture}(5663,3238)(289,-2698)
\put(2817,-2197){\makebox(0,0)[lb]{\smash{{\SetFigFont{11}{13.2}{\rmdefault}{\mddefault}{\itdefault}{$-1$}%
}}}}
\put(2397,-135){\makebox(0,0)[lb]{\smash{{\SetFigFont{11}{13.2}{\rmdefault}{\mddefault}{\itdefault}{$+1$}%
}}}}
\put(2817,352){\makebox(0,0)[lb]{\smash{{\SetFigFont{11}{13.2}{\rmdefault}{\mddefault}{\itdefault}{$m(\beta,B)$}%
}}}}
\put(5684,-1149){\makebox(0,0)[lb]{\smash{{\SetFigFont{11}{13.2}{\rmdefault}{\mddefault}{\itdefault}{$B$}%
}}}}
\put(3726,-485){\makebox(0,0)[lb]{\smash{{\SetFigFont{11}{13.2}{\rmdefault}{\mddefault}{\itdefault}{$T>T_c$}%
}}}}
\put(3341,-904){\makebox(0,0)[lb]{\smash{{\SetFigFont{11}{13.2}{\rmdefault}{\mddefault}{\itdefault}{$T=T_c$}%
}}}}
\put(1559,-382){\makebox(0,0)[lb]{\smash{{\SetFigFont{11}{13.2}{\rmdefault}{\mddefault}{\itdefault}{$T<T_c$}%
}}}}
\put(2825,-1845){\makebox(0,0)[lb]{\smash{{\SetFigFont{11}{13.2}{\rmdefault}{\mddefault}{\itdefault}{$-m_0$}%
}}}}
\put(2399,-510){\makebox(0,0)[lb]{\smash{{\SetFigFont{11}{13.2}{\rmdefault}{\mddefault}{\itdefault}{$+m_0$}%
}}}}
\end{picture}%

%% file: rc.pstex_t
\begin{picture}(0,0)%
\includegraphics{rc.pstex}%
\end{picture}%
\setlength{\unitlength}{3947sp}%
\begingroup\makeatletter\ifx\SetFigFont\undefined%
\gdef\SetFigFont#1#2#3#4#5{%
  \reset@font\fontsize{#1}{#2pt}%
  \fontfamily{#3}\fontseries{#4}\fontshape{#5}%
  \selectfont}%
\fi\endgroup%
\begin{picture}(3269,1149)(178,-523)
\put(1904, 19){\makebox(0,0)[lb]{\smash{{\SetFigFont{9}{10.8}{\rmdefault}{\mddefault}{\itdefault}{$R$}%
}}}}
\put(349, 51){\makebox(0,0)[lb]{\smash{{\SetFigFont{9}{10.8}{\rmdefault}{\mddefault}{\itdefault}{$I_r(t)$}%
}}}}
\put(3229, 22){\makebox(0,0)[lb]{\smash{{\SetFigFont{9}{10.8}{\rmdefault}{\mddefault}{\itdefault}{$C$}%
}}}}
\end{picture}%

%% file: linearresponse.pstex_t
\begin{picture}(0,0)%
\includegraphics{linearresponse.pstex}%
\end{picture}%
\setlength{\unitlength}{3947sp}%
\begingroup\makeatletter\ifx\SetFigFont\undefined%
\gdef\SetFigFont#1#2#3#4#5{%
  \reset@font\fontsize{#1}{#2pt}%
  \fontfamily{#3}\fontseries{#4}\fontshape{#5}%
  \selectfont}%
\fi\endgroup%
\begin{picture}(4952,3138)(139,-2398)
\put(1757,269){\makebox(0,0)[lb]{\smash{{\SetFigFont{9}{10.8}{\rmdefault}{\mddefault}{\itdefault}{$\left<W(t)\right>$}%
}}}}
\put(4249,-681){\makebox(0,0)[lb]{\smash{{\SetFigFont{9}{10.8}{\rmdefault}{\mddefault}{\itdefault}{$\left<W(\infty)\right>$}%
}}}}
\put(151, 40){\makebox(0,0)[lb]{\smash{{\SetFigFont{9}{10.8}{\rmdefault}{\mddefault}{\itdefault}{$\left<W(0^-)\right>$}%
}}}}
\put(4970,-1698){\makebox(0,0)[lb]{\smash{{\SetFigFont{9}{10.8}{\rmdefault}{\mddefault}{\itdefault}{$t$}%
}}}}
\put(544,-1304){\makebox(0,0)[lb]{\smash{{\SetFigFont{9}{10.8}{\rmdefault}{\mddefault}{\itdefault}{$\epsilon$}%
}}}}
\put(1725,-1501){\makebox(0,0)[lb]{\smash{{\SetFigFont{9}{10.8}{\rmdefault}{\mddefault}{\itdefault}{$F(t)$}%
}}}}
\end{picture}%

%% file: rc2.pstex_t
\begin{picture}(0,0)%
\includegraphics{rc2.pstex}%
\end{picture}%
\setlength{\unitlength}{3947sp}%
\begingroup\makeatletter\ifx\SetFigFont\undefined%
\gdef\SetFigFont#1#2#3#4#5{%
  \reset@font\fontsize{#1}{#2pt}%
  \fontfamily{#3}\fontseries{#4}\fontshape{#5}%
  \selectfont}%
\fi\endgroup%
\begin{picture}(3096,1131)(264,-815)
\put(2207,-397){\makebox(0,0)[lb]{\smash{{\SetFigFont{7}{8.4}{\rmdefault}{\mddefault}{\itdefault}{$C$}%
}}}}
\put(3173,-397){\makebox(0,0)[lb]{\smash{{\SetFigFont{7}{8.4}{\rmdefault}{\mddefault}{\itdefault}{$R$}%
}}}}
\put(328,-397){\makebox(0,0)[lb]{\smash{{\SetFigFont{7}{8.4}{\rmdefault}{\mddefault}{\itdefault}{$V(t)$}%
}}}}
\put(988,-92){\makebox(0,0)[lb]{\smash{{\SetFigFont{7}{8.4}{\rmdefault}{\mddefault}{\itdefault}{$R$}%
}}}}
\end{picture}%

%% file: transline.pstex_t
\begin{picture}(0,0)%
\includegraphics{transline.pstex}%
\end{picture}%
\setlength{\unitlength}{3947sp}%
\begingroup\makeatletter\ifx\SetFigFont\undefined%
\gdef\SetFigFont#1#2#3#4#5{%
  \reset@font\fontsize{#1}{#2pt}%
  \fontfamily{#3}\fontseries{#4}\fontshape{#5}%
  \selectfont}%
\fi\endgroup%
\begin{picture}(4190,1684)(461,-1433)
\put(4224,-625){\makebox(0,0)[lb]{\smash{{\SetFigFont{5}{6.0}{\rmdefault}{\mddefault}{\itdefault}{$V_r(t)$}%
}}}}
\put(4051,-176){\makebox(0,0)[lb]{\smash{{\SetFigFont{5}{6.0}{\rmdefault}{\mddefault}{\itdefault}{$R$}%
}}}}
\put(832,-141){\makebox(0,0)[lb]{\smash{{\SetFigFont{5}{6.0}{\rmdefault}{\mddefault}{\itdefault}{$R$}%
}}}}
\put(555,-625){\makebox(0,0)[lb]{\smash{{\SetFigFont{5}{6.0}{\rmdefault}{\mddefault}{\itdefault}{$V_r(t)$}%
}}}}
\put(2528,-1179){\makebox(0,0)[lb]{\smash{{\SetFigFont{5}{6.0}{\rmdefault}{\mddefault}{\itdefault}{$L$}%
}}}}
\end{picture}%